\documentclass[fleqn,usenatbib]{mnras}

\usepackage{newtxtext,newtxmath}
\usepackage{xspace}
\usepackage[T1]{fontenc}
\usepackage[normalem]{ulem}
\DeclareRobustCommand{\VAN}[3]{#2}
\let\VANthebibliography\thebibliography
\def\thebibliography{\DeclareRobustCommand{\VAN}[3]{##3}\VANthebibliography}

\newcommand{\teff}{\mbox{$T_{\rm eff}$}\xspace}

\newcommand{\logg}{\mbox{$\log(g)$}\xspace}

\newcommand{\vmic}{\mbox{$v_\mathrm{mic}$}\xspace}
\newcommand{\vsini}{\mbox{$v \sin i$}\xspace}

\usepackage{graphicx}	
\usepackage{amsmath}	
\usepackage{xcolor}
\usepackage{subcaption}
\usepackage{array}
\usepackage{makecell}
\usepackage{csvsimple}
\usepackage{pdflscape}
\usepackage{adjustbox}

\title[HARPS Abundances with Korg I]{HARPS Abundances with Korg I: 22 Element Abundances for 426 Red Giant Stars}

\author[Aquilina et al.]{
Sarah E. Aquilina$^{1}$,
Melissa Ness$^{1}$,
Adam J. Wheeler$^{2}$ and
Sven Buder$^{1}$\\
$^{1}$School of Astronomy and Astrophysics, The Australian National University, ACT 2611, Australia\\
$^{2}$Center for Computational Astrophysics, Flatiron Institute, 162 5th Avenue, Manhattan, NY, USA\\
}

\date{Accepted XXX. Received YYY; in original form ZZZ}

\pubyear{2025}

\begin{document}
\label{firstpage}
\pagerange{\pageref{firstpage}--\pageref{lastpage}}
\maketitle

\begin{abstract}
Large stellar surveys have revealed the global abundance structure of the Milky Way, but small high-fidelity spectral samples offer a critical complement of nucleosynthetic depth. We aim to access the encoded information in an ensemble of abundances by leveraging highest-quality spectra. We used HARPS spectra (R=115,000) to determine (\teff,~\logg,~[M/H],~\vmic~and~\vsini) and 22 element abundances (Na,~Mg,~Al,~Si,~K,~Ca,~Sc,~Ti,~V,~Cr,~Fe,~Ni,~Zn,~Sr,~Y,~Zr,~Mo,~Ba,~La,~Ce,~Nd~and~Eu) for 426 red giant stars at a median internal precision of $\sim$0.02~dex evaluated from analysing repeat observations of a subset of stars. Stellar parameters and line-by-line abundances were obtained using the modern spectral synthesis code \textsc{korg} -- the first time it has been used for HARPS. Comparisons with the literature reveal good overall agreement. A minor ~0.1 dex offset in metallicity and specific discrepancies in individual element abundances are attributed to local thermal equilibrium assumptions and inaccuracies in atomic data. We show that 22 individual elements can be collapsed into a generative 6-parameter latent-variable model of shared enrichment patterns expressed in different per-star fractions; this model accurately generates the abundances with a median $\chi_{reduced}^2 = 6$. We report element gradients with respect to selected elements from different nucleosynthetic families. These gradients are a measure of inter-element production efficiencies and indicate multiple r-process production sites. Our analysis shows that abundances occupy a low-dimensional subspace, but joint (gradient-based) information encodes nucleosynthetic signatures. We have developed a \textsc{korg}-pipeline to apply across evolutionary states on high-resolution spectra to provide our precision catalogue to serve as empirical constraints on chemical evolution and as a set of benchmark red giant abundance measurements.
\end{abstract}

\begin{keywords}
catalogues -- stars: abundances -- techniques: spectroscopic
\end{keywords}

\section{Introduction}
\label{Section: introduction}
The origin of the elements has been extensively studied in the last century but remains one of the primary questions of astrophysics. The foundations for our current understanding of galactic nucleosynthesis were established by \citet{burbidge_synthesis_1957}, who considered the synthesis of elements in stars to explain their formation at different timescales. Inter-element abundance correlations as well as spatial abundance gradients, alongside theoretical expectations of production timescales, have since been used to infer production sites \citep[eg.,][]{griffith_kpm_2024,nugent_rprocess_2025,rastinejad_grb_2024} and uncover the formation history of galaxies \citep[eg.,][and references therein]{buder_galahdr4_2024,ness_homogeneity_2022,griffith_2025,hayden_apogee_2015,nidever_2014,hunt_milky_2025}. Element abundance measurements have been shown to be correlated with stellar age and subsequently enabled age derivation using abundances as chemical clocks \citep{Iliadis_stars_2007,solderblom_2010,hayden_galah_clocks_2022,casali_gaia_clocks_2020,molero_2025}, and the stars that deviate from the typical abundance trends have enabled identification of anomalous populations and accreted stars \citep[eg.,][]{dasilva_metalpoor_2023,horta_halo_2023,giribaldi_chronology_2023,horta_alpha_2025,buck_impact_2023}. The plethora of abundance information we can derive from stars that connects to production timescales is key in reconciling chemical evolution and stellar evolution theory and observations. 
\\\\
Measurements from stellar spectra provide the essential data to reconstruct the formation and evolution of the Milky Way, as well as understand the origin of elements. Stellar parameters express the evolutionary state of a star, and the element abundances encode the conditions of its birth environment as well as subsequent internal evolution. Decoding the imprint of the enrichment history in individual stars requires precise chemical abundances for many elements that trace the ensemble of sources that contribute across Galactic history. Sources include type Ia supernovae \citep{keegans_1asne_2023}, core collapse supernovae \citep{nomoto_ccsne_2006}, Asymptotic Giant Branch (AGB) stars of a range of masses \citep{busso_agb_review_1999,herwig_agb_2005}, neutron star mergers \citep{kobayashi_neutron_2023}, magnetars \citep{patel_magnetar_2025} and Wolf-Rayet stars \citep{peng_wolfrayet_2019} -- many with mass and metallicity dependent yields -- and all producing different elements in varying  fractions over time-- notably each element is produced by multiple sources, with exception \citep[see][]{kobayashi_origin_2020,kobayashi_review_2025}. Element accessibility is determined by spectral signal-to-noise ratio (SNR), wavelength coverage, resolution and stellar temperature. Abundance derivation relies on absorption features with line depths that exceed the noise level. Line depths are characterized by well-determined laboratory measurements of oscillator strengths and excitation potentials, which govern line strengths under the physical conditions of the stellar photosphere.
\\\\
Large surveys (eg., GALAH, APOGEE) have provided vast samples of data and element measurements at moderate resolutions ($R=28,000$ and 22,500, respectively). These enable chemodynamical mapping of the Galaxy, the study of global trends, and uncover large samples of anomalous populations \citep{sayeed_Li_2024}. Surveys such as the Terra Hunting Experiment (THE) \citep{thompson_the_2016} instead prioritise exquisite repeat spectra of 40 stars at $R=115,000$ to search for variability induced by earth-sized planets. Benchmark studies of the highest-quality available spectra are key to complement the large survey data and provide the most informative, high-precision set of measurements for the detailed study of nucleosynthesis. Precision line-by-line abundance analysis further aid in refining theoretical models and their inputs, which have largely unknown uncertainties \citep{blanco_caveats_2019,jofre_blackbox_2017}. This deepens our understanding of the variability of abundances across different wavelength regions.
\\\\
Small high–fidelity stellar instruments like The High Accuracy Radial Velocity Planet Searcher \citep[HARPS,][]{barbieri_esoharps_2023} have not been fully realised as an experiment of element origins, despite its spectral resolution of $R=115,000$ and wide sky coverage in the solar neighbourhood. The high resolution and SNR required for exoplanet analyses typical of HARPS spectra \citep[eg.,][]{pepe_harps_2000,adibekyan_abundance_2016,adibekyan_chemical_2012,Bedell_2018,leonardi_taste_2024,borsato_small_2024} in general enables greater precision abundance measurements than moderate resolution large survey data. The HARPS spectrograph covers 3800-6900~\AA{} and in principle includes measurable features of the following elements: odd-Z elements Na, Al and K; $\alpha$-elements Mg, Si, Ca and Ti; iron-peak elements Co, Cu, Sc, V, Cr, Mn, Fe, Ni and Zn; s-process elements Sr, Y, Zr, Ba, La, Ce and Nd; r-process elements Mo and Eu. This is a greater range of neutron capture elements than APOGEE \citep{majewski_apache_2017} which measures elements (C, C I, N, O, Na, Mg, Al, Si, S, K, Ca, Ti, Ti II, V, Cr, Mn, Fe, Co, Ni, and Ce) from the infrared at median precision of 5\%. GALAH \citep{buder_galahdr4_2024} measures up to 30 elements in the optical, including many shared by HARPS, although at a moderate precision of 15\%. Therefore, even though the HARPS database is relatively small, its inclusion of a wide range of elements at a high precision provides a potentially high chemical discriminating power. 
\\\\
The balance between diversity in elements and precision by HARPS allows the construction of a comprehensive and self-consistent catalogue. There are a number of existing studies that infer element abundances from HARPS spectra with the goal of understanding galactic nucleosynthesis and galaxy formation \citep{adibekyan_chemical_2012, adibekyan_kgiants_2015, delgado_mena_chemical_2017}. However, there remains substantial opportunity to use derived element abundances from databases of $R=10,000$ stars as benchmark samples, and to understand the origin of the element \citep{mead_regression_2025}. 
\\\\
There are several approaches to derive stellar parameters using both photometry \citep[eg.,][]{Mucciarelli_2021}, spectroscopy \citep[eg.,][]{blanco_ispec_2014} and empirical relations \citep[eg.,][]{heiter_teff_2015} as well as for calculating element abundances through equivalent widths and synthetic spectral fitting \citep[eg.,][and references therein]{jofre_alpha_2015}. Each of these methods utilise different assumptions and approximations that typically produce inconsistent results \citep{jofre_alpha_2015,jofre_blackbox_2017,blanco_caveats_2019}. Further contradictions introduced by distinct stellar parameter and abundance scales across surveys can be resolved with data-driven models \citep{ness_cannon_2015,horta_lux_2025}. Such models create a common label scale given stars in common between surveys but rely on high-fidelity reference sets of stars and, with exception, their accuracy is inherited from theoretical models. It is therefore beneficial to analyse the highest quality data available to produce catalogues of consistently and carefully derived stellar abundances and parameters. This subsequently provides not only reference measurements for training data-driven models, but serves studies of the origin of elements and understanding inaccuracies of theoretical stellar models. 
\\\\
In this paper, stellar  parameters and abundances are derived using the spectral synthesis code \textsc{korg} \citep{wheeler_korg_2022,wheeler_korg_2023}. This is the first public derivation of stellar parameters using \textsc{korg}. \textsc{korg} uses two common approximations in spectral synthesis to minimise computation time: local thermal equilibrium (LTE), and representing model atmospheres as one-dimensional. \textsc{korg} differs from other spectral synthesis codes through its use of automatic differentiation to fit spectra significantly faster (up to two orders of magnitude) than older spectral synthesis codes, and its updated atomic and molecular data with all absorption sources. 
\\\\
We report the stellar parameter (\teff, \logg, [M/H], \vmic and \vsini) and abundance measurements for 426 red giant stars in HARPS spectra from \citet{barbieri_esoharps_2023}. Our catalogue contains line-by-line and per star abundances of 22 elements--odd-Z elements Na, Al and K; $\alpha$-elements Mg, Si, Ca and Ti; iron-peak elements Sc, V, Cr, Fe, Ni and Zn; s-process elements Sr, Y, Zr, Ba, La, Ce and Nd; r-process elements Mo and Eu--including 9 neutron capture elements and 2 additional ionised element abundances (Ti II and Zr II) with uncertainties.
\\\\
We compared our $\alpha$- and iron-peak abundances for 103 stars in common with the Local Thermodynamic Equilibrium (LTE) analysis of \citet{adibekyan_kgiants_2015}. However, few studies of red giant stars contain abundances across a wide range of metallicities \citep[eg.,][]{adibekyan_kgiants_2015,bensby_kgiants_2010,silva_subgiant_2015,souto_cluster_2018,jofre_giants_2015}. \citet{jofre_giants_2015} and \citet{silva_subgiant_2015} both further focus primarily on Ba and do not include the abundances of other neutron capture elements. We therefore compared to the LTE main sequence abundances trends derived from equivalent widths in \citet{adibekyan_chemical_2012}, \citet{delgado_mena_chemical_2017} and \citet{bensby_dwarf_2014} as well as those derived from synthetic spectral fitting in \citet{battistini_peak_2015}, \citet{battistini_neutron_2016}, \citet{zhao_k_2016} and \citet{mishenina_mo_2019}. \citet{ mishenina_mo_2019} also contains abundances of red giant stars. All these studies use high-resolution observations from a range of instruments/surveys spanning $R=40,000$ to 120,000. The fusion phases of red giant stars indicate that their abundances should match that of main sequence stars for all elements in the catalogue except possibly Na, Al, and Mg. However, deviations between main sequence and giant stars may be attributed to non-evolutionary effects such as inaccuracies in stellar models including non-LTE (NLTE) effects, and line selection differences \citep{smiljanic_na_al_2016}.
\\\\
This paper is structured as follows. In Section \ref{sec:method} we describe our selection and processing of high-fidelity HARPS spectra for red giant stars including our selection of main sequence stars that will be analysed in series II of this program. Stellar parameters (\teff, \logg, [M/H], \vsini and \vmic) and line-by-line abundances for 22 elements were derived for each of the stars using the spectral synthesis code \textsc{korg}. The abundances of individual absorption features were combined to obtain a median abundance for each star. Per-star element abundance trends are grouped by their nucleosynthetic families (odd-Z, $\alpha$, iron-peak, light s-process, heavy s-process and r-process) and compared to literature in Section \ref{sec:results}. In Section \ref{sec:discussion} we analysed the gradients between elements and the element-by-element contribution from different production timescales. This includes reproducing the per star abundances of every element in all stars with a latent model to demonstrate the low dimensionality of chemical space. The relationship between elements is summarised in Section \ref{sec:concl} with suggested extensions and improvements to the methodology.

\section{Methods}
\label{Section: method}
\label{sec:method}
\subsection{Data Selection}
\label{sec:method-data selection}
\begin{figure}
	\includegraphics[width=\columnwidth]{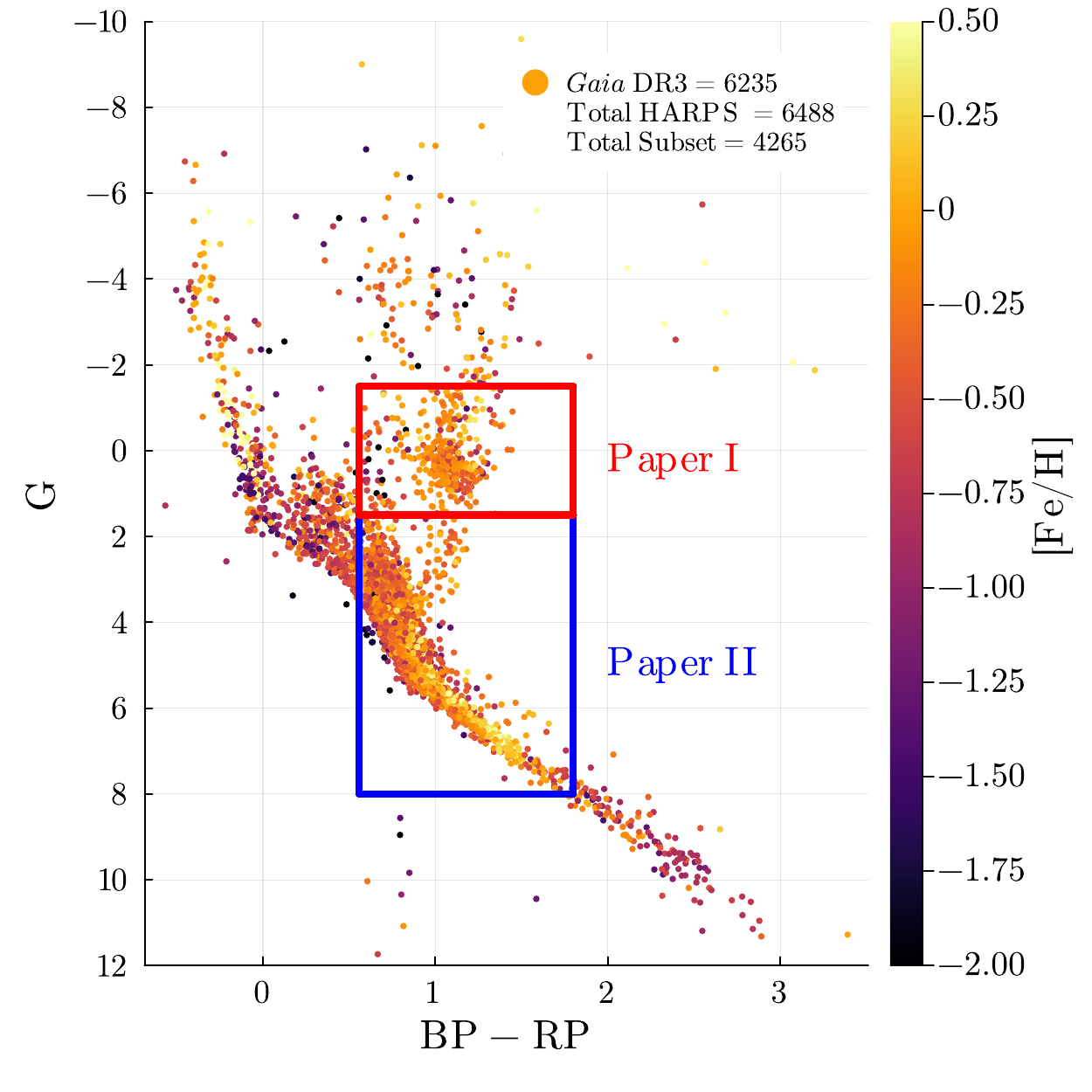}
    \caption{Colour-magnitude diagram of stars observed by HARPS using \textit{Gaia} DR3 colours and reddening corrections. Points are coloured by their corresponding [Fe/H] in \textit{Gaia} DR3. Selected red giant stars reside in the red box and FGK main sequence stars in the blue box from the 6488 HARPS stars in our crossmtached sample. The main sequence stars will be included in paper II of this series.}
    \label{fig:cmd selection}
\end{figure}

\begin{figure}
    \centering
    \begin{subfigure}[b]{0.45\textwidth}
        \centering
        \includegraphics[width=\textwidth]{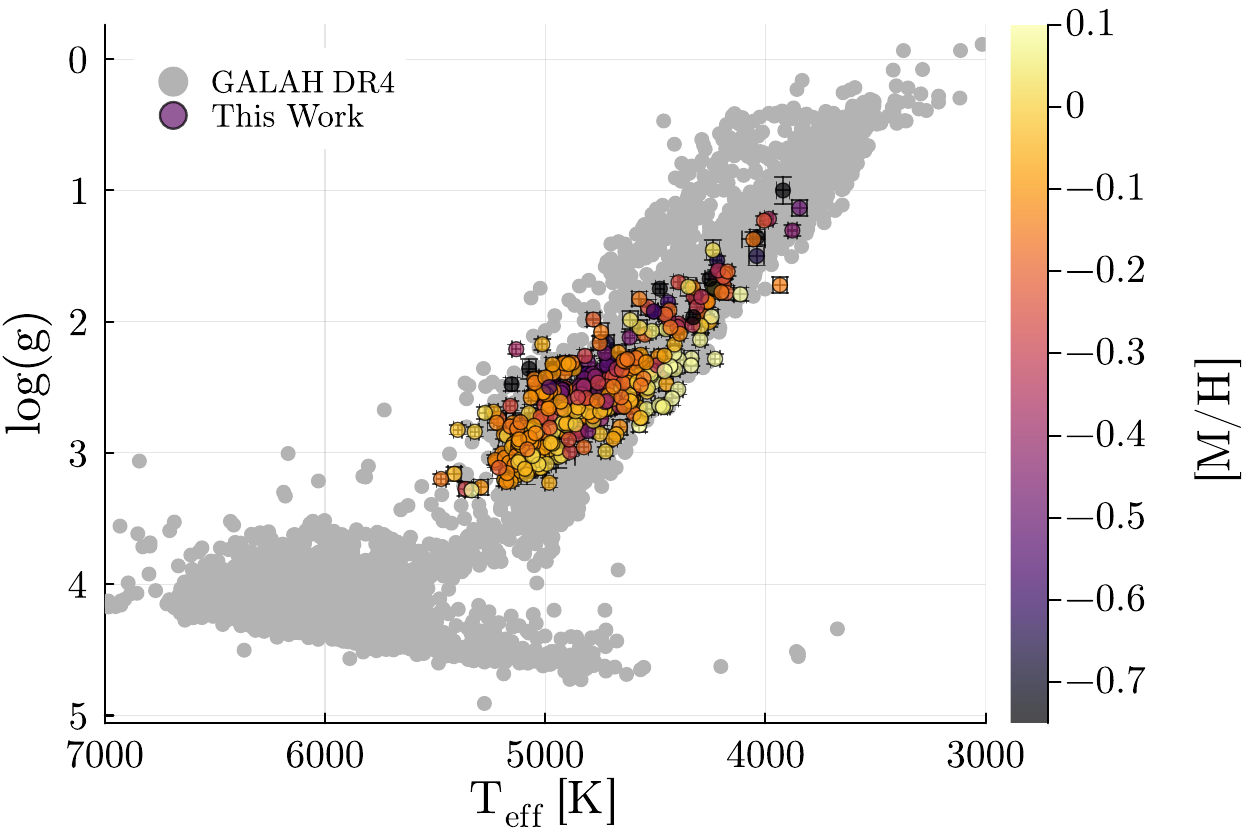}
    \end{subfigure}
    \hfill
    \begin{subfigure}[b]{0.45\textwidth}
        \centering
        \includegraphics[width=\textwidth]{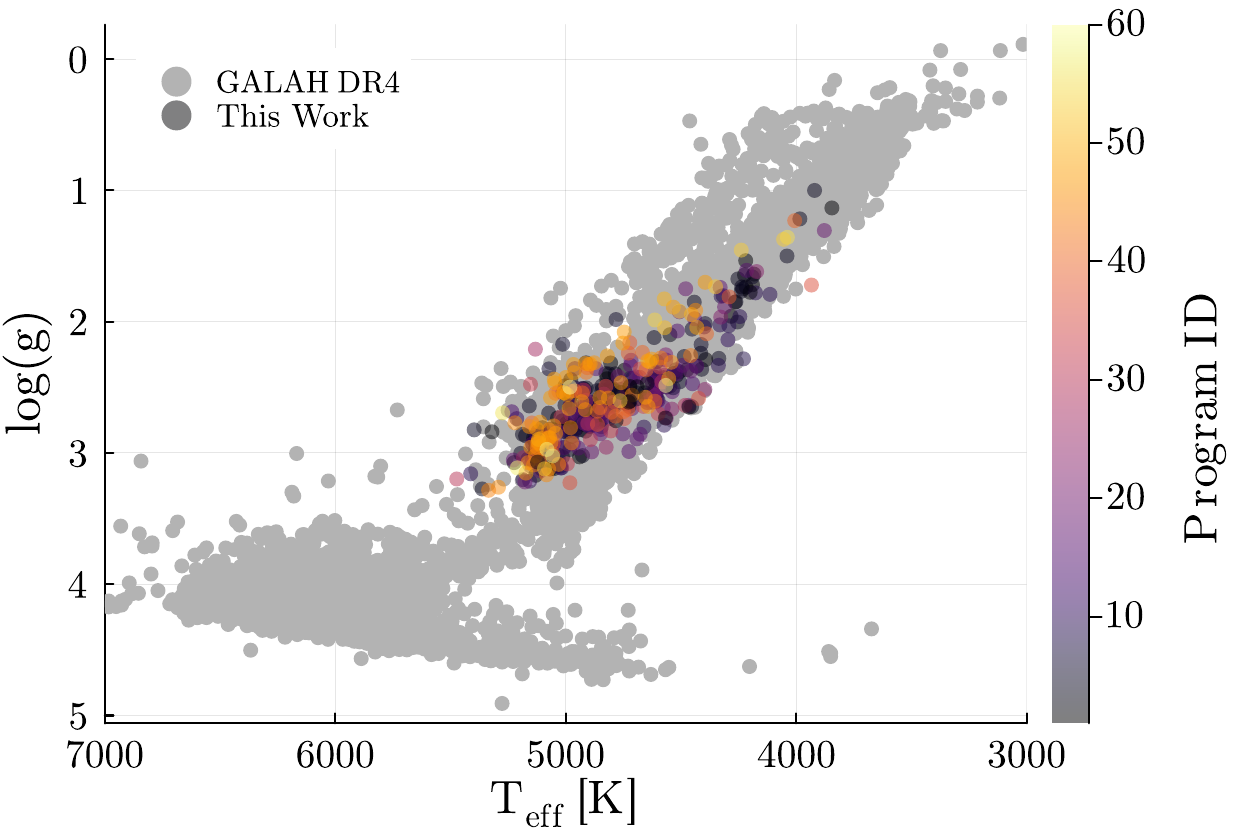}
    \end{subfigure}
    \caption{\textit{Top:} Kiel diagram showing our derived \logg, \teff and [M/H] for 426 red giant stars after selection cuts with the corresponding uncertainties. Parameters for unflagged GALAH DR4 stars are shown in grey to serve as comparison and validation these fall in appropriate \teff-\logg regions. \textit{Bottom:} Kiel diagram showing our derived \logg and \teff by the corresponding program ID of each star. There are 113 program IDs across our sample.} 
    \label{fig:kiel diagram}
\end{figure}

\begin{figure}
	\includegraphics[width=\columnwidth]{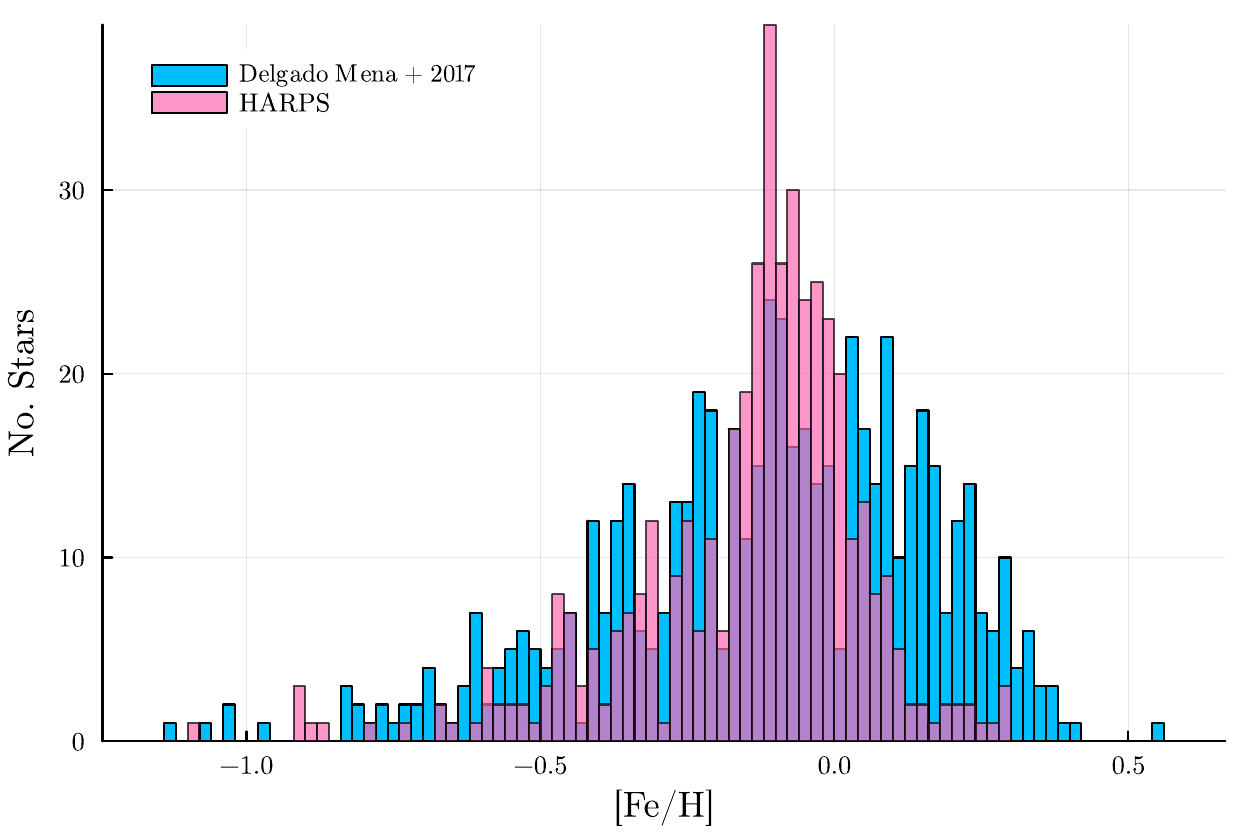}
    \caption{A comparison of our metallicity distribution of red giant stars observed by HARPS in pink to that of main sequence stars observed by HARPS in \citet{delgado_mena_chemical_2017} in blue. This highlights the fewer observed metal-rich ([Fe/H] $>$ 0.1~dex) red giant stars compared to the main sequence stars in HARPS.}
    \label{fig:mdf}
\end{figure}

\noindent
Stars were selected from the 289,843 HARPS observations of 6,488 unique objects collated by \citet{barbieri_esoharps_2023}. The spectra cover 3800~\AA{} to 6800~\AA{} for mostly main sequence stars in the solar neighbourhood with absolute V band magnitudes between 4 and 6. These stars were crossmatched by \citet{barbieri_esoharps_2023} with stars from SIMBAD \citep{wenger_simbad_2000} in a 30~arsecond radius, removing planets, extended sources, clusters, non–optical sources, and binaries. However, we performed our own crossmatch to both 2MASS \citep{Skrutskie_2mass_2006} and \textit{Gaia} DR3 \citep{gaia_2023} in a 10~arcsecond radius with \textsc{topcat} \citep{taylor_topcat_2011}. We also performed the crossmatch with a 2~arcsecond radius which provided 84 less stars. Given that the number of stars did not change significantly we decided to use a 10~arcsecond radius. To ensure that the crossmatched stars were the same we removed stars that had a B band magnitude in SIMBAD and BP magnitude in \textit{Gaia} DR3 that differed by more than 2. We crossmatched with GALAH DR3 \citep{buder_galah_2021} and DR4 \citep{buder_galahdr4_2024} but there were no stars in common with HARPS due to the different magnitude range of observable stars. Stars flagged as binaries and/or variable stars in SIMBAD were removed. Some stars were further flagged in \citet{barbieri_esoharps_2023} if their derived radial velocity was in disagreement with literature.
\\\\
From the above crossmatch, we have 940 stars. However, in this work, we focus on red giant stars and red clump candidates. These stars are sufficiently bright such that they have reliable observational measurements that allow us to obtain precise parameters and abundances. Choosing a small subset further minimises evolutionary changes to the observed abundance trends. We selected red giant stars from our crossmatched subset of HARPS spectra within BP-RP = 0.56 to 1.8 and absolute G band magnitudes from -1.5 to 1.5. \textit{Gaia} DR3 colours were used with reddening corrections applied. Figure \ref{fig:cmd selection} shows the colour-magnitude diagram with selected red giant stars in a red box. Selected FGK main sequence stars are in the blue box and will be analysed separately in the second paper of this series. We limited our selection to stars with a \textit{Gaia} DR3 effective temperature between 4500~K and 6000~K, but not all stars had a recorded temperature value. Stars without \textit{Gaia} temperatures were still included if they were inside our colour-magnitude selection cuts. These colour and temperature selection cuts identify 787 red giant stars from 113 different program IDs shown in Figure \ref{fig:kiel diagram}. This demonstrates that program IDs are not strongly associated with specific regions of stellar parameter space, with targets from different programs spanning the Kiel diagram. However, the primary goal of HARPS to measure radial velocities for exoplanet detection results in fewer red giant stars with [Fe/H] $> 0.1$~dex. This selection effect is demonstrated in Figure \ref{fig:mdf} where we show the metallicity distribution of our red giant stars from HARPS in pink and that of main sequence stars from HARPS in \citet{delgado_mena_chemical_2017} in blue.

\subsection{Radial Velocity Corrections and Combining Observations}
\label{sec:method-rv}
Every star has an average of 10 observations, each at $\mathrm{SNR} = 98$ that must be individually processed before being combined and continuum normalised. The catalogue by \cite{barbieri_esoharps_2023} contains radial velocity measurements but only those derived via the H$\alpha$ line are considered reliable. We calculated our own radial velocity corrections by generating per-star synthetic spectra with \textsc{korg} and cross-correlating that spectrum with the observed one. To create our synthetic spectrum with \textsc{korg} we used the \textit{Gaia} ESO line list \citep{heiter_linelist_2021}, interpolated MARCS model atmospheres \citep{gustafsson_grid_2008}, and \textit{Gaia} DR3 stellar parameters for each star. We subsequently produced a template synthetic spectrum of each star from 3800~\AA{} to 6800~\AA{}. It should be noted that CH absorption features in the \textit{Gaia} ESO line list are filtered by \textsc{korg} since we found presumably erroneous oscillator strengths ($\mathrm{log(gf)>-1.9}$) for these molecules unreasonably depressed the pseudo-continuum of the synthetic spectra. Any stars missing \textit{Gaia} stellar parameters were synthesised with solar parameters. In this procedure, \textsc{korg} and observed spectra were normalised by dividing by a version of themselves smoothed with a 1D Gaussian kernel of standard deviation 500 pixels. Smoothing ensured that the synthetic and observed spectra of each observation had a similar amplitude to perform a cross-correlation between 5000~\AA{} to 6800~\AA{} with radial velocities from $-250$~km/s to 250~km/s in 0.1 steps. Wavelengths below 5000~\AA{} were removed to exclude regions with large flux spikes that led to overestimated radial velocities compared to HARPS. The mean of a Gaussian fit to the cross-correlation was taken to be the radial velocity of the observation, which on average deviated from the HARPS radial velocity by up to 0.9 km/s.
\\\\
Radial velocity-corrected observations were interpolated onto a wavelength grid from approximately 4000~\AA{} to 6900~\AA{} with steps of 0.01 \AA. The lower limit of the wavelength grid was chosen to remove a large spectral feature that occurred in all observations, and the upper limit was determined by the observation with the largest radial velocity shift. The individual observations of each star were combined by summing the flux at each wavelength.

\subsection{Continuum Normalisation}
\label{sec:method-continuum}

\begin{figure}
    \centering
    \begin{subfigure}[b]{0.45\textwidth}
        \centering
        \includegraphics[width=\textwidth]{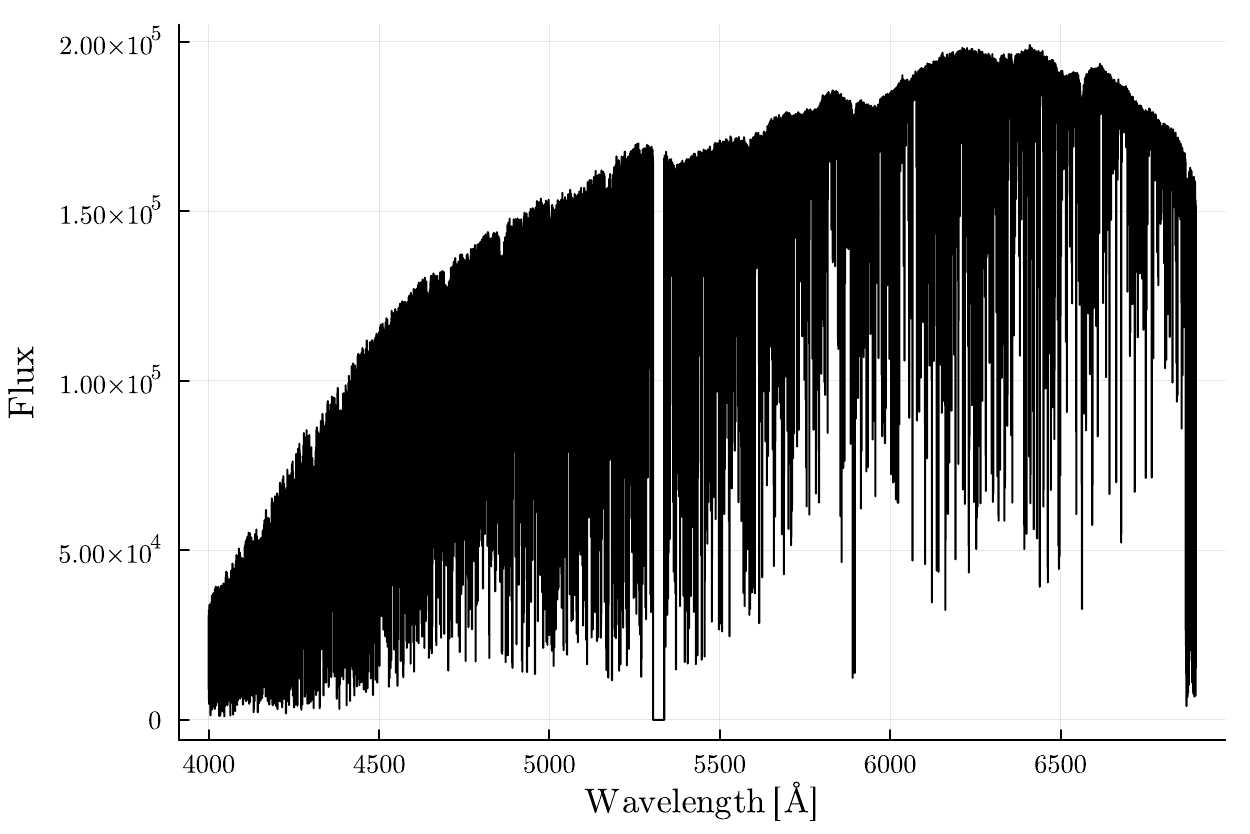}
        \label{fig:continuum plot}
    \end{subfigure}
    \hfill
    \begin{subfigure}[b]{0.45\textwidth}
        \centering
        \includegraphics[width=\textwidth]{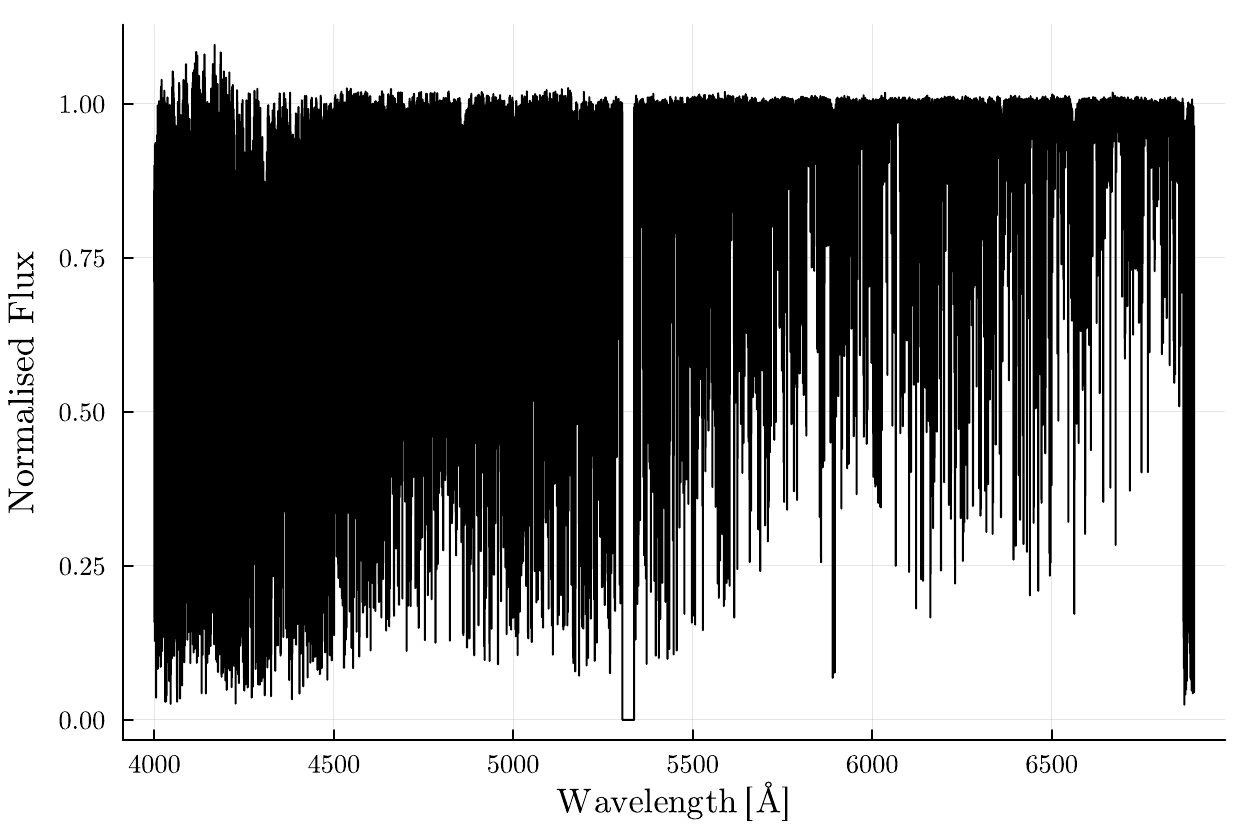}
        \label{fig:normalised spectra}
    \end{subfigure}
    \caption{\textit{Top:} Spectrum of $\lambda$~Pyxidis after stacking 5 observations. \textit{Bottom:} Continuum normalised spectrum of $\lambda$~Pyxidis using \textsc{suppnet}. There is a CCD gap from 5300 \AA{} to 5330 \AA{}.}
    \label{fig:continuum and norm}
\end{figure}

The combined spectrum of each star must be continuum normalised. We used \textsc{suppnet} \citep{tomasz_suppnet_2022} to perform our continuum normalisation since it can be applied directly to order-merged spectra and provides a consistent normalisation across our parameter space. \textsc{suppnet} is a fully convolutional deep neural network that predicts a pseudo-continuum by re-sampling the spectrum across sliding windows and calculating a weighted average over all predictions for each sample. The mean pseudo-continuum is further processed with a smoothing spline to account for noise. We chose to use the recommended sampling of 0.05 \AA{} and a smoothing of 1.0 since it performed well for both metal-poor and metal-rich stars. Figure \ref{fig:continuum and norm} shows the stacked spectrum and normalised spectrum for $\lambda$~Pyxidis. We estimated the normalised flux uncertainty per wavelength in every star by scaling the square root of the combined observed spectrum flux by the continuum function ($\mathrm{flux_{obs,\lambda}} / \mathrm{flux_{norm,\lambda}}$):

\begin{equation}
    \sigma_{\lambda} = \frac{\mathrm{flux_{norm,\lambda}}}{  \sqrt{\mathrm{flux_{obs,\lambda}}}}, 
\end{equation}

\noindent
where $\mathrm{flux_{obs,\lambda}}$ is the flux per wavelength of the combined observed spectrum in Analog to Digital Units and $\mathrm{flux_{norm,\lambda}}$ is the flux per wavelength of the normalised spectrum. 
\\\\
Some stars had hundreds of observations that when stacked produced SNR up to 2885. As we take Possion noise as our flux errors, this resulted in small uncertainties that made it difficult to optimise for stellar parameters and line-by-line abundances. We therefore inflated our flux errors for all stars when performing any optimisation by adding a factor of 0.003 in quadrature with $\sigma_{\lambda}$:

\begin{equation}
\label{eqn:inflated error}
    \sigma_{\mathrm{inflated},\lambda} = \sqrt{\sigma_{\lambda}^2 + 0.003^2}. 
\end{equation}

\noindent
This corresponds to a 0.003 standard deviation in continuum flux—a conservative estimate, given our measured 0.002 standard deviation across 0.1 \AA{} continuum regions.

\subsection{Stellar Parameters}
\label{sec:method-parameters}
\noindent
To determine the stellar parameters of \teff, \logg, [M/H], \vmic and \vsini of our 787 selected red giant stars, we used the synthetic spectral fitting in \textsc{korg}. Given the observed flux, flux errors and vacuum wavelengths as well as a line list, wavelength windows of fitting regions, an initial guess for each  parameter and resolution, we used \textsc{korg}'s \textsc{fit\_spectrum} method to fit a synthesised spectrum to the observed spectrum via $\chi^2$ minimisation with the limited memory Broyden–Fletcher–Goldfarb–Shanno (LBFGS) algorithm and automatic differentiation. We derived our stellar parameters from 0.5~\AA{} windows of 70 Fe I and 7 Fe II absorption features above 5000~\AA{} in the \textit{Gaia} ESO line list. These Fe absorption features are a subset of the 400 lines originally collated by \citet{griffith_untangling_2023} and was accessed via the \textsc{korg} repository. Fe lines from \citet{griffith_untangling_2023} with $\lambda < 5000$~\AA{} were excluded to avoid regions with line blanketing, and we removed Fe lines with poor fits in the majority of stars upon visual inspection. In Appendix \ref{appendix:line fits} we validated our choice of Fe lines by demonstrating their wide range of excitation potentials (0.86~eV to 6~eV) and equivalent widths (5~m\AA{} to 200~m\AA{}). The stellar parameters were fitted simultaneously across all Fe windows. We used 0.5~\AA{} windows to avoid automatic merging of nearby windows in \textsc{korg}, which in combination with the varying continuum-placement across our parameter space resulted in poor fits. Note that \textsc{korg}'s \textsc{fit\_spectrum} synthesises an additional $\pm 1$~\AA{} around all provided fitting windows. This allows users to validate the modelling of the continuum and line wings while optimising only the desired region.   
\\\\
The 77 Fe windows in each observed spectrum were simultaneously fitted to a corresponding \textsc{korg} spectrum in a 2 step optimisation process. The first step used \textsc{korg}'s \textsc{fit\_spectrum} algorithm to globally vary \teff, \logg, [M/H], \vsini, and \vmic with a fixed continuum. The second step performed the same optimisation but with a fixed \vmic, since we found this prevented the underestimation of [M/H] for stars known to have [M/H]$> 0.0$~dex. \vmic is calculated by inputting the previous iterations \teff and \logg into the equation from \citet{buder_galahdr4_2024}. In the absence of other abundance inputs, \textsc{korg} treats [M/H] as the abundance of all elements heavier than He. We therefore refer to [M/H] as the metallicity of a star.
\\\\
We successfully derived stellar parameters for 716 of our 787 selected red giant stars. The remaining 71 stars for which the stellar parameter optimisation failed were removed from our sample but are discussed in Section \ref{results:parameters}. From the stellar parameters of these 716 stars, we performed subsequent selection cuts listed below. These parameter-based cuts ensured we retained only high SNR stars with physically valid parameters and removed potential unidentified binaries. We obtained a final sample of 426 red giant stars after these cuts, which we considered to have reliable stellar parameters. We included these 426 stars in all subsequent analysis and no further selection cuts are made to our overall number of stars. 

\begin{itemize}
    \item Removal of unphysical stellar parameters using the Kiel diagram in Figure \ref{fig:kiel diagram}
    \item Removal of stars flagged as binaries or variable stars in SIMBAD (excluding wide binaries)
    \item SNR $\geq$ 100
    \item \vsini $<$ 15 $\mathrm{km}\,\mathrm{s}^{-1}$ to remove unidentified binaries.
    \item \vmic $<5$ $\mathrm{km}\,\mathrm{s}^{-1}$
\end{itemize}

\subsection{Line-by-line Abundances}
\label{sec:method-line abundances}
\noindent
Individual absorption features were selected via spectral synthesis for each of our 22 elements (including an additional 2 ionised abundances) using the \textsc{linemake} line list \citep{placco_linemake_2021} and \textit{Gaia} ESO line list. The \textsc{linemake} line list spans 4000~\AA{} to 4750~\AA{} and the \textit{Gaia} ESO linelist from 4750~\AA{} to 6800~\AA{}. Absorption features were chosen based on a \textsc{korg} spectrum and its corresponding gradient spectrum with respect to the desired element abundance $\left(\mathrm{\frac{dFlux}{dX}}\right)$ synthesised with the stellar parameters we derived for $\lambda$~Pyxidis. This allowed us to avoid saturated lines or blended lines where the gradient spectra did not correlate with abundance variation. Weak absorption features were removed by ensuring that the line had an identifiable trough in the \textsc{korg} spectrum and gradient spectrum as well as $\mathrm{\frac{dFlux}{dX}} < -0.01$~dex$^{-1}$. Blending was minimised by selecting absorption features that appeared within 0.1~\AA{} of the expected wavelength from the line list in both the \textsc{korg} and gradient spectrum as well as ensuring that no other features appeared within 0.05~\AA{} of the chosen lines. The latter requirement that no other lines appeared near the desired line was removed for elements that had no lines after these selection cuts. We used all absorption features for elements with less than 30 lines total in either line list. Lines with U or N flags in the \textit{Gaia} ESO line list were not used because the reliability of the atomic data was unknown or considered poor, respectively. However, a U flag Mg I line at 5712 \AA{} (vacuum rest wavelength) was used since no other Mg I lines were selected after cuts. The chosen U flag Mg I line has been reliably measured in GALAH DR4 \citep{buder_galahdr4_2024} with the same oscillator strength as that in the \textit{Gaia} ESO line list.
\\\\
Selected line abundances ([X/H]\footnote{$\mathrm{[X/H] = log_{10}(N_X/N_H)_* - log_{10}(N_X/N_H)_\odot}$. Solar abundances were taken from \citet{asplund_chemical_2021}.}) were measured individually in every star by synthesising a \textsc{korg} spectrum with our derived stellar parameters and either the \textsc{linemake} or \textit{Gaia} ESO line list depending on the wavelength of the desired line. The \textsc{korg} spectrum was fitted to the observed spectrum in 0.5~\AA{} windows around each line via $\chi^2$ minimisation with the LBFGS algorithm and automatic differentiation. The abundance of the desired element was allowed to vary while the stellar parameters (\teff, \logg, [M/H], \vmic and \vsini), and the continuum placement were fixed.
\\\\
Abundances of lines with significant blends or neighbouring strong lines were derived separately. We performed the optimisation for the element with the highest oscillator strength in the fitting window. The abundance of this element was fixed to the optimised value before performing the optimisation on the desired element. The optimisation was instead performed with a smaller window size if there is a neighbouring strong line that is not blended with the desired line. A combination of the two procedures was sometimes required to derive accurate line abundances. Note that this does not include elements affected by hyperfine structure splitting. For elements affected by hyperfine structure, Korg automatically models the splitting components provided in the \textit{Gaia} ESO line list.

\subsubsection{Line Selection Cuts}
\label{section:cayrel cuts}
The $\chi^2$ or residuals between theoretical spectra and observed spectra can be used to determine goodness of fit. We chose to use the rms to evaluate the goodness of fit for individual absorption features since the high SNR of some observations led to misleading $\chi^2$ values. Lines with poor fits (rms $\geq 0.1$), those that did not appear in the observed spectra of almost all stars, any line abundance trends enhanced above 1~dex or depleted less than $-1$~dex, as well as those with discontinuities in abundance trends with respect to temperature were removed. Table \ref{table: number of lines} contains the number of lines considered `good' for each element out of the measured lines. We measured abundances for an additional 26 elements not listed in Table \ref{table: number of lines} that were determined to have no absorption lines of sufficient quality. 
\\\\
Further selection cuts were made for individual stars to remove weak lines and/or lines where abundances ([X/Fe]) were $\pm 1$~dex.We used the Equation 7 from \citet{cayrel_1988} to obtain the minimum equivalent width per line given the SNR of each stellar spectrum. This required the full-width-half-maximum (FWHM) of each of our measured lines across all stars. The FWHM was obtained by fitting a Gaussian and a Voigt function. We preferentially used the FWHM of the Voigt function since this is a more accurate representation of absorption line profiles. However, nearby blends for $5\%$ of lines across all stars resulted in a better Gaussian fit, which we adopted instead. The equivalent width limits provided by the formula in \citet{cayrel_1988} were converted to per star abundance limits across all elements using \textsc{korg's} \textsc{ews\_to\_abundance} function. Measured abundances should be greater than this derived limit for detectability. An absorption feature was therefore considered weak if it had a measured abundance below the derived limit and removed. This quality selection criteria means that not all stars have the same number of lines per element, and that some stars may not have a recorded abundance for all elements. The number of stars with recorded abundances for each element are reported in Table \ref{table: number of lines}.
 
\begingroup{} 
    \setlength{\tabcolsep}{10pt} 
    \renewcommand{\arraystretch}{1.5} 
    \setlength{\extrarowheight}{2pt}    
    \begin{table}
        \centering
        \begin{tabular}{ c c c }  
            \hline 
            \textbf{Element} & \textbf{Number of Lines} & \textbf{Number of Stars}\\
            \hline     
            {Na I} & $1$ & $426$\\
            {Mg I} & $1$ & $253$\\
            {Al I} & $1$ & $426$ \\
            {Si I} & $3$ & $426$ \\
            {K I} & $1$  & $272$ \\
            {Ca I} & $4$ & $426$ \\
            {Sc I} & $3$ & $426$ \\
            {Ti I} & $14$ & $426$ \\
            {Ti II} & $1$ & $426$ \\
            {V I} & $7$ & $426$ \\
            {Cr II} & $1$ & $426$\\
            {Fe I} & $23$ & $426$ \\
            {Ni I} & $1$ & $426$ \\
            {Zn I} & $3$ & $426$ \\
            {Sr I} & $2$ & $403$ \\
            {Y II} & $5$ & $426$ \\
            {Zr I} & $4$ & $423$ \\
            {Zr II} & $3$ & $426$ \\
            {Mo I} & $2$ & $421$ \\
            {Ba II} & $2$ & $426$ \\
            {La II} & $1$ & $406$ \\
            {Ce II} & $4$ & $426$ \\
            {Nd II} & $2$ & $426$ \\
            {Eu II} & $2$ & $425$ \\
            
         \hline 
        \end{tabular}  
        \caption{Number of selected good lines used to determine the per star abundance for each element. Stars included fewer lines if the line was bellow the detection threshold (see Section \ref{section:cayrel cuts}) and/or had enhanced or depleted abundances. Some stars do not have abundances recorded for every element listed if all lines were removed. The number of stars per element with record abundances for out of the 426 red giant stars is included.}
        \label{table: number of lines}
    \end{table}
\endgroup{}

\noindent
\subsubsection{Correcting Systematic Offsets in line-by-line Abundances}
\label{sec:line corrections}
\begin{figure*}
    \centering
    \includegraphics[width =\textwidth]{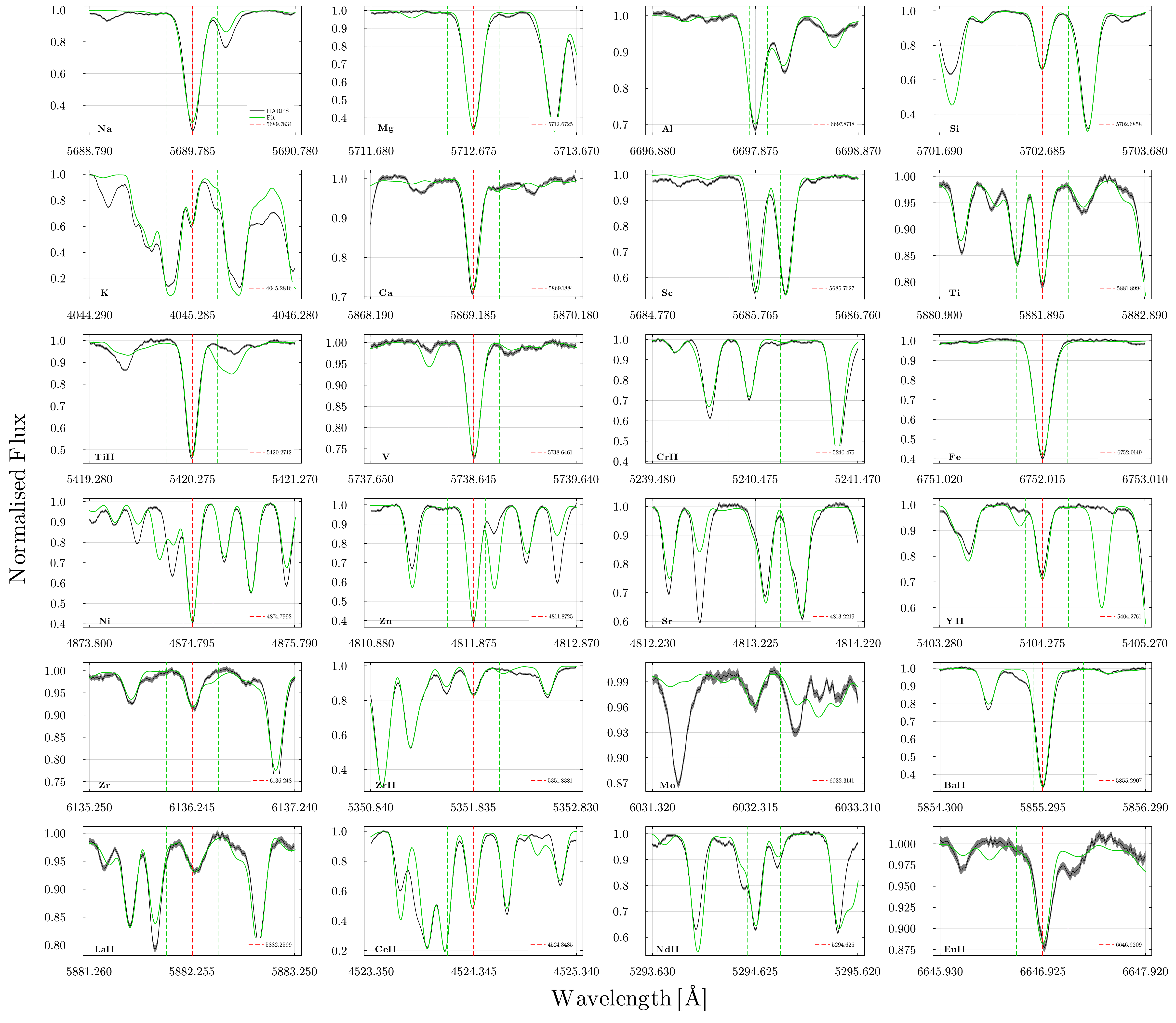}
    \caption{Reference line fits in $\lambda$~Pyxidis for all 22 elements including an additional 2 ionisation states. Elements are listed by increasing atomic number. The wavelength of the line is shown as a red dashed line, the green line is the fit, the black line is the continuum normalised HARPS spectrum, and the flux errors are the black shaded region. The optimised wavelength region is indicated by the dashed green lines. Note that the fitted spectrum is synthesised with the optimised abundance for the desired element and Solar abundances from \citet{lodders_2025} for all other elements. All wavelengths are in the vacuum rest frame.}
    \label{fig:oddz fits}
\end{figure*}

\begin{figure}
    \centering
    \begin{subfigure}[b]{0.45\textwidth}
        \centering
        \includegraphics[width=\textwidth]{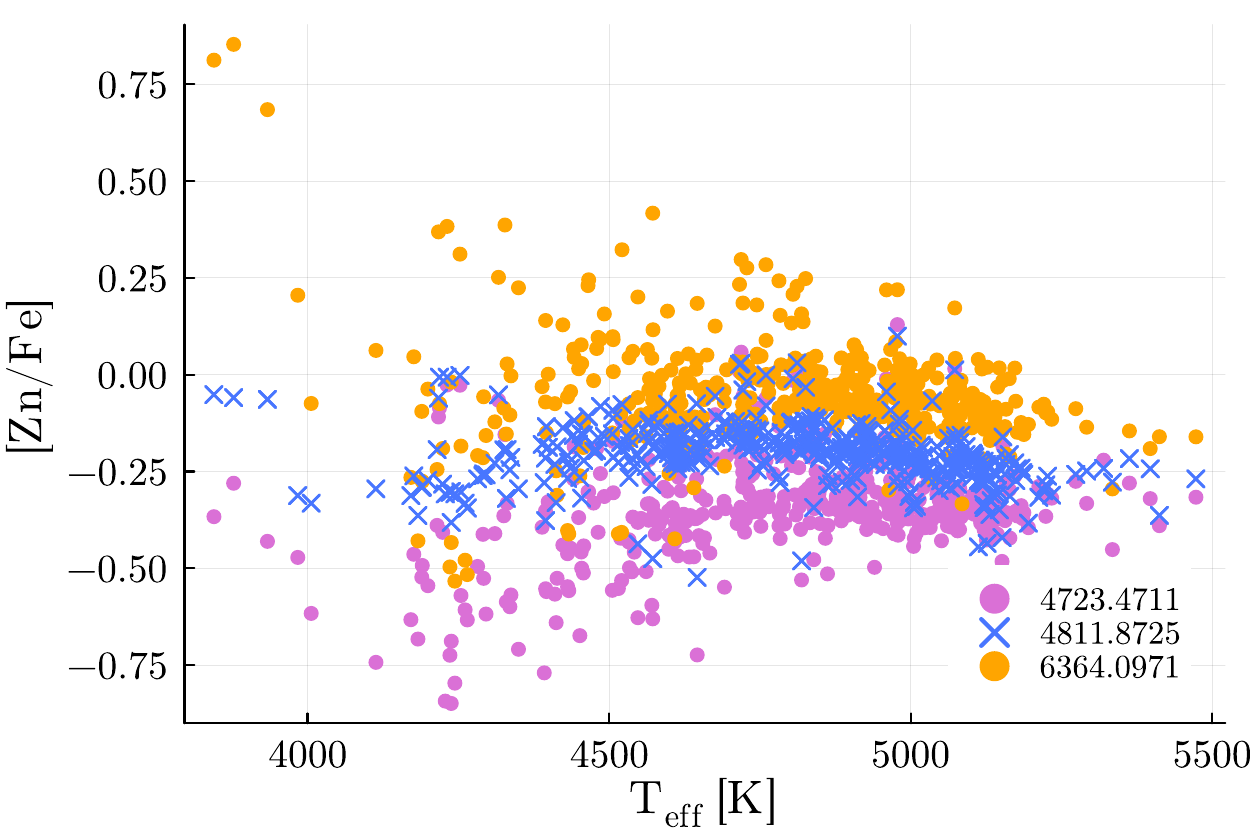}
    \end{subfigure}
    \hfill
    \begin{subfigure}[b]{0.45\textwidth}
        \centering
        \includegraphics[width=\textwidth]{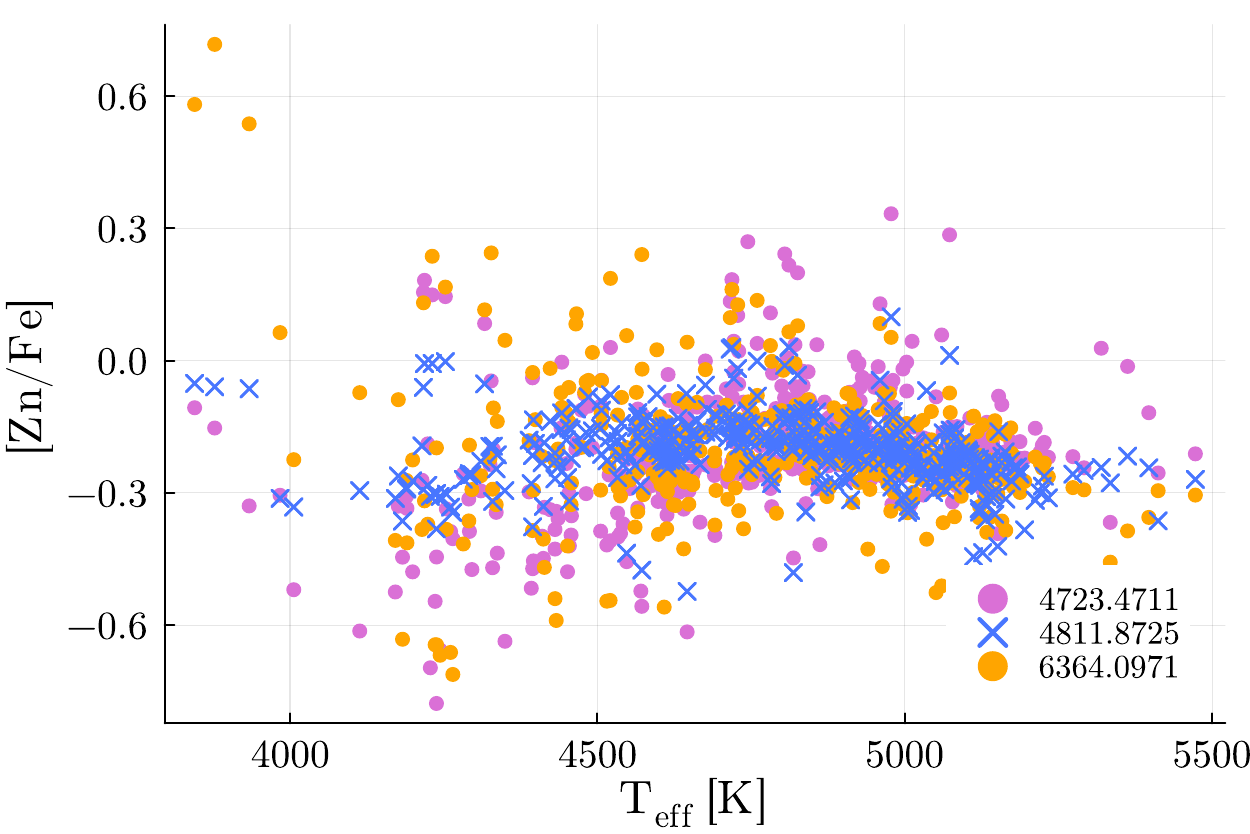}
    \end{subfigure}
    \caption{\textit{Top:} Distribution of [Zn I/Fe] measurements across stellar temperature for each of the three Zn I lines as indicated in different colours (reference line at 4811~\AA{} in crosses). \textit{Bottom:} Distribution of the three Zn I line abundances after applying corrections. All wavelengths are in the vacuum rest frame. We expect some scatter in line abundances since higher [Zn I/Fe] abundances occur with decreasing metallicity.}
        \label{fig:temp shift example}
\end{figure}

More than one absorption feature was used to measure each abundance for 16 of our 24 elements, including 2 ionisation states. For any element, different lines should, in principle, yield the same abundance. However, abundances typically differ per-line for each element due to uncertain atomic and molecular data as well as the simplified physical approximations of stars in the synthesis of \emph{ab-initio} spectra. Additionally, wavelength dependent SNR and resolution as well as telluric and instrumental artifacts can interfere with the fidelity of abundances. We found and corrected for systematic differences in our per-line abundances for a single element using their relationship with \teff. We did not want to remove physical relationships between stellar temperature and abundance, but we wanted to ensure each line had a consistent relationship with temperature. This correction prevents inconsistencies between lines manifesting as systematic offsets and inflated abundance errors, while maintaining physical abundance-temperature trends of individual absorption features.
\\\\
We determined our line corrections by first selecting a reference line for each element. This was selected as the line least subject to blending and with the highest per-line SNR (line depth). Reference lines for each element in $\lambda$~Pyxidis are shown in Figure \ref{fig:oddz fits} with the normalised HARPS spectrum in black and the optimised spectrum in green. The green dashed lines indicate the window provided when optimising the desired abundance. We fitted a second-order polynomial to the abundance of the reference line with stellar temperature and assumed this as the fiducial trend for that element. We similarly fitted the abundance-temperature trend for the remaining lines. The difference between the two parameterisations of the trend was used to shift individual stars to be consistent with the reference line;

\begin{equation}
    \Delta x_i = x_{i,ref} - x_{i,poly}
\end{equation}

\begin{equation}
    x'_i = x_i + \Delta x_i,
\end{equation}

\noindent
\\
where $x_{i,ref}$ is the second-order polynomial fit of the reference line, $x_{i,poly}$ is the second-order polynomial fit of the remaining lines, $x_i$ is the calculated line abundances of the $i^{th}$ star, $\Delta x_i$ is the corresponding line shifts, and $x'_i$ is the final shifted line abundances. An example of the temperature shifts is provided in Figure \ref{fig:temp shift example}. Figure \ref{fig:temp shift example} shows the abundance trends of the Zn I line with temperature after its lines were shifted to that of the reference line at 4811~\AA{} in the vacuum rest frame displayed as crosses. Note that abundance-temperature trends of the individual Zn I lines remains the same before and after applying our offsets. The total abundance for each element in every star is the median of the shifted line abundances which we converted to [X/Fe] using the corresponding median [Fe/H] of that star.

\subsection{Parameter and Abundance Uncertainties}

\begin{figure}
    \centering
    \includegraphics[width=\columnwidth]{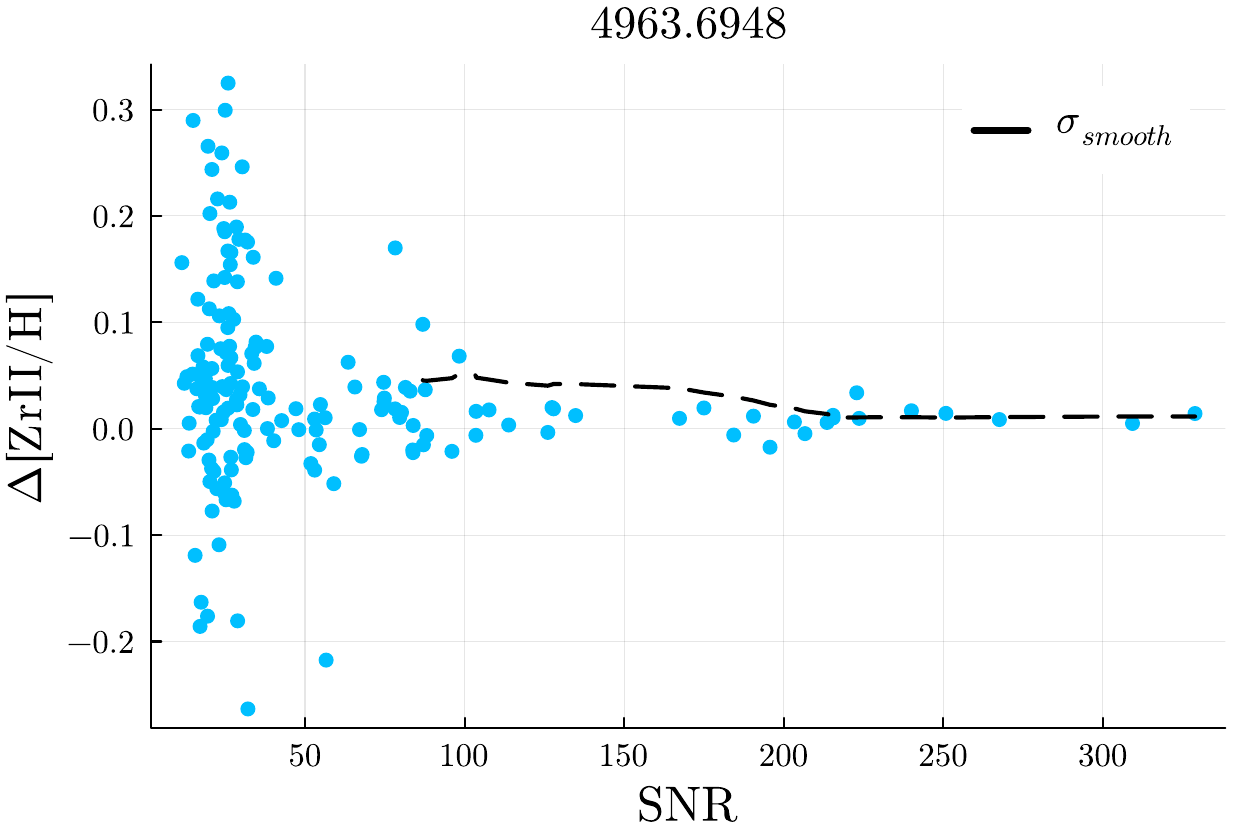}
    \caption{Difference between the Zr II 4963 \AA{} line abundances of individual observations and the combined spectrum of the 33 randomly selected stars with respect to the SNR of those observations. A smoothed 1-$\sigma$ standard deviation of these differences across 10 stars provided a line abundance uncertainty model. We only consider stars SNR $> 85$ since all stars have SNR $>100$ after selection cuts.}
    \label{fig:Zn uncertainty SNR}
\end{figure}

We used repeat observations for a subset of stars to calibrate the internal precision of our reported stellar parameters and abundances across SNR. This empirical approach quantifies the repeatability of the analysis pipeline, capturing the combined effect of observational noise and the abundance determination procedure. These empirical uncertainties do not capture systematic errors that affect all visits equally, such as any inaccuracies in the line list or model atmospheres. We therefore adopted these repeat-observation statistics as our primary estimate of the measurement precision of our sample. In Appendix \ref{appendix: accuracy test}, we also report the stellar parameter sensitivity to the fit and propagate these sensitivities to the individual abundance measurements. The parameter sensitivities are complementary to the repeat-observation analysis, providing an estimate of the internal model-fitting uncertainties associated with the adopted stellar parameters.
\\\\
To calculate the parameter uncertainties (\teff, \logg, [M/H], \vmic and \vsini) we derived the stellar parameters for observations of 33 stars -- selected across the full parameter space range --  using their individual exposures, in addition to the combined spectra. The individual exposures cover an SNR range of 15 to 333. Note that our flux error factor of 0.003 in Equation \ref{eqn:inflated error} corresponds to a maximum SNR of 333 for our combined spectra. As our repeat observations cover a range of SNR, this enables an evaluation of the scatter of the parameters across SNR. The processing and derivation of parameters for the individual observations was the same as for the combined spectra in Section \ref{sec:method-parameters}. We calculated the standard deviation of the difference between the parameters of the individual observations and the corresponding combined spectrum, in bins of width 30 across SNR for the set of stars. The smoothed trend of this standard deviation (smoothed across 10 stars) was used to provide a model for the uncertainty of each parameter as a function of SNR. Note that using repeat observations to assign uncertainties incorporates not only statistical noise, but also other random error-sources that vary between visits, such as short-term instrumental fluctuations and measurement variability. However, it does not capture systematic errors common to all visits--it measures precision not accuracy.
\\\\
The same methodology to derive the parameter uncertainties was implemented to obtain line-by-line abundance uncertainties. The randomly selected stars and the parameters of their individual observations were used to re-derive the line-by-line abundances of all elements following the optimisation described in Section \ref{sec:method-line abundances}. We calculated the 1-$\sigma$ standard deviation of the difference between the parameters of the individual observations and the corresponding combined spectrum in bins of width 5 across SNR $> 85$. We start the trend at a higher SNR to ensure that scatter at lower SNR measurements do not bias our uncertainties, since we only include stars with SNR $\leq 100$. The smoothed trend of this standard deviation (smoothed across 10 stars) was used to provide a model for the uncertainty for each element as a function of SNR. This relationship was utilised to assign all stars an uncertainty for that line in [X/H] based on their SNR. An example of this process is shown for the Zr II 4963~\AA{} line in Figure \ref{fig:Zn uncertainty SNR}, with the black dashed line denoting the smoothed trend of the standard deviation starting at SNR = 85.

\noindent
\\\\
We combine the line uncertainties to obtain uncertainties on the per star abundance. The line uncertainties for each element were combined as:

\begin{equation}
    \sigma_{jk} = \left(\frac{1}{\sum_{i}^{N} u_{ijk}^2}\right)^{-1/2},
\end{equation}

\noindent
where $u_{ijk}$ is the line uncertainties, $\sigma_{jk}$ is the combined abundance uncertainty ($\sigma$[X/H]) for element $j$ in star $k$ and $N$ is the total number of lines for that element. We added in quadrature the combined abundance uncertainty of each element with $\sigma_{Fe,jk}$ and multiplied by $\sqrt{\pi/2}$ to convert our combined uncertainties to median abundance uncertainties with respect to Fe ($\sigma$[X/Fe]).
\\\\
For elements where we have at least 5 absorption features (Ti I, V I, Fe I \& Y II), we also report the 1-$\sigma$ standard deviation of the combined individual line measurements, after correcting for systematic abundance offsets as described above. The standard deviation incorporates both the propagation of the Poisson noise in the flux as well as systematic uncertainties in the \emph{ab-initio} model descriptions of different absorption features.

\section{Results}
\label{sec:results}
We derived stellar parameters and line-by-line abundances of 22 elements (including 2 additional ionisation states) for 426 red giant stars. In this Section, we discuss sources of error inflation from line selection and blends for line-by-line abundances. Derived \teff, \logg and [Fe/H] were compared to those in \citet{luck_2015} and the Benchmark stars in \citet{soubrian_gbs_2024}, \citet{jofre_mh_2014} and \citet{heiter_teff_2015}. We also provide interpretations in the behaviour of \vsini and \vmic across \logg considering the limitations of 1D LTE spectral synthesis. Star-by-star $\alpha$- and iron-peak abundances were compared with \citet{adibekyan_kgiants_2015}. Individual abundance trends of the measured elements are shown with measurements for  main sequence stars from an ensemble of studies \citep{bensby_dwarf_2014,battistini_peak_2015,battistini_neutron_2016,adibekyan_chemical_2012,delgado_mena_chemical_2017,zhao_k_2016,mishenina_mo_2019}. We found that our individual element abundance trend were overall consistent with these 8 samples. We discuss the origin of any differences, which in general may arise from incorrect atomic data, evolutionary effects, NLTE effects, abundance variations with \teff, as well as differences between abundance derivation procedures.

\subsection{Stellar Parameters}
\label{results:parameters}
\begin{figure*}
    \centering
    \includegraphics[width=\textwidth]{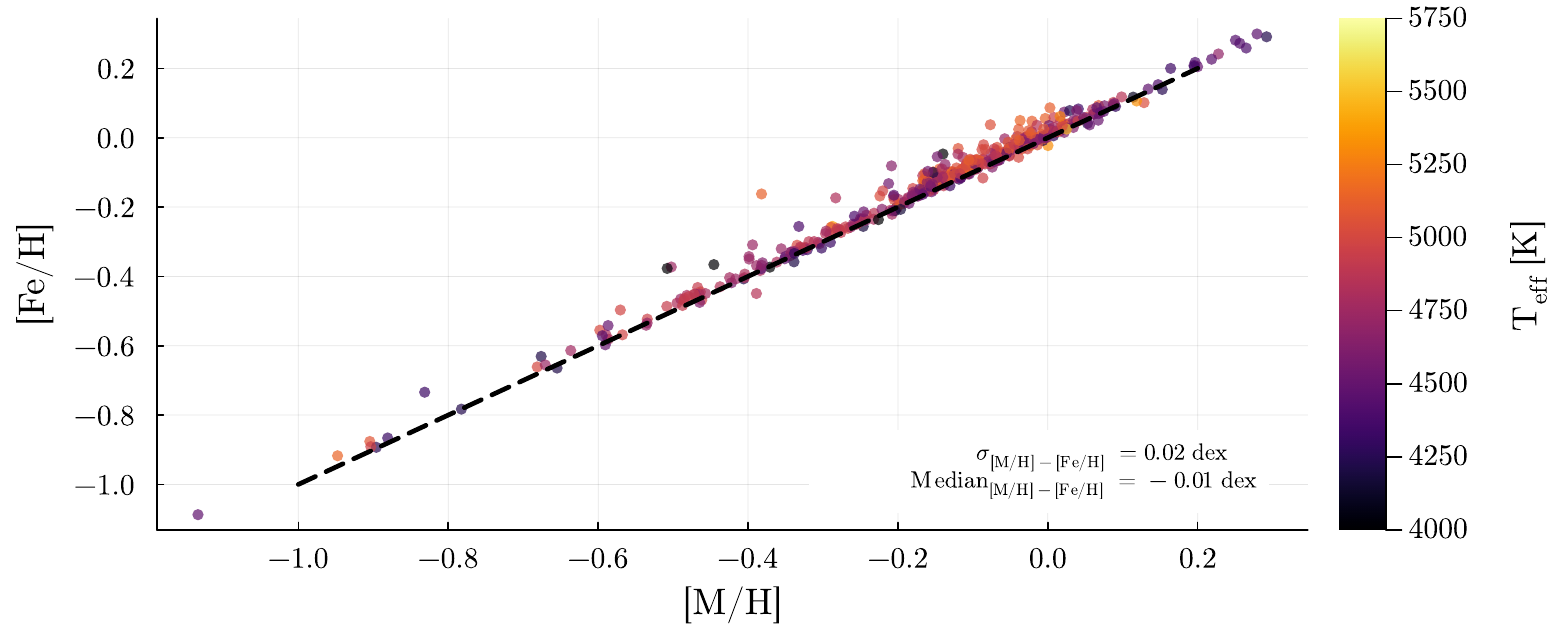}
    \caption{The [Fe/H] abundance measurements compared to the metallicity ([M/H]) measurements, coloured by \teff. The median difference between [M/H] and [Fe/H] is $-0.01$~dex with a standard deviation of 0.02~dex. Uncertainties are included but are on the order of 0.01~dex in both [Fe/H] and [M/H]. Most points lie along the black dashed one-to-one line demonstrating that they are correlated across our parameter space.}
    \label{fig:median FeH vs MH}
\end{figure*}

\noindent
Our analysis choices successfully label 91\% of the analysed stars with stellar parameters. The optimiser failed to converge for the remaining 9\%. Stars with derived parameters from HARPS spectra are displayed as a Kiel diagram in Figure \ref{fig:kiel diagram} coloured by overall metallicity [M/H]. Table \ref{table:parameters} contains the stellar parameters and their uncertainties for each star (the full Table can be found in supplementary materials). GALAH DR4 parameters for unflagged stars shown as grey points. We removed stars from Figure \ref{fig:kiel diagram} with either non-physical parameters or those that deviated from the red giant branch. These stars were not included in our analysis. Unphysical stellar parameters are a likely consequence of using the same optimisation windows across the entire sample for stars that have particularly hot or cold temperatures where alternative selections may be necessary. In particular, fixed optimisation windows do not accommodate cool stars with broad wings for some Fe lines. This contributed to some stars being labeled with inflated \logg values of $\sim$5.4~dex and created a degeneracy between \teff and [M/H] to fit the line depth. Larger windows are required to include more continuum pixels in these cooler stars. Hot stars with returned parameters $\teff \gtrsim$ 7000~K and [M/H] $\lesssim -1.25$ showed no absorption in some of the selected Fe windows. These cool and hot stars were excluded from our abundance trends. 
\\\\
The [Fe/H] measurement calculated from line-by-line abundances is approximately one-to-one with the [M/H] derived in Section \ref{sec:method-parameters} as part of our stellar parameters. This is highlighted in Figure \ref{fig:median FeH vs MH}, where our [Fe/H] and [M/H] values lie along the dashed one-to-one line. The median difference between [Fe/H] and [M/H] is 0.02~dex, with both [Fe/H] and [M/H] having median measurement uncertainties of 0.01~dex. 
\\\\
Note that we used different Fe absorption features to derive the parameters and abundances due to the distinct requirements for each phase of the analysis. The global parameter derivation is robust to individual line scatter, as all 77 Fe lines used were optimised simultaneously. However, deriving line-by-line abundances required fitting each feature individually. This individualised approach demands low scatter for each Fe line abundance and consistency across the parameter space to achieve a high precision. We therefore removed any Fe lines with large scatter in abundance upon visual inspection or those with strong temperature trends (since we do not correct out temperature trends, we only correct for offsets in Section \ref{sec:method-line abundances}). These additional selection cuts as well as those described in Section \ref{sec:method-line abundances} culled the 286 Fe lines with measured abundances to 23 lines. As a result, only five of these 23 lines were used when optimising the stellar parameters. Consequently, our requirement for high-precision abundances resulted in only five of these 23 lines overlapping with the set used for global parameter optimisation.

\begin{figure}
    \centering
    \begin{subfigure}[b]{0.45\textwidth}
        \centering
        \includegraphics[width=\textwidth]{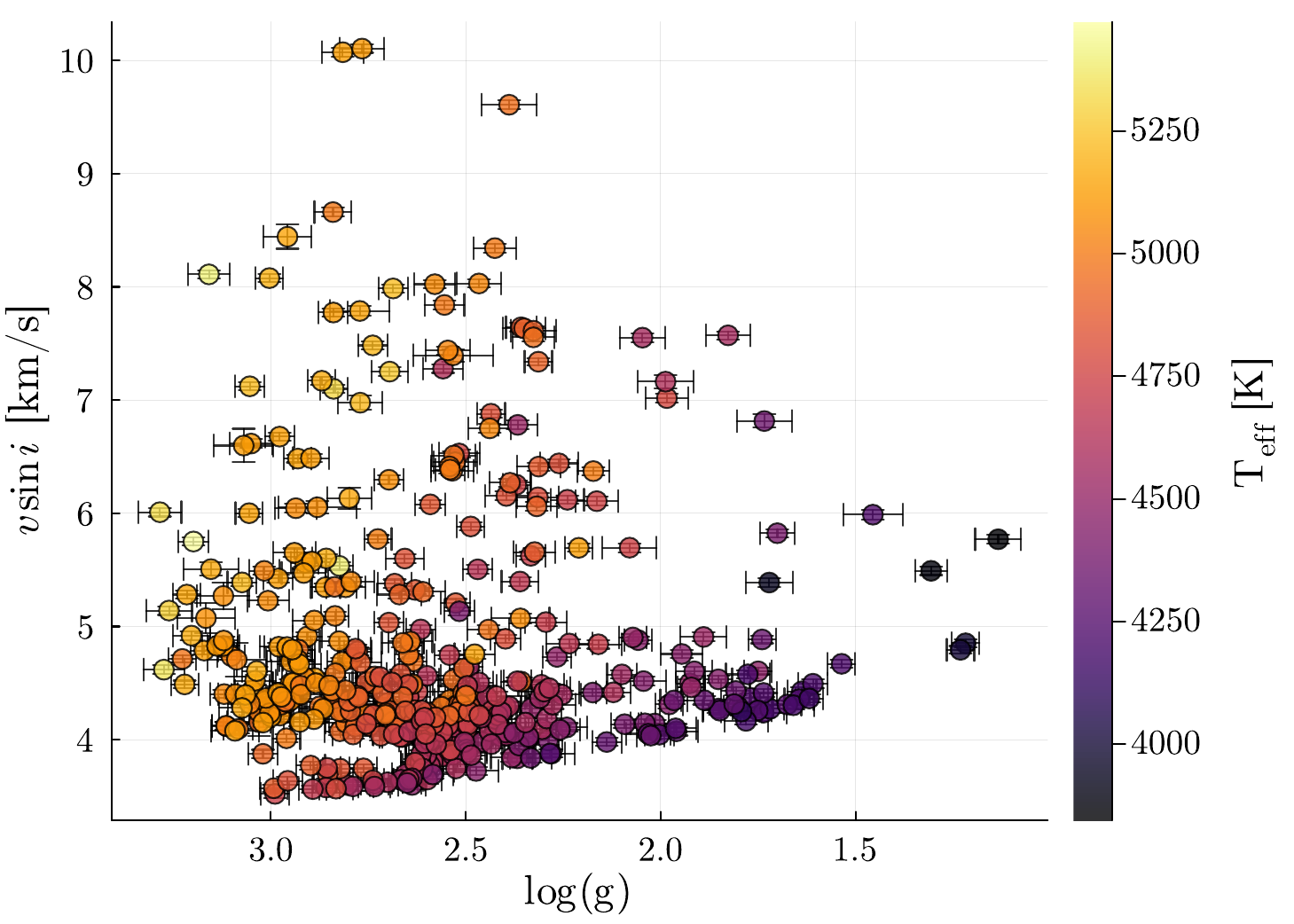}
        \label{fig:fig1}
    \end{subfigure}
    \hfill
    \begin{subfigure}[b]{0.45\textwidth}
        \centering
        \includegraphics[width=\textwidth]{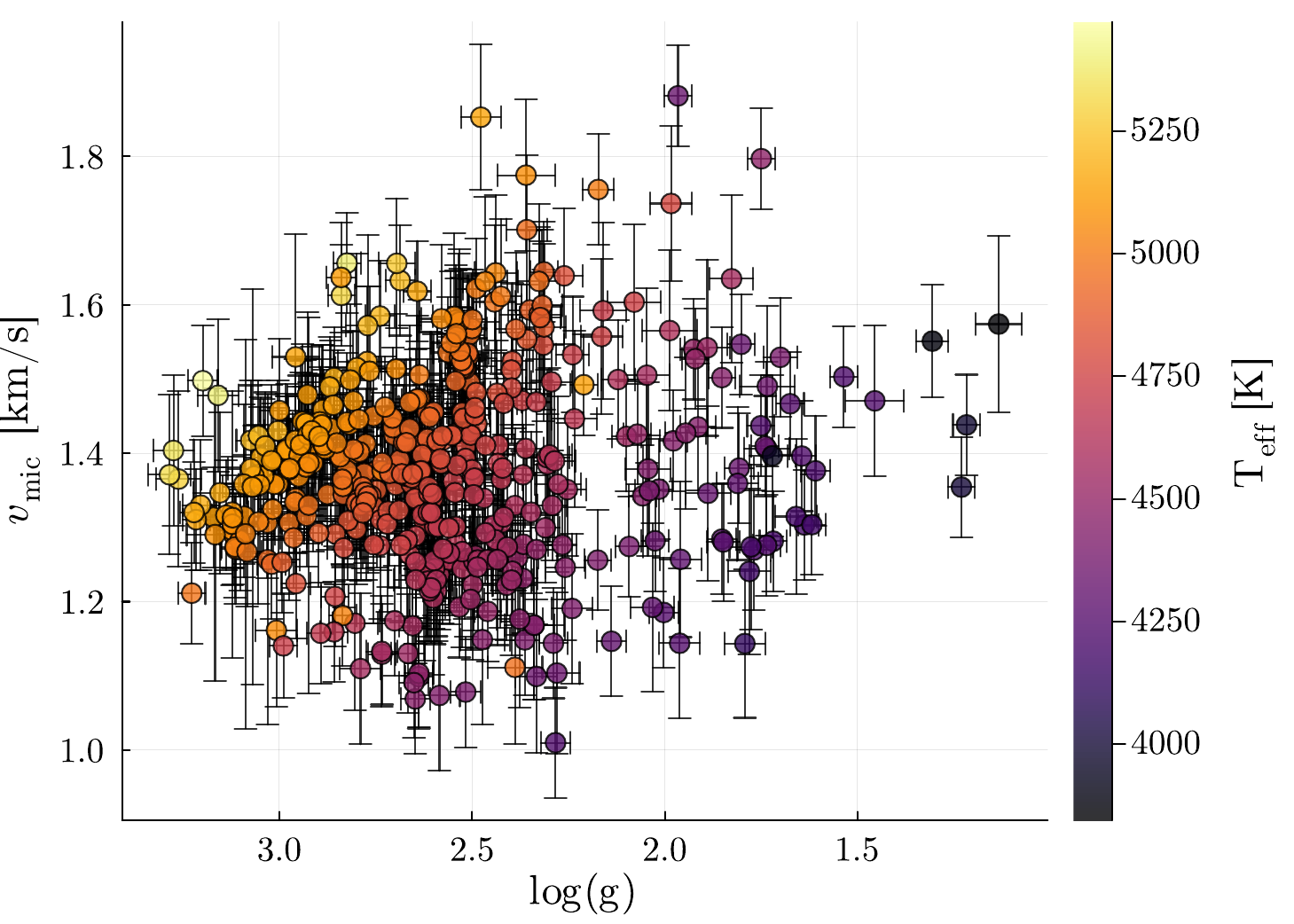}
        \label{fig:fig2}
    \end{subfigure}
    \caption{\textit{Top:} Variation in our \vsini with \logg derived from HARPS spectra. \textit{Bottom:} Variation in our \vmic with \logg derived from HARPS spectra. All points are coloured by the effective temperature. The decreasing \vsini with \logg highlights their evolution across the red giant branch. Stars with \vsini $>5$~km/s are likely an artifact from imperfect model spectra and optimisation. Uncertainties in \vsini are included but are on average 0.02~km/s.}
    \label{fig:vmic vsini}
\end{figure}

\noindent
\\\\
\textsc{korg} does not yet support fitting for macroturbulent velocity ($v_\mathrm{mac}$), meaning that \vsini and \vmic are the only Doppler-broadening parameters included in out fits. \vsini encapsulates the rotational broadening and \vmic is a non-physical correction in 1D models to account for microturbulence in 3D atmospheres of stars. However, any large scale velocity fields in the stellar atmospheres (macroturbulence) are also absorbed into these parameters, meaning that neither should be interpreted straightforwardly. Measured values of \vsini with respect to \logg are displayed in Figure \ref{fig:vmic vsini} (top) coloured by \teff. Stars with a derived \vsini greater than 10~km/s are not included in Figure \ref{fig:vmic vsini} since most are removed by selection cuts described in Section \ref{sec:method-parameters}. In particular, stars with \vsini~$> 60$~km/s are likely unidentified binaries. Measurement effects and limitations of the 1D model probably further contribute to \vsini $> 5$km/s, since such values are unexpected for red giant stars \citep{gray_vsini_1982}. The 1D approximation can in some cases lead to an increase in \vsini instead of \logg or \vmic when fitting the line shape. An artificial inflation of \vsini and \logg may also be the result of our fixed \vmic optimisation when deriving stellar parameters in Section \ref{sec:method-parameters}. The trend between \logg and \vmic enforced during optimisation is shown in Figure \ref{fig:vmic vsini} (bottom).

\begin{table*}
\centering
\begin{tabular}{ c c c c c c c c c c c } 
 \hline 
  \textbf{star} & \textbf{teff} & \textbf{u$\_$teff} & \textbf{logg} & \textbf{u$\_$logg} & \textbf{[M/H]} & \textbf{u$\_$[M/H]} & \textbf{vmic} & \textbf{u$\_$vmic} & \textbf{vsini} & \textbf{u$\_$vsini}\\
 \hline
  11$\_$Lib                      & 4744 & 20   & 2.37 & 0.04 & -0.50 & 0.01 & 1.47 & 0.07 & 6.25 & 0.03\\ 
  18$\_$Del                      &  5050& 20   & 3.12 & 0.04 & -0.08 & 0.01 & 1.29 & 0.07 & 4.40 & 0.03\\
  ...                         &  ... & ... & ...  & ...  & ...   & ...   & ...  & ...  & ...  & ...  \\
 \hline 
\end{tabular}
\caption{Sample Table containing the stellar parameters derived from the simultaneous fitting of Fe lines with \textsc{korg} in every star. The corresponding uncertainty for each parameter is provided.}
\label{table:parameters}
\end{table*}

\subsection{Element Abundances}
We derived abundances for 426 stars after applying the selection cuts listed in Section \ref{sec:method-line abundances}, of which 254 are red clump candidates identified from the over-density of stars in Figure \ref{fig:kiel diagram}. This contains 8 confirmed planet hosting giant stars 18 Del, HD11977, 81 Cet, HD47366, HD005891, HIP63242, NGC2423No3 and 39 Hya as well as 1 candidate NGC3680No41 identified using data from the extra-solar planets encyclopedia\footnote{https://exoplanet.eu/}. Condensation temperature trends for these planet-hosting stars are unremarkable compared to other stars not identified as planet hosts. Table \ref{table:line abundances} and \ref{table:median abundances} contain a sample of the line-by-line and median abundances of each element in every star. All figures use the median per-star abundances. Note that all wavelengths provided and stated in the following sections are in the vacuum rest frame.

\begin{table*}
\centering
\begin{tabular}{ c c c c c } 
 \hline 
  \textbf{star} & \textbf{element} & \textbf{central wavelength} & \textbf{[X/H]} & \textbf{u$\_$[X/H]} \\
 \hline
  11$\_$Lib                      & Na & 5689.7834 & -0.20 & 0.01\\ 
  18$\_$Del                      & Na & 5689.7834 & 0.15 & 0.01\\
  ...                         & ...&  ...     & ...  & ... \\
 \hline 
\end{tabular}
\caption{Sample Table containing the line-by-line abundances and uncertainties for each element in every star. Elements are listed in order of increasing atomic mass. The vacuum wavelength of each line is provided.}
\label{table:line abundances}
\end{table*}

\begin{table*}
\begin{tabular}{ c c c c c c c c }
\hline
\textbf{star}                  & \textbf{{[}Na/Fe{]}}          & \textbf{u\_{[}Na/Fe{]}}       & \textbf{std\_{[}Na/Fe{]}} & \textbf{{[}Mg/Fe{]}}            & \textbf{u\_{[}Mg/Fe{]}}    & \textbf{std\_{[}Mg/Fe{]}}    & \textbf{...}   \\
\hline
11$\_$Lib               & 0.15   & 0.02  & -999 & 0.12     & 0.01 & -999 & ... \\
18 Del                & 0.07   & 0.02 & -999  & -0.02    & 0.01 & -999 & ... \\
...                & ...   & ...  & ... & ...  & ...  & ... & ... \\
 \hline 
\end{tabular}
\caption{Sample Table containing the per star abundances and uncertainties for each element for every star. Elements with more than 5 lines (Ti I, V I, Fe I, Y II, Nd II) have a per star standard deviation. We use -999 as a placeholder value for stars and/or elements with no recorded values.}
\label{table:median abundances}
\end{table*}

\subsection{Uncertainties}
Our stellar parameter uncertainties show no trends with \teff, \logg, [M/H], \vsini or \vmic. Stars within our Kiel diagram selection in Figure \ref{fig:kiel diagram} have median uncertainties of $\sigma{\teff} = 21$~K, $\sigma\logg = 0.04$~dex, $\mathrm{\sigma{[M/H]} = 0.01}$~dex, $\sigma{\vmic} = 0.07$~km/s and $\sigma{\vsini} =0.04$~km/s. Uncertainties in \vmic and \vsini remain consistent across the Kiel diagram. The uncertainty in \logg is the dominant source of error in our parameters. The error in \logg may be due to the saturation of chosen Fe absorption features from \citet{griffith_untangling_2023}, as they focused on a narrow region in parameter space ($-2.1~\mathrm{dex} \leq \mathrm{[Fe/H]} \leq -1.0~\mathrm{dex}$ and $3.3~\mathrm{dex} < \mathrm{\logg} < 4~\mathrm{dex}$) when selecting absorption features for sub-giants and we chose to use the same Fe absorption features across our wider parameter space. 
\\\\
Abundance uncertainties across our parameter space are typically within 0.05~dex for elements with less than 5 lines and 0.02~dex for elements with 5 or more lines (Ti I, V I, Fe I and Y II in Table \ref{table: number of lines}). Median abundance uncertainties are $\leq 0.04$~dex for all elements. Uncertainties $> 0.04$~dex occur for some stars in element lines where an additional optimisation step was required to account for blends and/or nearby strong lines. In some cases, these blends produced a depressed continuum to one-side of the measured line. This resulted in uncertainties up to 0.1~dex for La II, as the continuum placement was fixed during our line-by-line abundance optimisation. Reduced window sizes to mitigate the affect of neighbouring strong lines further reduced the number of continuum pixels and resulted in an uneven fit to the absorption feature on the left and right of the line. There are additional cases where an adaptive window size across our parameter space might have mitigated blends (e.g. Ni I 4874 \AA{}). We may investigate this in future work for the next catalogue of main sequence stars.

\subsection{Comparison of Parameters and Abundances with Literature}
\label{sec: param comp}

We compared stellar parameters of 176 stars in common with \citet{luck_2015} (top) and to 35 stars in common with the \textit{Gaia} benchmark stars (bottom) in Figure \ref{fig:param lit comp} \citep{jofre_mh_2014,heiter_teff_2015,soubrian_gbs_2024}. We also compared the stellar parameters of 103 stars in common with \citep{adibekyan_kgiants_2015} in Figure \ref{fig:adibekyan comp} and their corresponding abundances for 11 shared elements in Figure \ref{fig:adibekyan comp abundances}. It should be noted that the comparison to literature abundances uses [Fe/H] which we calculated from line-by-line abundances, although we optimised for [M/H] when deriving parameters. Both [Fe/H] and [M/H] produced similar abundance trends, but [Fe/H] is the value most reported in literature and definitions of [M/H] may vary.
\\\\
The number of stars included for each element varies due to the star-by-star abundance cuts mentioned in Section \ref{sec:method-line abundances}. All element abundance trends are reported with respect to [Fe/H], and coloured by effective temperature in Figures \ref{fig:median oddz}$-$\ref{fig:median r}. The elements are grouped by their nucleosynthetic families with increasing atomic number: odd-Z elements (Na I, Al I, K I), $\alpha$-elements (Mg I, Si I, Ca I, Ti I/II), iron-peak elements (Sc I, V I, Cr II, Ni I, Zn I), light s-process elements (Sr I, Y II, Zr I/II), heavy s-process (Ba II, La II, Ce II, Nd II) and r-process elements (Mo I, Eu II). We note that while we assigned each element into a family, each is created from multiple sources. In particular, we categorised Mo I as an r-process element but it has 27.5\% r-process and 49.7\% s-process contributions in Sun-like stars \citep{prantzos_nc_contributions_2020}. Literature LTE abundances for FGK main sequence stars are included as grey points from \citet{adibekyan_chemical_2012}, \citet{delgado_mena_chemical_2017}, \citet{bensby_dwarf_2014}, \citet{battistini_peak_2015}, \citet{battistini_neutron_2016} and \citet{zhao_extreme_2023}. \citet{zhao_k_2016} also included NLTE abundances, which we did not compare to. These main sequence star catalogues span our abundance space, although there are some smaller catalogues of LTE red giant star abundances \citep[eg.,][]{adibekyan_kgiants_2015,bensby_kgiants_2010,silva_subgiant_2015,souto_cluster_2018,jofre_giants_2015}. The catalogue of Mo I abundances provided by \citet{mishenina_mo_2019} was the only literature we compared to that includes LTE abundances for both main sequence and giant stars.

\subsubsection{Comparison of Parameters with Literature}
\begin{figure*}
    \centering
    \includegraphics[width=\textwidth]{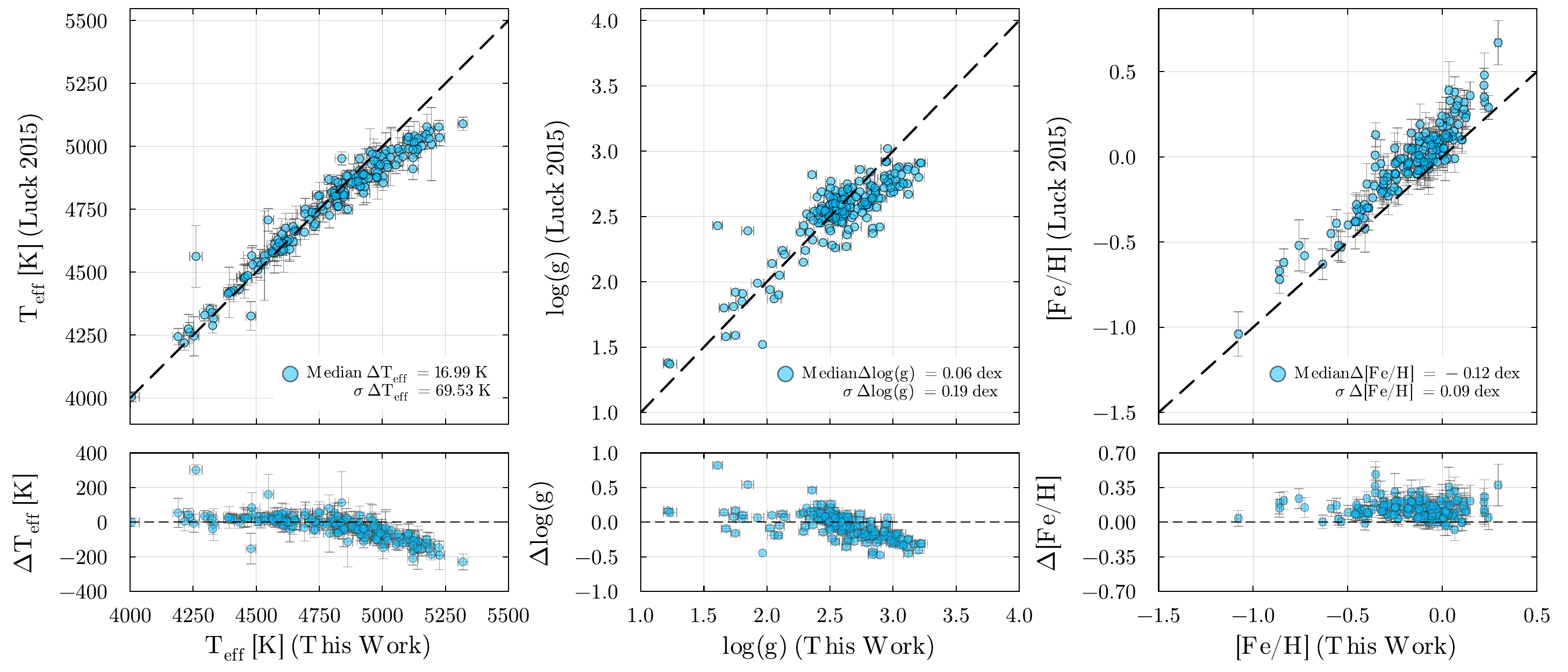}

    \vspace{0.1cm} 
    \includegraphics[width=\textwidth]{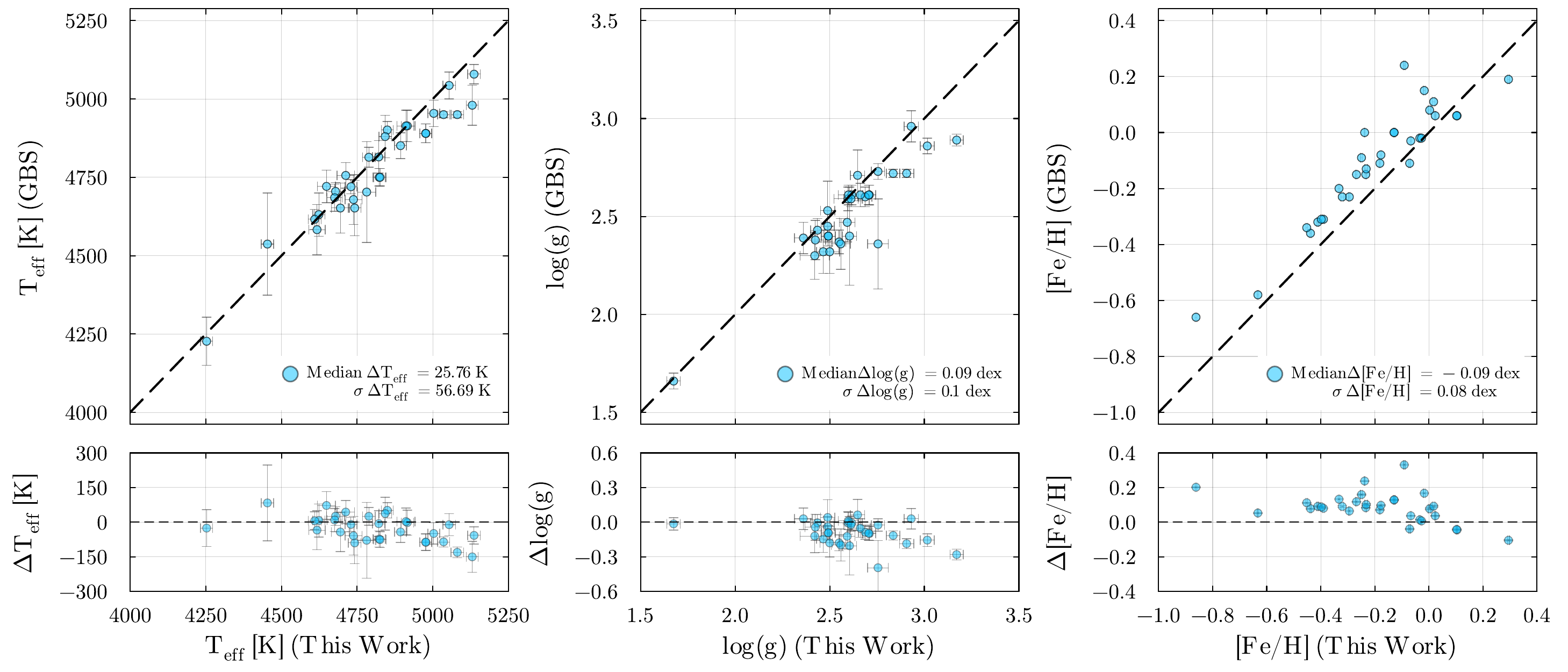}

    \caption{\textit{Top:} A direct comparison of our derived \teff, \logg, and [Fe/H] for 176 stars in common with \citet{luck_2015} with a 1:1 dashed line drawn for reference. We include residuals (\citet{luck_2015} $-$ This Work) in sub-panels at the bottom for each parameter. \textit{Bottom:} A direct comparison of our derived \teff, \logg, and [Fe/H] for 35 \textit{Gaia} benchmark stars \citep{jofre_mh_2014,heiter_teff_2015,soubrian_gbs_2024} with a 1:1 dashed line drawn for reference. We include residuals (\textit{Gaia} $-$ This Work) in sub-panels at the bottom for each parameter. The median difference between values and the corresponding standard deviation is included for each parameter.}
    \label{fig:param lit comp}
\end{figure*}

Overall, we find good agreement between the parameters of our 35 benchmark stars and \citet{heiter_teff_2015} and \citet{soubrian_gbs_2024} but with a constant bias of 26~K in \teff, 0.09~dex in \logg and $-0.09$~dex in [Fe/H] in Figure \ref{fig:param lit comp} (bottom). These deviations between spectroscopic (both from spectral synthesis and equivalent widths) and empirical results have been previously observed by \citet{heiter_teff_2015} with only a 1\% difference in \teff but a 14\% difference in \logg for Arcturus. As the stellar parameters of benchmark stars are derived empirically, the discrepancies between our parameters and those of \citet{heiter_teff_2015} and \citet{soubrian_gbs_2024} lie in the physical limitations of spectroscopic derivations. These have been acknowledged by previous works to cause offsets \citep{heiter_teff_2015,blanco_caveats_2019}. Such limitations that likely limit our analysis include uncertain inputs such as atomic data in line lists required by spectral synthesis codes, and the 1D approximation implemented by \textsc{korg}. NLTE departures may contribute to such disagreements, but are typically negligible relative to abundance errors for FGK giants \citep{jofre_mh_2014}. All these features of spectral synthesis codes have largely unknown uncertainties that lead to contradictions when comparing spectroscopic results (see \citet{blanco_caveats_2019} \& \citet{jofre_blackbox_2017}). 
\\\\
Good agreement is shown between our parameters and \citet{luck_2015} in Figure \ref{fig:param lit comp} (top), with median deviations of 17 K in \teff, 0.06 dex in \logg, and -0.12 dex in [Fe/H]. However, we find that these offsets are temperature dependent. \citet{luck_2015} noted a similar temperature dependence in their \logg values for stars with \teff~$>$~5000~K. The \logg values of \citet{luck_2015} were derived via ionisation balance, meaning they vary \logg to enforce equal abundances from Fe I and Fe II equivalent widths. \citet{luck_2015} concluded that significant line broadening caused large uncertainties in these equivalent widths, driving the temperature dependence. Broadened features also make ionisation balance sensitive to the interpolated model atmosphere. The choice of interpolated model atmosphere can also affect ionisation balance, but \citet{luck_2015} ruled this out as the primary cause. Using both ATLAS9 \citep{kurucz_atlas9_2017} and MARCS \citep{gustafsson_grid_2008} grids, they recovered the same temperature trend when comparing their spectroscopic gravities against physical gravities derived from evolutionary masses and parallaxes.

\subsubsection{Star-by-Star Comparison of Abundances with Literature}
\label{section:avles comp}
\begin{figure*}
    \centering
    \includegraphics[width=\textwidth]{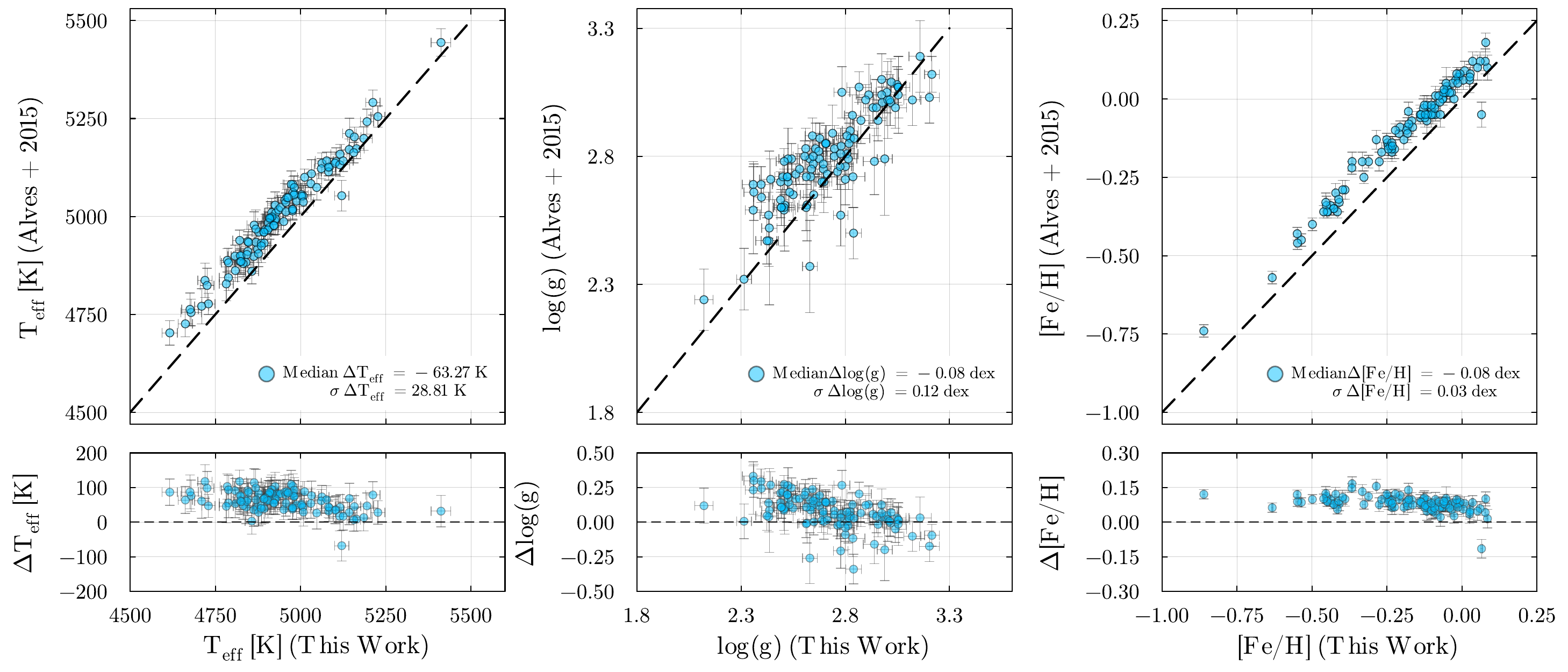}

    \caption{A direct comparison of our derived \teff, \logg, and [Fe/H] for 103 stars in common with \citet{alves_giant_params_2015} from the \citet{tsantaki_linelist_2013} line list with a 1:1 dashed line drawn for reference. We include residuals (\citet{alves_giant_params_2015} $-$ This Work) in sub-panels at the bottom for each parameter. Parameters from \citet{alves_giant_params_2015} were adopted by \citet{adibekyan_kgiants_2015}.}
    \label{fig:adibekyan comp}
\end{figure*}

\begin{figure*}
    \centering
    
    \includegraphics[width=\textwidth]{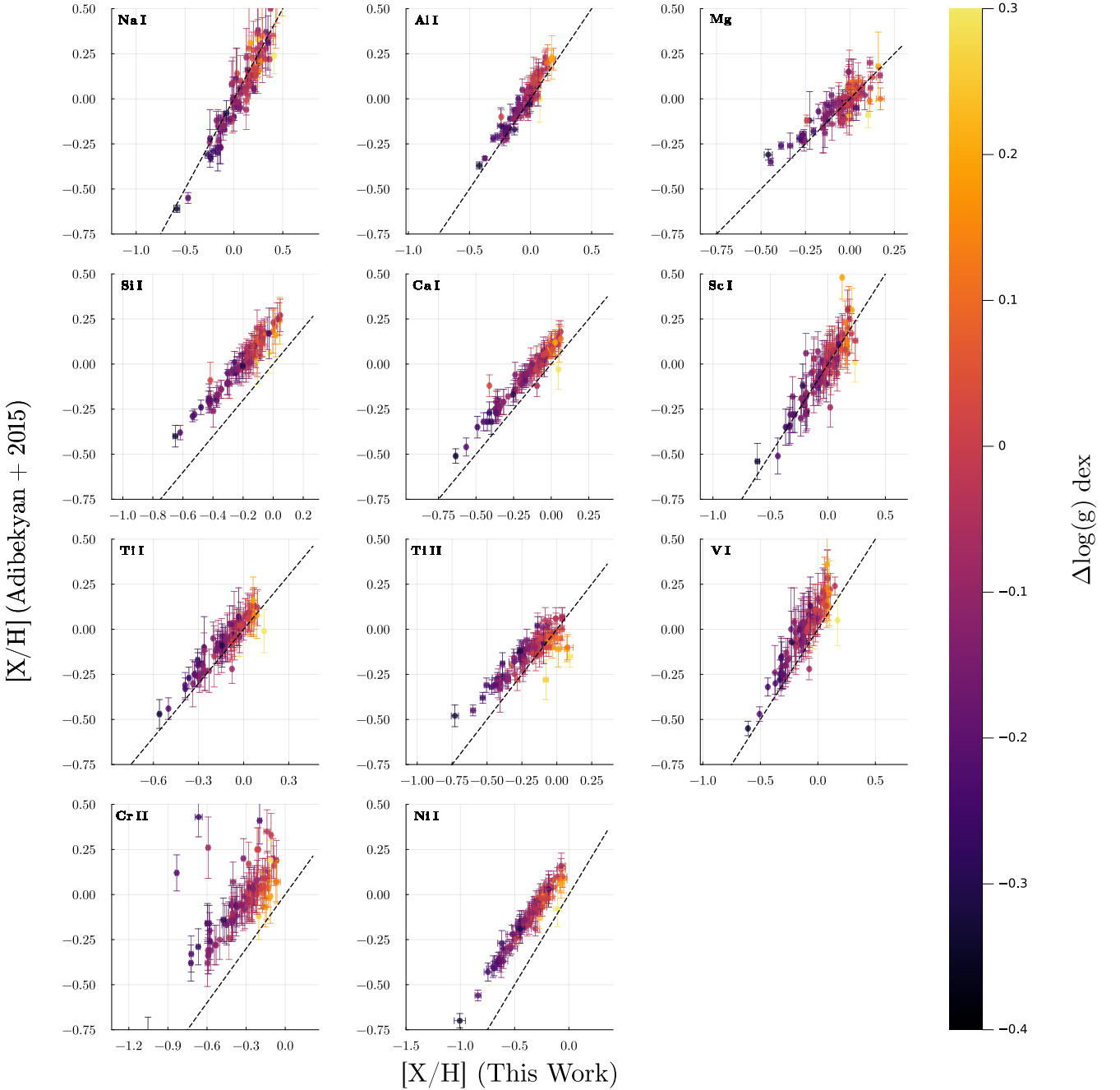}

    \caption{Comparison of our individual element abundances in common with \citet{adibekyan_kgiants_2015} for 103 red giant stars. The points in the abundance comparison are coloured by the difference between our \logg and that derived from the \citet{tsantaki_linelist_2013} line list in \citet{alves_giant_params_2015} to highlight its impact on Ti II compared to the other neutral species (see Section \ref{section:avles comp}).}
    \label{fig:adibekyan comp abundances}
\end{figure*}

We compared our element abundances to those measured in \citet{adibekyan_kgiants_2015} for 103 giant stars in common between our catalogues. This includes Na I, Al I, Mg I, Si I, Ca I, Sc I, Ti I, Ti II, V I, Cr II and Ni I. We report an average systematic offset of $-0.08$~dex in [Fe/H] for all stars compared to those from \citet{alves_giant_params_2015} adopted by \citet{adibekyan_kgiants_2015}. This metallicity offset has propagated to most elements, except Na I, Al I, Mg I and Sc I. We observe the largest deviations from the one-to-one line for Cr II in Figure \ref{fig:adibekyan comp abundances}. \citet{adibekyan_kgiants_2015} also found that the Cr II abundances for these specific stars deviated from previous literature they compared to. Our stellar parameters compared to \citet{alves_giant_params_2015} and abundances compared to \citet{adibekyan_kgiants_2015} are displayed in Figure \ref{fig:adibekyan comp} and \ref{fig:adibekyan comp abundances}, respectively. Note that \citet{alves_giant_params_2015} derived parameters using three different line lists and \citet{adibekyan_kgiants_2015} chose to adopt those obtained with \citet{tsantaki_linelist_2013}. We therefore discuss parameters determined from the \citet{tsantaki_linelist_2013} line list. The difference in our Ti II abundance trend to \citet{adibekyan_kgiants_2015} is probably due to deviations between our \logg values from spectral synthesis compared to that from the ionisation balance of Fe equivalent widths in \citet{alves_giant_params_2015}. Ionised species are more sensitive to \logg than neutral species because their number densities are dependent on the electron pressure \citep{Gray_2005}. This means that a lower surface gravity from \citet{alves_giant_params_2015} would result in the reduced Ti II abundances in \citet{adibekyan_kgiants_2015} we observe. We coloured points in Figure \ref{fig:adibekyan comp abundances} by the difference between our \logg values and \citet{adibekyan_kgiants_2015} to show the impact of \logg on the abundances of ionised and neutral species. 

\begin{figure*}
    \centering
    \includegraphics[width =\textwidth]{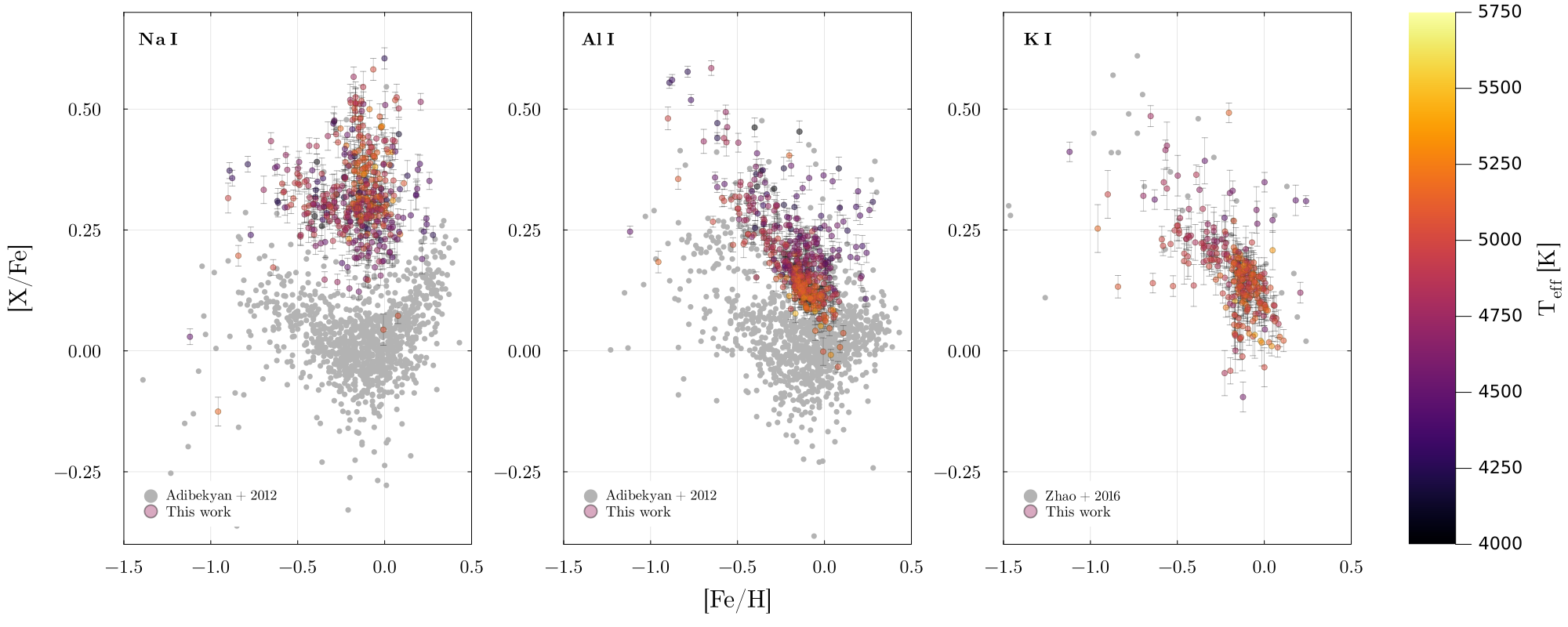}
    \caption{Element abundance trends of odd-Z elements with respect to [Fe/H] and coloured by effective temperature. LTE abundances of FGK main sequence stars for Al I from \citet{adibekyan_chemical_2012} and from \citet{zhao_k_2016} for K I are shown as grey points. Uncertainties in [Fe/H] and [X/Fe] are included for stars in our sample.}
    \label{fig:median oddz}
\end{figure*}

\begin{figure*}
    \centering
    \includegraphics[width =\textwidth]{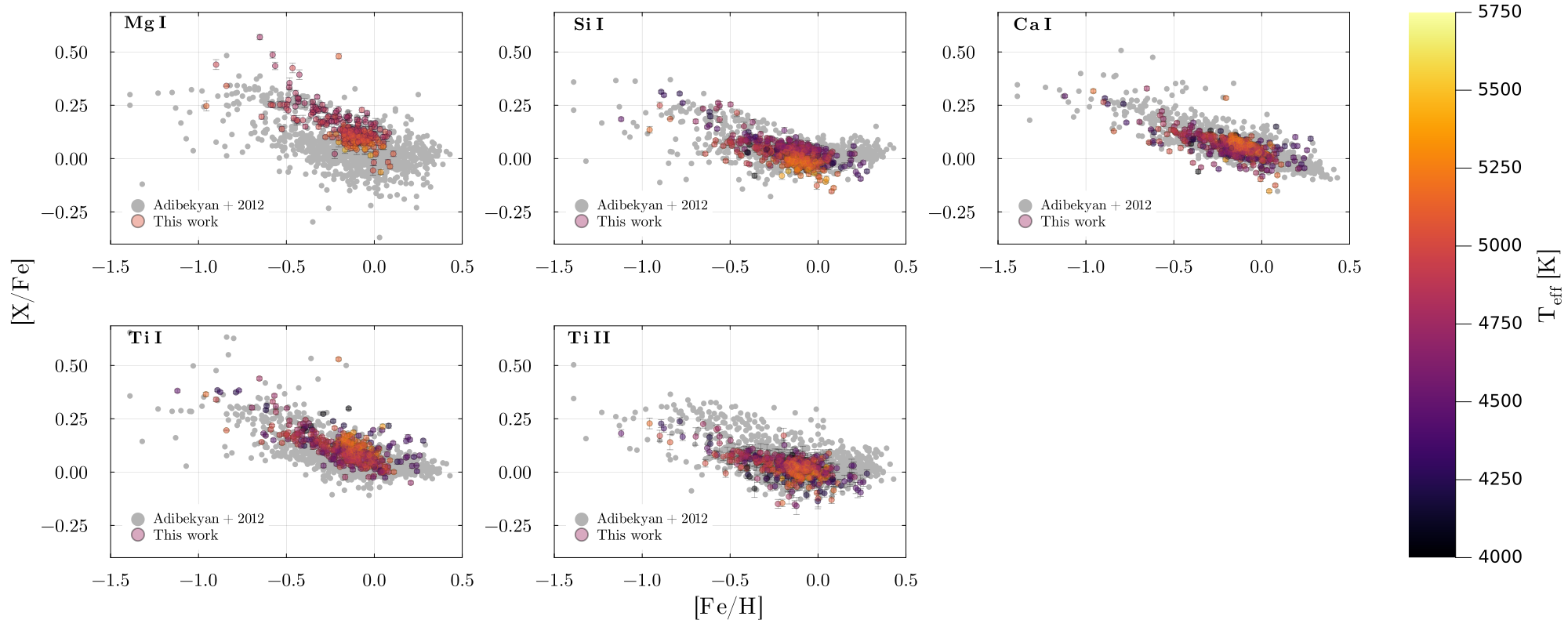}
    \caption{Element abundance trends of $\alpha$-elements with respect to [Fe/H] and coloured by effective temperature. \citet{adibekyan_chemical_2012} HARPS LTE abundances of FGK main sequence stars are shown in grey for each $\alpha$-element. Uncertainties in [Fe/H] and [X/Fe] are included for all stars in our sample.}
    \label{fig:median alpha}
\end{figure*}

\begin{figure*}
    \centering
    \includegraphics[width=\textwidth]{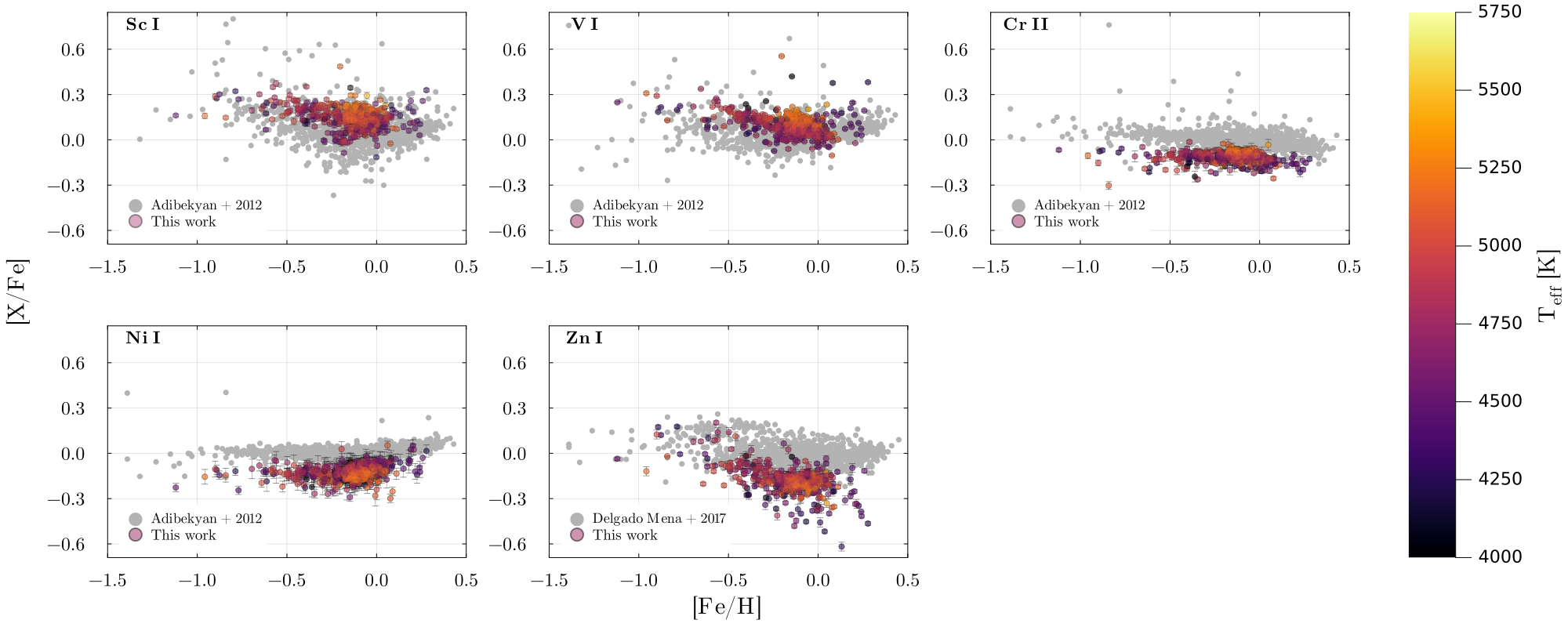}
    \caption{Star abundances of iron-peak elements with respect to [Fe/H] and coloured by effective temperature. \citet{adibekyan_chemical_2012} HARPS LTE abundances of FGK main sequence stars are shown in grey for Sc I, V I, Cr II, Ni I. We compare to the HARPS abundances of \citet{delgado_mena_chemical_2017} for Zn I. Uncertainties in [Fe/H] and [X/Fe] are included for all stars in our sample.}
    \label{fig:median fe}
\end{figure*}

\begin{figure*}
    \centering
    \includegraphics[width=\textwidth]{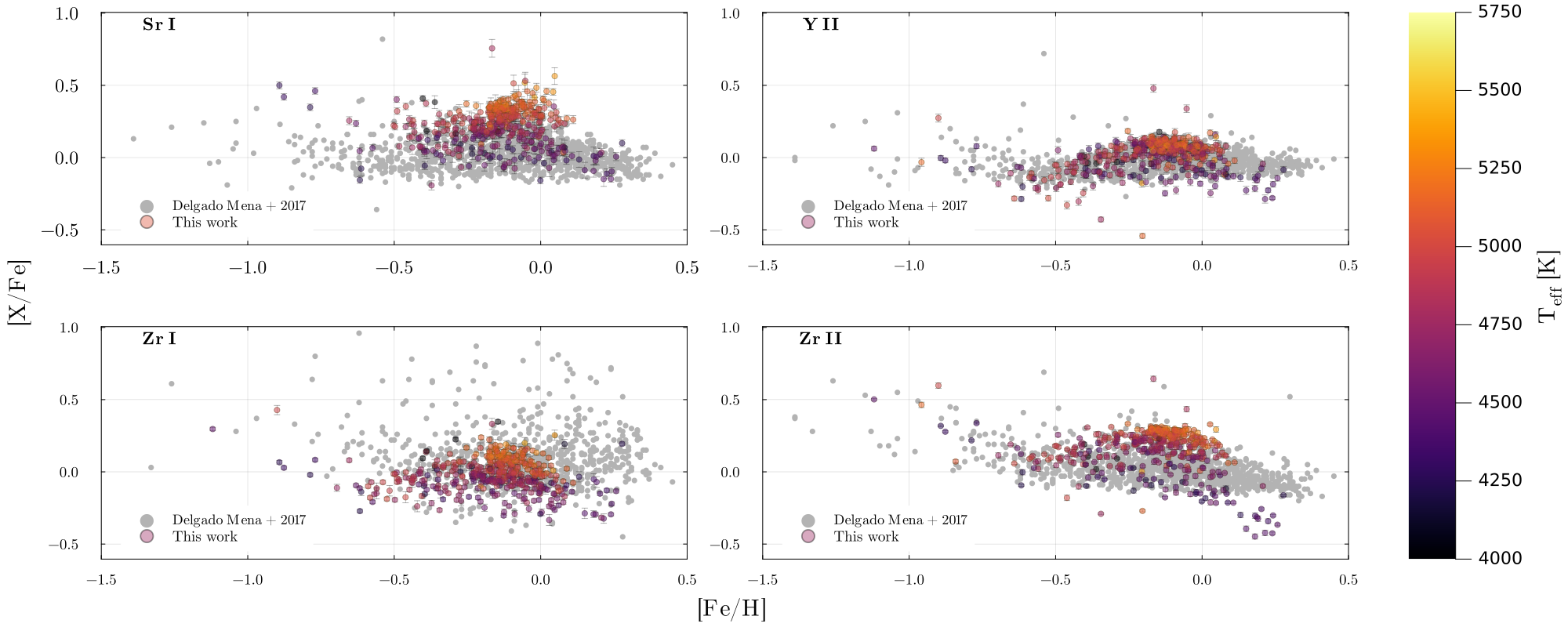}
    \caption{Element abundance trends of light s-process elements with respect to [Fe/H] and coloured by effective temperature. \citet{delgado_mena_chemical_2017} HARPS LTE abundances of FGK main sequence stars are shown in grey for Sr I, Y II, Zr I, Zr II. Uncertainties in [Fe/H] and [X/Fe] are included for all stars in our sample.}
    \label{fig:median ls}
\end{figure*}

\begin{figure*}
    \centering
    \includegraphics[width = \textwidth]{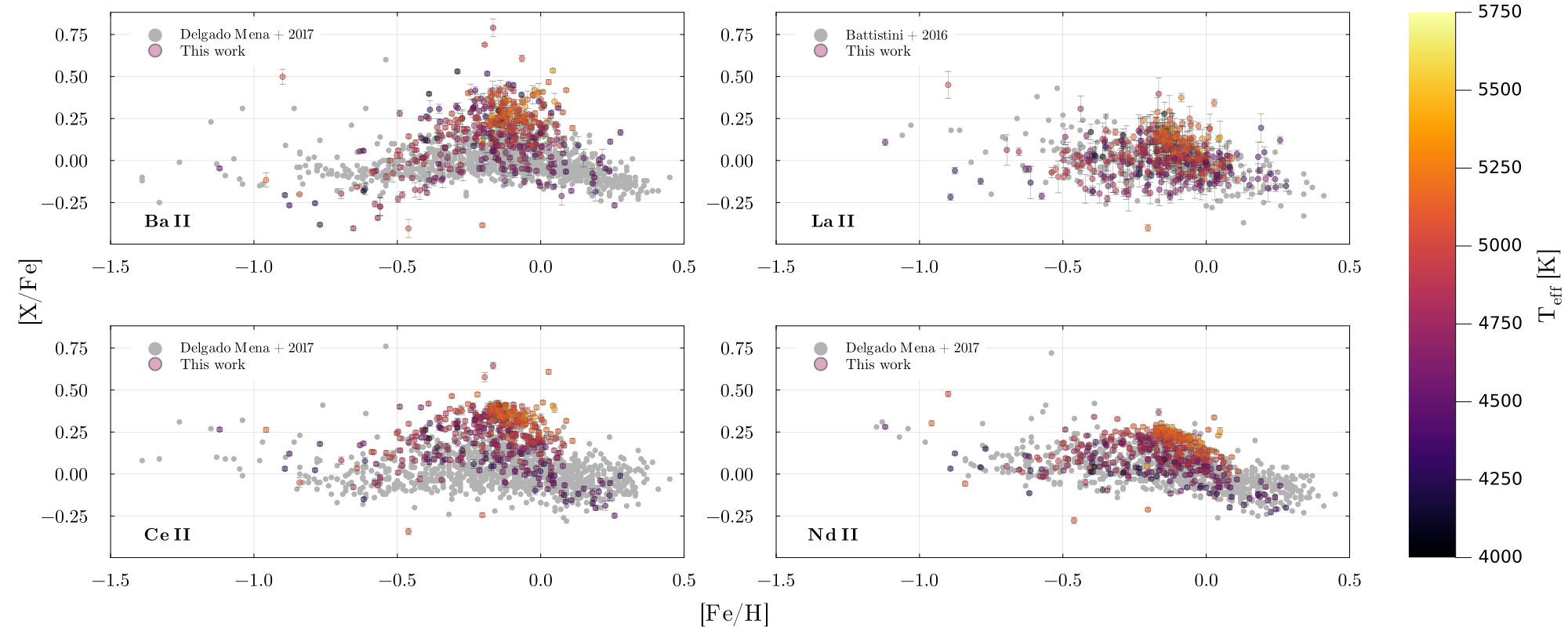}
    \caption{Element abundance trends of heavy s-process elements with respect to [Fe/H] and coloured by effective temperature. We include abundances from \citet{delgado_mena_chemical_2017} for Ba II, Ce II and Nd II as well as \citet{battistini_neutron_2016} for La II in grey. Uncertainties in [Fe/H] and [X/Fe] are included for all stars in our sample.}
    \label{fig:median hs}
\end{figure*}

\begin{figure*}
    \centering
    \includegraphics[width = \textwidth]{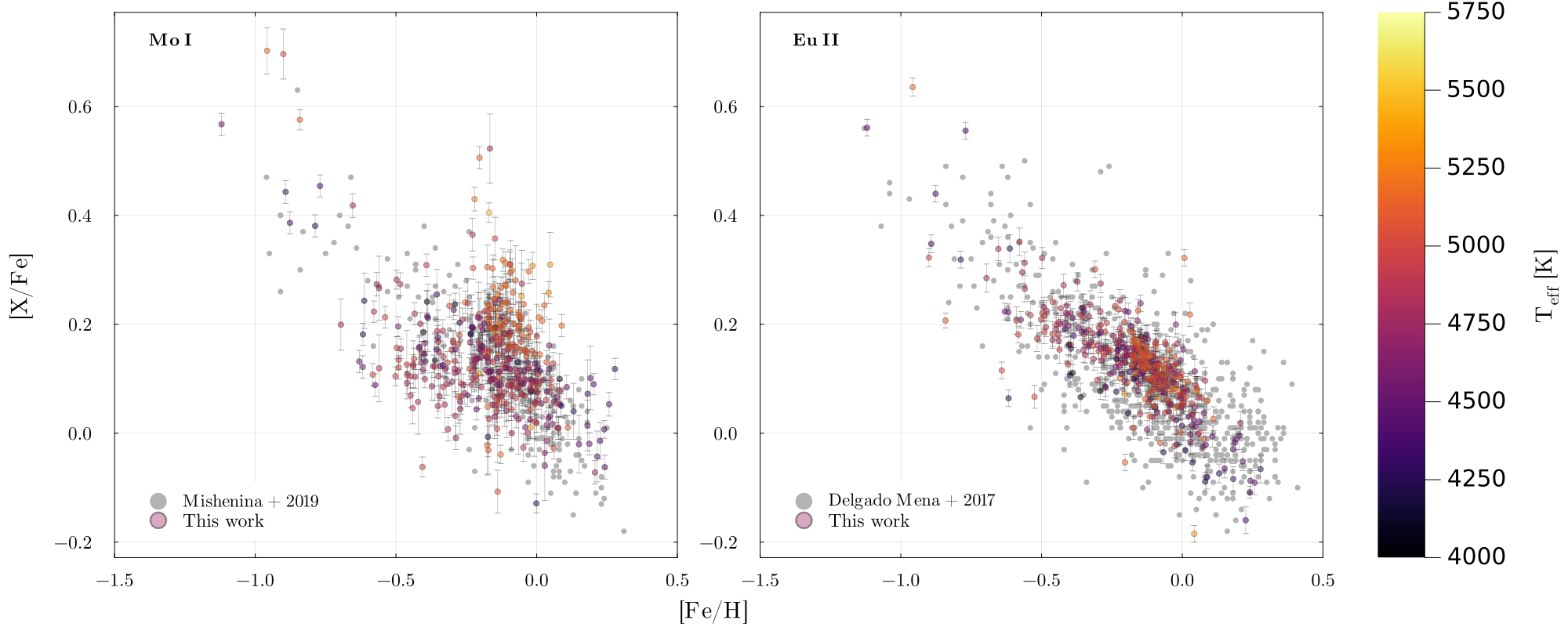}
    \caption{Element abundance trends of r-process elements with respect to [Fe/H] and coloured by effective temperature. For Mo I we compare to the main sequence and giant stars of \citet{mishenina_mo_2019} shown in grey. Eu II abundances are compared to \citet{delgado_mena_chemical_2017}. Uncertainties in [Fe/H] and [X/Fe] are included for all stars in our sample. Note that while we group Mo I as a r-process element it also has considerable s-process contribution.}
    \label{fig:median r}
\end{figure*}

\subsubsection{Odd-Z elements (Na I, Al I \& K I)}
\label{results:oddz}
Odd-Z elements refers to elements with an odd atomic number. These elements are expected to have an increasing abundance with increasing metallicity due to their dependence on a free neutron source. Excess neutrons are produced in proportion to the initial metallicity of the star via ${}^{14}$N($\alpha$,$\gamma$)${}^{18}$F($\beta^+\nu$)${}^{18}$O($\alpha$,$\gamma$)${}^{22}$Ne during He fusion \citep{pagel_galactic_1997}. An increase from free neutron sources is observed at [Fe/H] $>$ 0~dex for stars coloured with \teff $<$ 5000~K in our Na I and Al I abundances in Figure \ref{fig:median oddz}. This trend is also observed in \citet{adibekyan_chemical_2012} shown in grey in Figure \ref{fig:median oddz}. We are unable to observe an increase in free neutron sources for K I since it lacks abundance measurements at [Fe/H] $>$ 0~dex. However, there is an initial decrease in the abundances of Na I, Al I and K I at [Fe/H] $> -1$~dex because of Fe production via Type Ia supernovae \citep{kobayashi_origin_2020}. We further discuss Figure \ref{fig:median oddz} in the following element summaries of Na I, Al I and K I.

\paragraph*{Na I:}
Our Na I abundance trends as a function of [Fe/H] for red giant stars follows that of main sequence stars in \citet{bensby_dwarf_2014} and \citet{adibekyan_chemical_2012}, albeit with a 0.25~dex offset. These Na I trends with [Fe/H] are displayed in Figure \ref{fig:median oddz} with our abundances coloured by our derived \teff and \citet{adibekyan_chemical_2012} in grey. \citet{adibekyan_kgiants_2015} also find a 0.2~dex enhancement in Na I for giants compared to the dwarf sample in \citet{adibekyan_chemical_2012}. An enhancement in Na is somewhat expected due to advanced hydrogen burning in red giant stars and internal mixing processes \citep{salpeter_1955}. However, Na I NLTE corrections may reconcile this difference \citep{alexeeva_2014}. Another possible cause for the discrepancy is an incorrect oscillator strength since we only include one `good' Na line and were unable to perform line corrections. The Na line we included at 5689 \AA{} had an oscillator strength of $-0.404$ in the GAIA ESO line list, differing by $-0.048$ from VALD \citep{Ryabchikova_vald3_2015} and by $-0.22$ from the \citet{neves_linelist_2009} line list used by \citet{adibekyan_chemical_2012}. The scatter in abundances of \citet{adibekyan_chemical_2012} at a similar 0.2~dex enhancement is likely a result of the temperature dependence bellow 5000~K for their two chosen Na lines.

\paragraph*{Al I:}
Our Al I abundances across [Fe/H] in Figure \ref{fig:median oddz} are consistent with both the main sequence abundances of \citet{bensby_dwarf_2014} and \citet{adibekyan_chemical_2012}. Figure \ref{fig:median oddz} includes the \citet{adibekyan_chemical_2012} Al I abundances in grey and our abundances are coloured by our derived \teff for that star. Red giant stars are expected to have a constant surface Al abundance, as no Mg fusion occurs during core H fusion \citep{weiss_mixing_2004} and NLTE effects around 6697~\AA{} are minimal for Al in stars with \teff $<$ 6500~K \citep{smiljanic_na_al_2016}. 

\paragraph*{K I:}
Our K I abundances decrease with increasing [Fe/H] in Figure \ref{fig:median oddz}. This is a similar trend to $\alpha$-elements due to contributions from Type II supernovae above [Fe/H] $> -1$~dex \citep{zhao_k_2016,kobayashi_origin_2020}. \citet{zhao_k_2016} observed a similar trend in FG main sequence stars shown in grey in Figure \ref{fig:median oddz}, but their small sample makes it difficult to conclude if our K I abundances are slightly depleted or in agreement. There are only 3 stars with \teff $<4500$~K that have reported K I abundances due to the decrease in line strength.

\subsubsection{$\alpha$-elements (Mg I, Si I, Ca I, \& Ti I/II):}
$\alpha$-elements (eg., Mg, Si, Ca and Ti) are those produced via a sequence of $\alpha$-capture reactions. These elements are primarily produced in Type II supernovae \citep{pagel_galactic_1997}. The ejection of these elements typically leads to a flat trend before decreasing once the production of Fe from Type Ia supernovae becomes significant at approximately $\mathrm{[Fe/H]} =-0.4$~dex \citep{Rojas-Arriagada_alpha_2019}. We see this transition between the production timescales of these supernovae in Mg I, Si I, Ti I and Ti II in Figure \ref{fig:median alpha}, but we require more stars at lower [Fe/H] for clarity. There is also a bifurcation in the \citet{adibekyan_chemical_2012} $\alpha$-element abundances that highlight the high- and low-$\alpha$ disc. The bimodality in the abundance trends is less prominent in Ca I and Si I because of additional explosive nucleosynthesis as well as Type Ia supernovae contributions \citep{kobayashi_origin_2020}. We further discuss Figure \ref{fig:median alpha} in the following element summaries of Mg I, Si I, Ca I and Ti I/II.

\paragraph*{Mg I:}
Our Mg I abundances increase with decreasing temperature and \logg similar to the abundances of GK giants in \citet{adibekyan_kgiants_2015}. However, \citet{adibekyan_chemical_2012} did not find a significant temperature trend for Mg I in main sequence stars. Our Mg I abundances with [Fe/H] are shown coloured by \teff in Figure \ref{fig:median alpha} and \citet{adibekyan_chemical_2012} Mg I abundances in grey. The \teff trend present in our red giant stars is unlikely to be caused by the increased production of Al I from Mg I with temperature since low-mass red giant stars do not display sufficient deep mixing \citep{souto_cluster_2018}. The temperature trend we observe could be attributed to surrounding CN bands that broaden the wings of the Mg I line as well as strengthen neighbouring lines for stars with \teff $<$ 4750~K. These cool stars in our sample have enhanced Mg I abundances up to 0.57~dex compared to the trend of our hotter stars that follow the main sequence star abundances from \citet{bensby_dwarf_2014}. We therefore removed stars with \teff $<4750$~K from our Mg I abundance trend in Figure \ref{fig:median alpha}. Note that the Mg I absorption feature used at 5712 \AA{} has a U flag in the \textit{Gaia} ESO line list and although the oscillator strength is the same as in the \citet{neves_linelist_2009} implemented in \citet{adibekyan_chemical_2012}, incorrect atomic data may be the cause of discrepancies. 

\paragraph*{Si I:} 
We discuss our Si I abundances with [Fe/H] in Figure \ref{fig:median alpha} coloured by \teff and compare to the abundances of \citet{adibekyan_chemical_2012} in grey. We observe similar Si I abundances between our red giant stars and the main sequence stars in \citet{adibekyan_chemical_2012}. Stars with Si I abundances up to 0.31~dex at [Fe/H] $< -0.5$~dex may be thick discs stars because they follow the same trend as those identified by \citet{adibekyan_chemical_2012}. It is unlikely that Si I abundances up to 0.31~dex are caused by NLTE effects. Si I absorption features in the optical are insensitive to NLTE effects because they are formed in the inner regions of the photosphere \citep{Tan_sinlte_2016}. For example, Sun-like stars show deviations on the order of $-0.05$~dex \citep{Sukhorukov_sun_2012}.

\paragraph*{Ca I:}
In Figure \ref{fig:median alpha}, our Ca I abundances coloured by \teff agree well with the main sequence abundance trends across [Fe/H] of \citet{adibekyan_chemical_2012} in grey. The four Ca I lines used all had abundances consistent with \citet{adibekyan_abundance_2016} and mean line abundances within 0.04~dex of the chosen reference line. We expect red giants to have similar abundances to main sequence stars because diffusion will decrease Ca I by only 0.01~dex as a star transitions from the main sequence turn-off to sub-giant branch \citep{onehag_diffusion_2014}. However, \citet{souto_cluster_2018} found Ca I abundance differences of up to 0.2~dex between red clump stars and solar-twins in M67. \citet{souto_cluster_2018} argued that this difference cannot be reconciled by NLTE corrections and indicate the presence of chemical diffusion.

\paragraph*{Ti I \& Ti II:}
Ti I and Ti II have an average abundance difference of 0.08~dex across our parameter space and follow the trend of \citet{bensby_dwarf_2014} and \citet{adibekyan_chemical_2012} across [Fe/H]. Both Ti I and Ti II abundance trends are included in Figure \ref{fig:median alpha} coloured by \teff and with \citet{adibekyan_chemical_2012} in grey. Our Ti I abundances do have scatter up to 0.16~dex for stars [Fe/H]~$>0.1$~dex. This scatter is not present in our Ti II abundance trend in Figure \ref{fig:median alpha}. The increased scatter for these metal-rich stars may be due to blends in the weaker Ti I lines that are not present for the strong Ti II line measured at 5240~\AA{}. However, \citet{adibekyan_kgiants_2015} also find a 0.1~dex abundance difference between Ti I and Ti II for red giant stars. This discrepancy between ionisation states is possibly due to the the LTE approximation. LTE spectral synthesis codes tend to underestimate Ti I by 0.05$-$0.1~dex, depending on the implementation of inelastic collisions in statistical equilibrium \citep{bergemann_ionisation_2011}. NLTE Ti I and Ti II abundances for Sun-like stars are within 0.01~dex \citep{bergemann_ionisation_2011}. \citet{bergemann_ionisation_2011} recommended disregarding Ti I absorption features when comparing main sequence and giant stars due to the current overestimation of NLTE effects in main sequence stars and the underestimation of ionisation in giant stars. Further investigation into Ti I NLTE effects in other stars is required \citep{sitnova_2016}.

\subsubsection{Iron-peak elements (Sc I, V I, Cr II, Ni I \& Zn I)}
Iron-peak elements are formed through both Type Ia and Type II supernovae, creating abundance ratios that are approximately constant with metallicity at [Fe/H] $> -1$~dex \citep{kobayashi_origin_2020}. We observe this flat trend across [Fe/H] for Sc I, V I, Cr II and Ni I in Figure \ref{fig:median fe} with an increase at [Fe/H] $\gtrsim 0$~dex. The abundances of Sc I and V I both increase with [Fe/H] since as odd-Z elements they are dependent on a free neutron source (see Section \ref{results:oddz}). Ni I abundances increase as its production is dependent on the metallicity of both Type Ia and Type II supernovae \citep{kobayashi_origin_2020}, while Zn I abundances decrease similar to $\alpha$-elements due to metallicity dependent Type II supernovae contributions \citep{kobayashi_2009}. We need more metal-rich stars in our sample to observe any change in the Cr II trend, but its abundances are mostly flat in \citet{adibekyan_chemical_2012}. We further discuss Figure \ref{fig:median fe} in the following element summaries of Sc I, V I, Cr II, Ni I and Zn I.

\paragraph*{Sc I:}
Our Sc I trend across [Fe/H] agrees with \citet{battistini_peak_2015} and \citet{adibekyan_chemical_2012}. This agreement is displayed in Figure \ref{fig:median fe}, with our abundances coloured by \teff and \citet{adibekyan_chemical_2012} in grey. A slight decrease in the Sc I abundance with Fe is expected with the increase of Type Ia supernovae. Sc is mainly produced in core-collapse supernovae but its dependence on the initial metallicity of the star for free neutrons prevents the decrease we observe for $\alpha$-elements with increasing [Fe/H]. We expect Sc I abundances to be the same between main sequence and giant stars as shown in \citet{adibekyan_kgiants_2015}. Our Sc I abundances show less scatter than \citet{adibekyan_chemical_2012} with abundances up to 0.5 $\pm$ 0.01~dex compared to those up to 0.75 $\pm$ 0.08~dex, respectively at [Fe/H]~$<-0.5$~dex.

\paragraph*{V I:}
We originally chose 5728.64~\AA{} as our reference line for V I since it is the deepest line with the least scatter. However, \citet{jofre_alpha_2015} reported that this line was difficult to measure in over 50$\%$ of their FGK giants so we instead used the V I line at 6041~\AA{}. The mean abundance of the two lines differed by 0.01~dex so our choice of reference line did not effect the per star V I abundances. 
\\\\
Our V I abundance trends across [Fe/H] are consistent with \citet{adibekyan_chemical_2012} in Figure \ref{fig:median fe}, although \citet{adibekyan_abundance_2016} has an enhanced population up to 0.5~dex prior to applying their temperature corrections. The temperature dependence of V I at \teff $< 5300$~K in their sample is not present in our red giant stars. The larger V I abundances in \citet{adibekyan_abundance_2016} prior to temperature corrections could be caused by their 0.16~dex uncertainties for cool stars in comparison to our $<$ 0.01~dex V I abundance uncertainties. 
\\\\
We agree with the \citet{battistini_peak_2015} V I abundances across all [Fe/H]. A flattening trend for V I is predicted for [Fe/H] $< -1$~dex when considering contributions from neutrino processes, explosive nucleosynthesis in Type II supernovae and nuclear statistical equilibrium in massive stars \citep{kobayashi_origin_2020}. \citet{battistini_peak_2015} show that the decrease in V I from $-1$~dex~[Fe/H]~$<-0.4$~dex is caused by thick disk stars. This indicates that while our sample consists of mostly thin disk stars, our metal-poor stars are likely predominantly thick disk stars.

\paragraph*{Cr II:}
Our Cr II abundance trend across [Fe/H] coloured by \teff in Figure \ref{fig:median fe} agrees with that of \citet{adibekyan_chemical_2012} shown in grey, albeit with a 0.15~dex offset. Our systematically lower Cr II abundances may stem from blending with a Cr I line; the LTE Cr I and Cr II abundances are known to be offset by ~0.15~dex in the Sun \citep{bergmann_cr_2010} and red giant stars \citep{hawkins_2016}. There is not a large offset between ionisation states when comparing NLTE derived Cr I and Cr II abundances in the Sun \citep{bergmann_cr_2010} and in GCE models \citep{kobayashi_origin_2020}. Furthermore, there is no expectation from theory that Cr II abundances should vary between main sequence and red giant stars. Previous observations by \citet{adibekyan_kgiants_2015} have found no significant differences between their K giant and main sequence population.

\paragraph*{Ni I:}
Our Ni I abundance trend across [Fe/H] (Figure \ref{fig:median fe}) matches that of the main-sequence stars analyzed by \citet{adibekyan_chemical_2012}. However, our absolute Ni I abundances are systematically lower than their values by $\sim$0.1 dex. This contrasts with the Ni I abundances reported by \citet{adibekyan_kgiants_2015}, who found no offset between their red giant stars and the main sequence from \citet{adibekyan_chemical_2012}. The systematic offset we observe may be influenced by NLTE effects--which have not been extensively studied for Ni I lines in the optical--though recent work has suggested that such corrections for optical Ni I lines are relatively small \citep{eitner_2023}. A more probable source of the discrepancy lies in the local continuum mismatch during spectral optimisation. The single Ni I line utilised in this study (4874.8 \AA{}) sits on the Stark-broadened wing of the H$\beta$ feature, a region characterised by a depressed pseudo-continuum in the observed HARPS spectra (see Figure \ref{fig:oddz fits}). 

\paragraph*{Zn I:}
Our Zn I abundance trend colored by \teff in Figure \ref{fig:median fe} shows agreement with \citet{delgado_mena_chemical_2017} (grey), albeit with a 0.2 dex offset despite measuring the same Zn I features. This offset is driven by differing oscillator strengths; our reference Zn I line at 4811 \AA{} has log(gf) = -0.16 in the \textit{Gaia} ESO line list, whereas the \citet{neves_linelist_2009} list implemented by \citet{delgado_mena_chemical_2017} adopts log(gf) = $-0.137$. We selected the stronger 4811 \AA{} line (average depth of 0.5, with a nearby blend mitigated by reducing the window size) over the theoretically higher-quality 6364 \AA{} line. While the 6364 \AA{} line ("Y" flag in \textit{Gaia} ESO) produced a trend that perfectly overlapped with \citet{delgado_mena_chemical_2017}, it yielded anomalously high Zn I abundances (up to 0.5~dex) due to a a V I blend for stars with \teff $<$ 4500~K and [Fe/H] $< -0.5$~dex. Such values are typically only expected from weak s-process, Type Ia supernovae, or hypernovae contributions at much lower metallicities ([Fe/H] $< -1$~dex) \citep{kobayashi_2009,saito_2009}. 

\subsubsection{Light s-process elements (Sr I, Y II, \& Zr I/II)}
The slow neutron-capture process (s-process) is where the timescale for beta decay is much shorter than the time to capture a neutron. Light s-process element abundances decrease from $-1.0$~dex~$<$[Fe/H]~$< -0.5$~dex in Figure \ref{fig:median ls} with the number of free neutrons from ${}^{13}$C($\alpha$,n)${}^{16}$O and ${}^{22}$Ne($\alpha$,n)${}^{25}$Mg in low- and high-mass stars, respectively \citep{pagel_galactic_1997}. The production of s-process elements is therefore dependent on the stellar mass. The low neutron flux of 10$^{-7}$cm$^{-3}$ \citep{pagel_galactic_1997} required to synthesise these light s-process elements allows for notable contributions from both low- and intermediate-mass AGB stars \citep{kobayashi_origin_2020}. All light s-process element abundances are compared to that of \citet{delgado_mena_chemical_2017} in Figure \ref{fig:median ls}, which we discuss in the following element summaries of Sr I, Y II and Zr I/II. Note that \citet{delgado_mena_chemical_2017} include temperature corrections to abundance trends of neutron capture elements and cuts due to blends with some lines for stars with \teff $< 5300$~K. While we correct for systematic offsets between line-by-line abundances of each element in Section \ref{sec:line corrections}, we do not remove/correct the abundance-temperature trends of individual features. We did not correct the abundance-temperature trends since we found all measured lines of each element had similar temperature trends. We also visually inspected fits for cool stars and perform selection cuts to exclude particular temperature regions if there were blends that were not captured. Few stars are present at [Fe/H] $>0.1$ due to HARPS selection effects and were not removed by selection cuts.

\paragraph*{Sr I:}
Here we discuss our Sr I abundance trend displayed in Figure \ref{fig:median ls} in comparison to that of \citet{delgado_mena_chemical_2017} shown in grey. Our Sr I trend across [Fe/H] follows \citet{delgado_mena_chemical_2017} with a 0.5~dex decrease from $\mathrm{-1~dex} < $[Fe/H] $< \mathrm{-0.5~dex}$ before increasing. This is enhanced by $\sim0.2$~dex compared to the trend of \citet{delgado_mena_chemical_2017} possibly due to NLTE effects. Our analysis utilised the Sr I lines at 4813~\AA{} and 4963~\AA{}, which form deep in the photosphere where higher density largely enforces LTE. In contrast, \citet{delgado_mena_chemical_2017} relied on the 4607~\AA{} resonance line. This Sr I resonance line has been found to have significant NLTE corrections of 0.21~dex in Arcturus \citep{bergemann_sr_2012}, as 1D LTE models ignore the radiation field in the upper photosphere. By neglecting this radiation, the models fail to account for the severe over-ionisation of Sr I into Sr II. However, the apparent decrease in Sr I for stars [Fe/H] $> 0.1$~dex is likely due to an emerging Ni I blend.

\paragraph*{Y II:}
Our Y II abundances show an increasing trend with [Fe/H], consistent wtih \citet{delgado_mena_chemical_2017}, as shown in Figure \ref{fig:median ls} coloured by \teff and in grey, respectively. The only common line between our analysis and \citet{delgado_mena_chemical_2017} was 5404~\AA{} for Y II. This line in our analysis and our chosen reference line at 5730 \AA{} both have a mean abundance of 0.01~dex. Selecting either line as our reference line produces a similar median Y II abundance trend, but the line at 5404~\AA{} has a Fe I blend that we attempted to account for. We did include the Y II line 4856~\AA{} which \citet{delgado_mena_chemical_2017} discarded due to its strong temperature trend. We found that the temperature trend for this line was the same as for all the other measured Y II lines, and did not remove it.
\\\\
There is a single star HD5891 that deviates from the expected Y II trend with [Y II/Fe] $= -0.49$~dex. This star has a weaker line depth of 0.95 compared to the average depth of 0.9, leading to a change in line profile from an emerging Ni II blend to the right. We also derived a higher effective temperature for this star of 5131~K compared to 4825~K from \citet{mortier_2013}, which would preferentially broaden the wings of the line in the \textsc{korg} spectrum fit to match the blend instead of increasing the line depth.

\paragraph*{Zr I \& Zr II:}
Zr I and Zr II show similar abundance trends, decreasing from $-1$~dex~$<$~[Fe/H]~$<$~$-0.5$~dex and increasing at higher metallicities. These trends are both displayed in Figure \ref{fig:median ls} coloured by \teff and mostly agree with that of \citet{delgado_mena_chemical_2017} shown in grey. We used the same Zr I absorption features as \citet{delgado_mena_chemical_2017} as well as 1 common Zr II absorption feature. However, Zr II has stars [Fe/H] $> 0$~dex with depleted Zr II abundances compared to \citet{delgado_mena_chemical_2017}, \citet{battistini_neutron_2016} and our Zr I abundances. Zr II is weaker in these cool (\teff $<$ 4500~K) metal-rich stars and blended with a strong Fe I line, whereas Zr I becomes stronger in these stars and has weaker blends that do not affect the line profile. As a result, Zr I abundances are more reliable than Zr II at [Fe/H] $> 0$~dex.

\subsubsection{Heavy s-process elements (Ba II, La II, Ce II, \& Nd II)}
Heavy s-process element abundances in Figure \ref{fig:median hs} increase with [Fe/H] but to a lesser extent than the light s-process elements in Figure \ref{fig:median ls}. The distinction between light and heavy s-process elements is the result of a decreasing neutron flux from light nuclei with small neutron capture cross-sections \citep{pagel_galactic_1997}. This leads to a greater production of light s-process elements at higher metallicities than heavy s-process elements \citep{vitali_slopes_2024}. The greater neutron flux required to produce heavy s-process elements means they are primarily formed in intermediate-mass AGB stars. Note that heavy s-process elements are still produced in low-mass AGB stars but typically to a lesser extent. Nd II contains a flatter trend than the other heavy s-process elements included in this catalogue due to contributions from massive stars \citep{adibekyan_chemical_2012}. We further discuss Figure \ref{fig:median hs} in the following element summaries of Ba II, La II, Ce II and Nd II. Note that \citet{delgado_mena_chemical_2017} include temperature corrections to abundance trends of neutron capture elements and cuts due to blends with some lines for stars with \teff $< 5300$~K. While we correct for systematic offsets between line-by-line abundances of each element in Section \ref{sec:line corrections}, we do not remove/correct the abundance-temperature trends of individual features. We also do not perform any selection cuts based on temperature. All measured lines of each element had similar abundance-temperature trends and their fits visually inspected across the parameter space. Few stars are present at [Fe/H] $>0.1$ due to HARPS selection effects.

\paragraph*{Ba II:}
The Ba II trend across [Fe/H] matches that of \citet{delgado_mena_chemical_2017}, with the exception of 8 stars at [Fe/H] $< -0.5$~dex that show a Ba II depletion below $-0.25$~dex. This apparent depletion is likely an artifact of fixing the continuum during line abundance optimisation. Hyperfine structure splitting broadened and depressed the wings of the Ba II lines (5855 \AA{} and 6143 \AA{}) in our \textsc{korg} synthetic spectra compared to the observed HARPS spectra. Fixing the continuum caused slight deviations in the line wing fits for both Ba II lines, although the core fits remained consistent across the parameter space. Allowing the continuum to vary produced a better fit to these wings and increased the resulting Ba II abundances by 0.3~dex. This would reconcile the observed deviation from the Ba II abundances of \citet{delgado_mena_chemical_2017} at [Fe/H] $< -0.5$~dex. We encountered a similar fitting challenge for 4 stars with [Ba II/Fe] $> 0.5$~dex; in these cases, the hyperfine broadening reduced the number of available continuum pixels in the fitting windows and nearby strong Fe I lines prevented us from increasing the window size. This made a fixed-continuum approach similarly problematic.

\paragraph*{La II:}
In Figure \ref{fig:median hs}, our La II trend coloured by \teff across [Fe/H] shows reasonable agreement with the measurements of \citet{battistini_neutron_2016} shown in grey. The \citet{cayrel_1988} line depth cut removed 16 stars from our La II trend. There were also an additional 2 stars we removed that remained within the scatter of  \citet{battistini_neutron_2016} but with [La II/Fe] $>0.4$~dex. These abundances are approximately 0.2~dex higher than most stars in our sample due to poor line fits upon visual inspection. The poor fits were the result of a large Fe blend and a nearby absorption feature depressing the continuum to the left of the line. The fits of all other stars were visually inspected and deemed reasonable.

\paragraph*{Ce II:}
Our Ce II abundance trend across [Fe/H] agrees with both \citet{delgado_mena_chemical_2017} and \citet{battistini_neutron_2016}. Figure \ref{fig:median hs} shows our Ce II abundances coloured by \teff with [Fe/H] compared to that of \citet{delgado_mena_chemical_2017} in grey. We used three Ce II lines that all had abundances within the mean of \citet{delgado_mena_chemical_2017} and \citet{battistini_neutron_2016} prior to applying line corrections to account for systematic offsets. The two lines at wavelengths $< 5000$~\AA{} had mean line abundances within 0.06~dex. However, blends in these lines led to the enhancement of [Ce II/Fe] above 0.3~dex that was not present in the third line at 5275~\AA{}. This resulted in a 0.15~dex line correction for the latter line.
\\\\
The three Ce II lines we measured were included in the analysis of \citet{delgado_mena_chemical_2017}. So as to remove cool stars with significant blending, \citet{delgado_mena_chemical_2017} applied temperature cuts to exclude stars with \teff $< 5300$~K for the line at 4524~\AA{}. We did not observe significant blending that impacted the fits of this line across our parameter space, and therefore do not apply any temperature cuts to our Ce II line abundances. We also noted that each of our three Ce II lines had the same abundance trend with temperature. The minimal impact of blending in our analysis is likely due to our use of spectral synthesis fitting to derive abundances, which is less susceptible to blending than equivalent widths (see \citet{jofre_blackbox_2017} \& \citet{blanco_caveats_2019}) implemented by \citet{delgado_mena_chemical_2017}.

\paragraph*{Nd II:}
Our Nd II abundances across [Fe/H] show strong agreement with the abundances of \citet{delgado_mena_chemical_2017} in Figure \ref{fig:median hs}. The median abundance of the 2 measured Nd II lines at 5294.63~\AA{} and 5742.41~\AA{} differed by only 0.02~dex. These lines were not included in the analysis of \citet{delgado_mena_chemical_2017} and we do not find a significant temperature dependence in our Nd II line abundances that they observed at \teff $<5500$~K for some lines. There is a Fe I blend to the left of the former line, which we mitigate by reducing the window size. We also note that both measured lines have hyperfine structure splitting that Korg and the \textit{Gaia} ESO line list include.

\subsubsection{r-process elements (Mo I \& Eu II)}
The rapid neutron-capture process (r-process) is where the timescale to capture a neutron is much shorter than the time to $\beta$ decay. The only confirmed production site for these elements are neutron star mergers but it has been hypothesised that Type II supernovae may be an additional contributor \citep{kobayashi_origin_2020,kobayashi_neutron_2023}. Both Eu II and Mo I display an $\alpha$-like abundance trend with [Fe/H] in Figure \ref{fig:median r} similar to Ca I in Figure \ref{fig:median alpha}. We further discuss Figure \ref{fig:median r} in the following element summaries of Mo I and Eu II. Note that we refer to Mo I as an r-process element but Mo I is 49.7$\%$ s-process in Sun-like stars \citep{prantzos_nc_contributions_2020}. 

\paragraph*{Mo I:} 
Our Mo I abundances coloured by \teff agree with \citet{mishenina_mo_2019} in grey across [Fe/H] in Figure \ref{fig:median r}. We measured Mo I from 2 lines at 5508~\AA{} and 6032~\AA{}. A strong Fe line to the right of 5508~\AA{} depressed the continuum on the left of the line. The uneven continuum to the left and the right of the Mo I line was difficult to fit in some stars given the fixed continuum placement when optimising for abundances.
\\\\
Other works have suggested that there may be deviations between red giant stars and main sequence stars. \citet{hansen_mo_2014} found a flat trend for Mo~I at [Fe/H] $< -0.63$~dex in giant stars in contrast to the decreasing trend of main sequence stars. However, \citet{hansen_mo_2014} did not include giant stars at higher metallicities. NLTE corrections are also expected to be negligible since the chosen Mo I lines form deep in the atmosphere, but there are no NLTE corrections for Mo I \citep{mishenina_mo_2019}.

\paragraph*{Eu II:}
The Eu II element abundance trend and individual absorption feature abundances agree with \citet{delgado_mena_chemical_2017} across [Fe/H]. This agreement is shown in Figure \ref{fig:median r} with our Eu II abundances coloured by \teff and \citet{delgado_mena_chemical_2017} in grey. We included two Eu II lines for which no line correction was required since both had the same mean abundance of 0.15~dex across all stars. There are no blends with other elements in the windows of either absorption feature. However, hyperfine structure splitting is present which \textsc{korg} accounts for by modeling both components provided in the line list. The absorption feature at 6646~\AA{} was also included in \citet{delgado_mena_chemical_2017}. 

\section{Discussion}
\label{sec:discussion}
\begin{figure}
    \centering
    \includegraphics[width=\columnwidth]{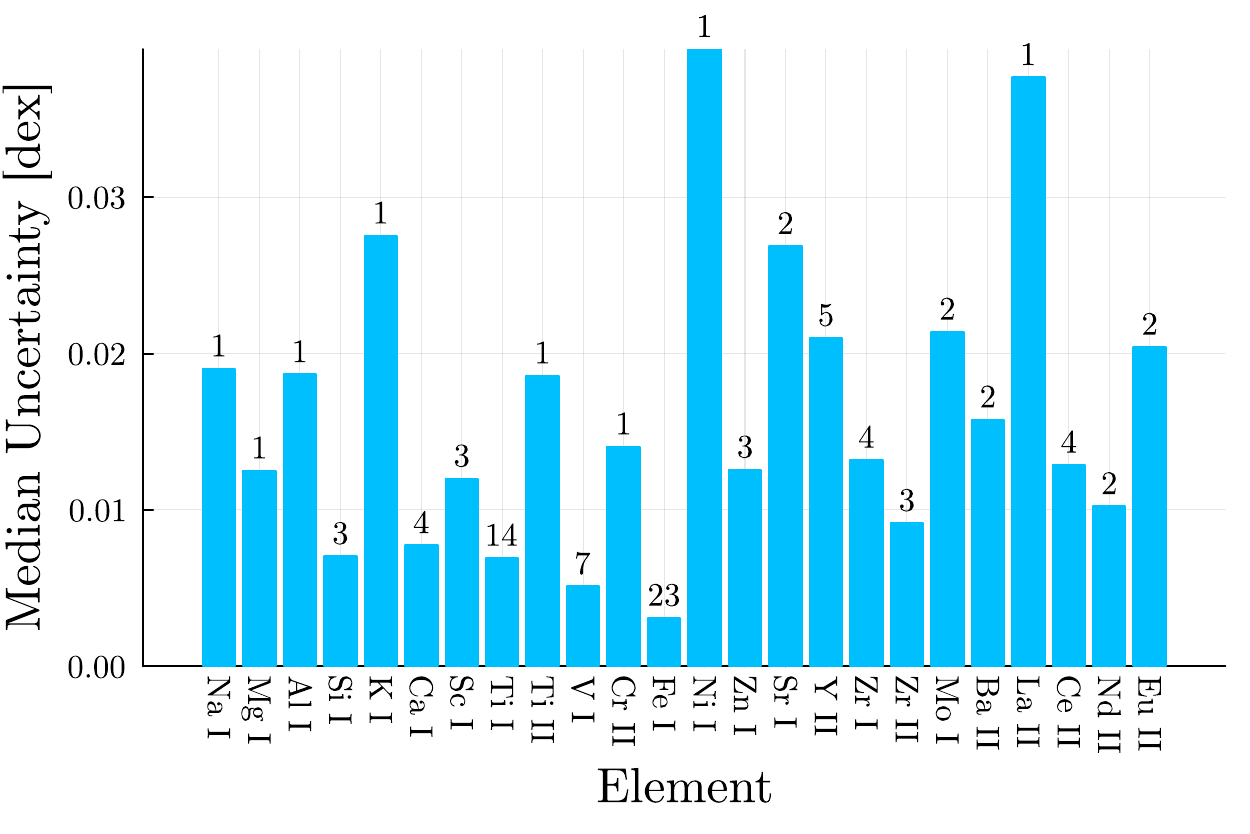}

    \vspace{1cm} 

    \includegraphics[width=\columnwidth]{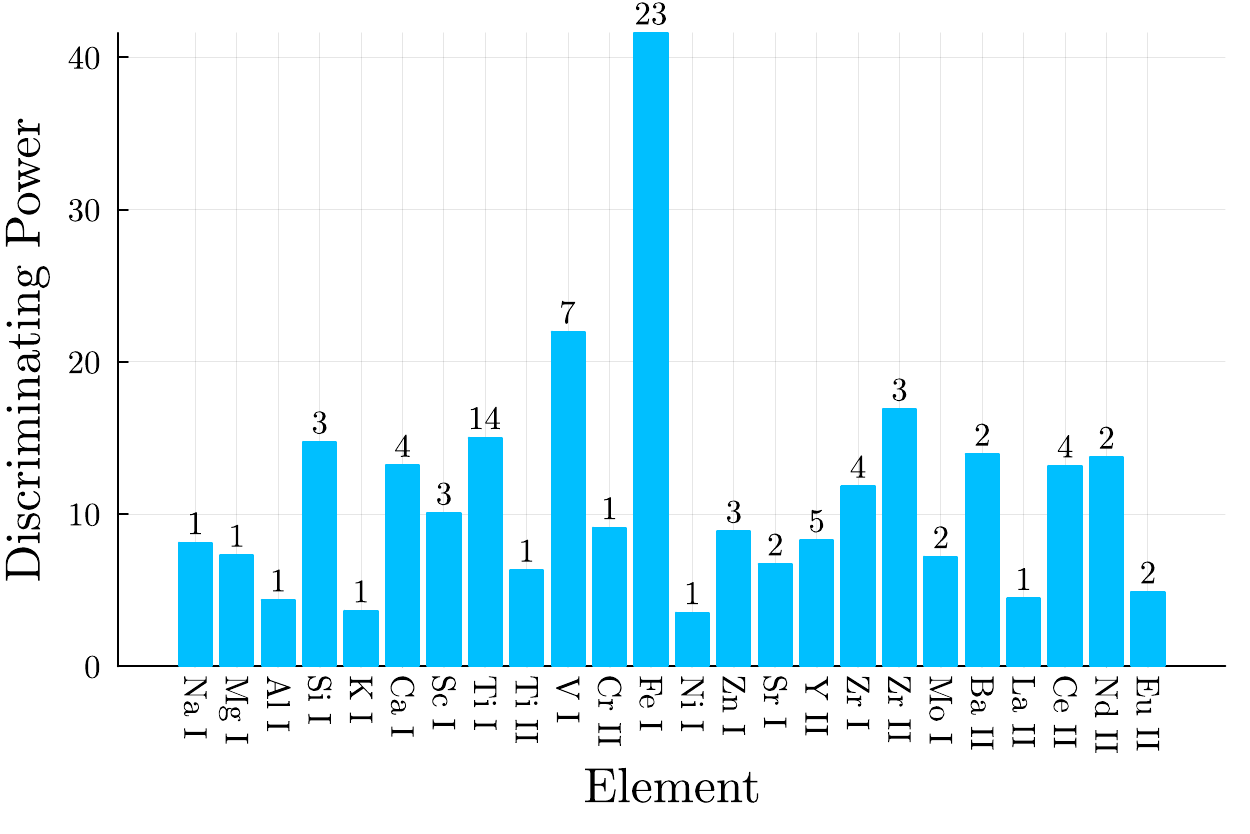}

    \caption{\textit{Top:} Distribution of the median uncertainty across all stars with recorded abundances per element. \textit{Bottom:} Distribution of the discriminating power across all stars with recorded abundances per element. The discriminating power is the ratio of the standard deviation of abundances and the median uncertainty. The number of lines used to calculate the overall abundance for each element is indicated above the contribution of each element.}
    \label{fig:median uncertainty and power}
\end{figure}

Our catalogue contains the stellar parameters (\teff, \logg, [M/H], \vmic and \vsini) and 22 element abundances (with an additional 2 ionised element abundances) of 426 red giant stars in HARPS. These are primarily low-alpha red giant stars which allow us to probe their unique chemical evolution in future work through age-abundance gradients and the intrinsic dispersion of elements conditioned on a subset. In this work, we identified which element measurements are potentially most powerful for investigating the chemical evolution of the low-alpha disk using their median abundance uncertainty and discriminating power across all stars shown in Figure \ref{fig:median uncertainty and power}. 
\\\\
The ratio of elemental dispersion to measurement uncertainty quantifies the number of bins of information contained in that measurement; more bins means more discriminating power between stars for that element. At a given SNR, the measurement uncertainty of an abundance is roughly inversely proportional to the square root of the number of lines available, and it also depends on the sensitivity given by the slope of the curve of growth for those lines. The intrinsic dispersion of the element is determined by the chemical evolution of the Milky Way and the sample selection function. The [Fe/H] measurement has the greatest discriminating power. Our sample spans stars across $-1$~dex~$<$~[Fe/H]~$< 0.3$~dex and the [Fe/H] measurement has the smallest median abundance uncertainty of 0.003~dex. The [Zr II/H] measurement has the highest discriminating power of the light s-process elements and [Ba II II/H] of the heavy s-process elements. This implies that [Zr II/H] and [Ba II/H] from this catalogue would be useful tracers for the production of elements in their respective families. However, Y II and Ce II have the largest number of measured absorption features among the light s-process and heavy s-process elements, respectively. Y II and Ce II also show trends across [Fe/H] that better follow the main sequence literature abundances we compared to in Section \ref{sec:results} than Zr II and Ba II which had blends. [Y II/H] and [Ce II/H] may therefore be better at encapsulating physical trends. [Ti I/H] also shows greater discriminating power than that of [Mg I/H] which is usually used to trace the production timescales of Type II supernovae \citep{ness_homogeneity_2022,griffith_kpm_2024,mead_regression_2025}. This is a consequence of using only a single Mg I absorption feature compared to 16 Ti I absorption features. The discriminating power of our [Ti I/H] measurements may improve constraints on its nucleosynthetic sources. This can help evaluate the extent to which jet-driven, multi-dimensional explosion mechanisms are needed to resolve the Ti underproduction seen in Galactic Chemical Evolution (GCE) models that rely solely on standard 1D Type II supernovae \citep{kobayashi_origin_2020}.

\subsection{Gradients between Element Families}
\label{sec:gradients}
\begin{center}
    \begin{figure*}
        \centering
    	\includegraphics[width=\textwidth]{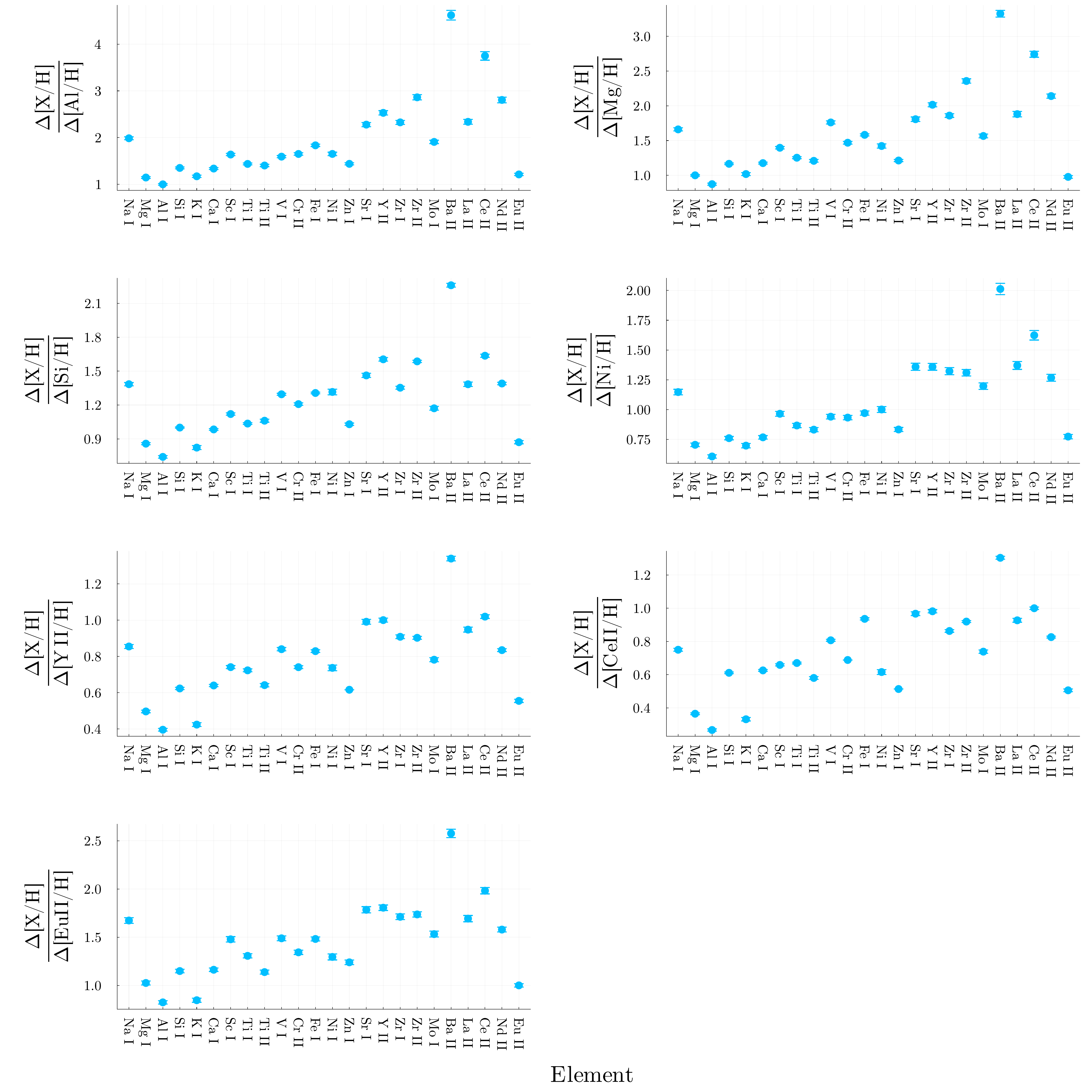}
    	\caption{Gradients between each elements and a chosen reference element for each family: Al I (odd-Z), Si I (explosive $\alpha$), Ni I (iron-peak), Y II (light s-process), Ce II (heavy s-process) and Eu II (r-process). Gradients of each element with respect to the reference elements were calculated using 214 stars with abundances recorded for every element. Uncertainties in the gradients are included but are not visible in most cases.}
        \label{fig:family slopes}
    \end{figure*}
\end{center}

A strong constraint on theoretical chemical evolution models is element abundance gradients $\left(\frac{\Delta [X_1 / \mathrm{H}]}{\Delta [X_2 / \mathrm{H}]} = \frac{\Delta A(X_1)}{\Delta A(X_2)}\right)$. Gradients are one way of parameterising the relative efficiency of different production sources. We derive gradients between all our elements and chosen reference elements of each nucleosynthetic family: Al I (odd-Z), Mg I (hydrostatic $\alpha$), Si I (explosive $\alpha$), Ni I (iron-peak), Y II (light s-process), Ce II (heavy s-process) and Eu II (r-process). All gradients are derived with abundances [X/H]. The inter-element gradients are calculated using the 214 stars with recorded abundances for each of the 22 elements. Gradients of each element with respect to the reference elements are derived using a Markov chain Monte Carlo sampler emcee \citep{Mackey_emcee_2013} with uniform priors (for the slope $-10 < m < 10$ and intercept $-20 < b < 20$) and a likelihood function based on the normalised orthogonal distance between points and the predicted line \citep{hogg_data_2010}. We included uncertainties in both the x- and y-axis, under the assumption they are uncorrelated. In practice, x- and y-axis uncertainties are likely to be correlated given that both abundances are derived from the same spectra and analysis procedures. However, we adopted the simplifying assumption of independence. Inter-element gradients are shown in Figure \ref{fig:family slopes} and are summarised in Table \ref{table:gradients}. These figures show each reference element on the denominator and every other element, shown along the x-axis ordered by increasing atomic number, is the numerator in the gradient calculation. For validation of our linear element gradients, gradient uncertainties and the sample of 214 stars see Appendix \ref{appendix:gradient validation}.
\\\\
We interpret the inter-element gradients in Figure \ref{fig:family slopes} such that gradients from 0.8-1.2 indicate similar production efficiencies, gradients $<$~0.8 indicate less efficient production and $>$~1.2 more efficient production relative to the production efficiency of the chosen reference element. Mg I has a gradient $\mathrm{\Delta [Mg~I/H]/\Delta [X_{ref}/H]} < 0.8$ with respect to Al I, Ni I, Y II and Ce II. This implies that the production efficiency of hydrostatic $\alpha$-elements is lower than odd-Z, iron-peak and s-process elements. However, Mg I is the only hydrostatic $\alpha$-element included in our catalogue. All neutron capture elements (excluding Eu II) have gradients $\mathrm{\Delta [n/H]/\Delta [X/H]_{Al~I, Eu~II}}>1.2$. Higher gradients for neutron capture elements are possibly due to greater contrast between production timescales relative to these reference elements. Neutron capture element gradients with respect to Y II and Ce II ($\mathrm{\Delta [n/H]/\Delta [X/H]_{Y~II,Ce~II}}$) are within 0.8-1.2 since there is little contrast in the timescales. We observed higher gradients for Ba II with respect to Y II and Ce II but this is likely due to scatter in our Ba II abundance trend at [Fe/H]~$< -0.25$~dex (see Figure \ref{fig:median hs}). Another exception is $\mathrm{\Delta [Eu~II/H]/\Delta [X/H]_{Y~II,Ce~II}}$, which shows a lower gradient value (see the sub-Figures in \ref{fig:family slopes} with Y II and Ce II on the denominator). The similar production timescales of Eu II to the hydrostatic $\alpha$, explosive $\alpha$ and odd-Z families implies that Eu II has multiple production sites. A similar conclusion was reached by \citet{mead_regression_2025} who found in their modeling of abundances that the inclusion of an r-process element in their predictive model did not improve the predictive power for any element, r-process or other, beyond their model of element prediction built on Mg I, Fe I, Y~II and Ce II. From this they inferred that it is not produced in lockstep with the ensemble of other elements.
\\\\
In Figure \ref{fig:family slopes} we note peaks in the inter-element gradients located around each element family (eg., around Si I, Sc I and Ba II). The peaks occur at the same element within each family. These peaks may reflect a varying metallicity and mass dependence in the production of individual elements as well as the diverse range of sources that contribute to some elements. For example, the elements Mo and Nd both have appreciable s- and r-process contributions in Sun-like stars with 49.7$\%$ and 61.5\% being s-process, respectively \citep{prantzos_nc_contributions_2020}. Mo I and Nd II also both have similar gradients with respect to Ni I, Y II, Ce II and Eu II; their gradients only imply different production efficiencies with respect Si I. Mo I displays a similar efficiency to the production of Si I, while the gradient of Nd II indicates a faster production efficiency. This difference in gradients may be because Mo I has a larger r-process contribution than Nd II, leading to a more $\alpha$-like Mo I abundance trend in Figure \ref{fig:median r}. The higher gradients of elements with greater s-process contributions with respect to Si I are potentially due to the similar production timescales of Type II supernovae and AGB stars. It could further be highlighting the lower efficiency of r-process production compared to the prompt production of $\alpha$-elements since Eu II has a similar gradient with respect to Si I as Mo I. This implies that the production efficiency of Mo I is mostly driven by its s-process contribution except in the case of the Si I where the contrast in timescales becomes significant.

\begin{table*}
\centering
\begin{tabular}{ c c c c c c c c c c c c c c c } 
 \hline 
  \textbf{Element} & \textbf{X\_Al} & \textbf{u$\_$X\_Al} & \textbf{X\_Mg} & \textbf{u$\_$X\_Mg} & \textbf{...} \\
 \hline
  Na & 1.99 & 0.03 & 1.66  & 0.02   & ... \\ 
  Mg & 1.15 & 0.02 & 1.00  & 0.01   & ... \\
  ...& ...  & ...   & ...   & ...     & ...  \\
 \hline 
\end{tabular}
\caption{Sample Table containing the gradients between every element and reference elements of each nucleosynthetic family. The uncertainty from the MCMC is included.}
\label{table:gradients}
\end{table*}

\begin{center}
    \begin{figure}
        \centering
    	\includegraphics[width=\columnwidth]{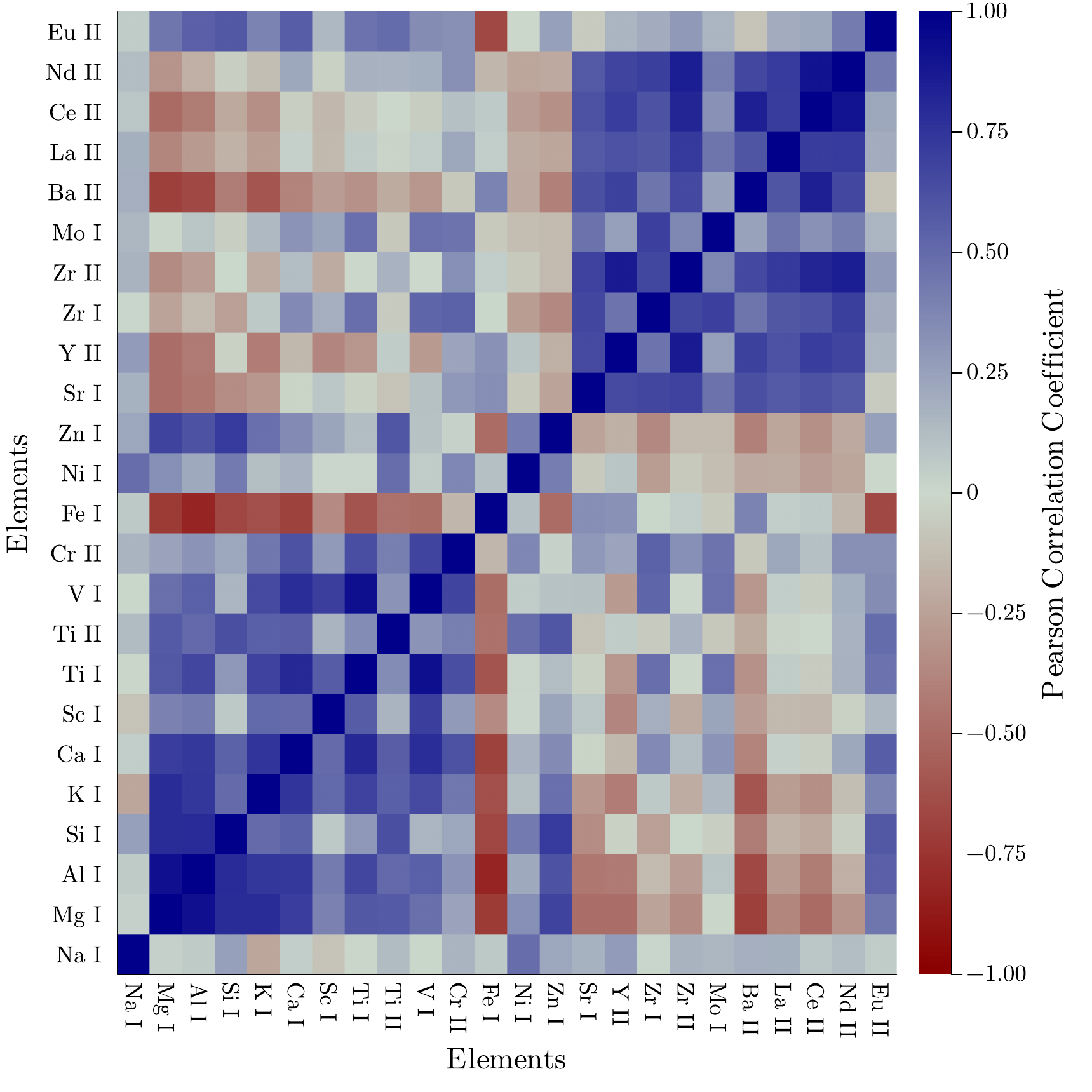}
    	\caption{Pearson correlation coefficients between all elements using the 105 stars with abundances recorded for each of the 22 elements, including the additional 2 ionisation states.}
        \label{fig:correlation heatmap}
    \end{figure}
\end{center}

\noindent
In addition to reporting the inter-element gradients, which have direct physical meaning with the relative change in one element per unit change in another, we show a map of the inter-element Pearson correlation coefficients for each element with respect to H. This latter metric provides a statistical summary of the tightness of the relationship between each element. The Pearson correlation coefficients between each element ([X/H])  is shown in Figure \ref{fig:correlation heatmap} with red indicating anti-correlations.

\subsection{Latent Model of Abundances}
\label{sec:latent model}
We used the element abundances of the 214 stars with all elements measured to construct a latent model representation with non negative matrix factorisation described in Ness et al. (submitted). This parameterises the abundances for the 214 stars as 
\begin{equation}
\label{eqn: latent model}
    X = f \times P
\end{equation}
where $P$ is a subset of latent sources (shared between the population) and $f$ the fractional contribution of each source (per star). This model uses non-negative matrix factorisation of the [X/H] abundances to solve for the per-star fractional contribution ($f$) of the so-called latent `channels' ($P$) that are shared among the population in Equation \ref{eqn: latent model}. Each channel represents the enrichment fingerprint of one or more sources. In Ness et al. (submitted), they found that $m=4$ latent patterns generates 16 element abundances of (O, Mg, Al, Si, S, K, Ca, Ti, V, Cr, Mn, Fe, Co, Ni, Ce) with an overall reduced $\chi^2_{mode}\sim 1.2$ for the population across all element abundances. They find a marginal improvement again for m=5. At $m = 6$, the $\chi^2$ value drops below 1 in their study, indicating that using more latent variables than $m = 5$ leads to overfitting. However, the optimal number of latent variables determined using the $\chi^2$-metric is subject to accurate measurement uncertainties. In general, increasing $m$ allows the model to better reproduce the data, but this approach provides a way to identify a small number of latent factors that can reasonably capture the observed abundance patterns without overfitting. We found that 4 channels produced a median $\chi_{reduced}^2 = 6.8$ of the 214 stars when comparing the predicted to our measured HARPS abundances. At $m=6$, we obtained a median $\chi_{reduced}^2 = 5.8$. The need for additional channels than what was reported in Ness et al. (submitted) presumably is due to the larger diversity of elements in our study, in particular the s- and r- process elements Sr I, Y II, Zr I, Zr II, Mo I, Ba II, La II, Ce II, Nd II and Eu II. We used the reduced $\chi^2$ as a diagnostic to assess how many channels are required to recover the overall element abundance behaviour. We do not interpret it as a strict statistical goodness-of-fit metric. Because our abundance measurements have very small uncertainties, even small absolute differences between the data and the model lead to large $\chi^2$ values. In this regime, $\chi^2$ is highly sensitive to modest model imperfections, and is therefore used comparatively to evaluate improvements in the reconstruction rather than to define a formal optimum.
\\\\
The reconstruction of 22 element abundances with 6 channels illustrates that these elements lie close to a low-dimensional subspace. {Nevertheless, imperfections in the reconstructed abundances suggest that individual elements encode higher-order information not captured by a 6-dimensional model. This demonstrates that subsequent relationships between these elements (characterised in the correlation coefficients and the inter-element gradients) offer critical constraints on chemical evolution and the origin of elements. These constraints can be compared and applied to chemical evolution models. 
\\\\
The generated abundances from the latent model in Equation \ref{eqn: latent model}, obtained by multiplying the solved matrix $P$ by matrix $f$ for a selection of our HARPS stars, is displayed in Figure \ref{fig:latent model comp}. These eight stars were randomly selected to span our parameter space in \teff, \logg, [M/H] and SNR. The model performance is uniform across variations in parameter space and uncertainties, demonstrating its effectiveness with $m=6$ latent patterns.

\begin{figure*}
    \centering
    \includegraphics[width=\textwidth]{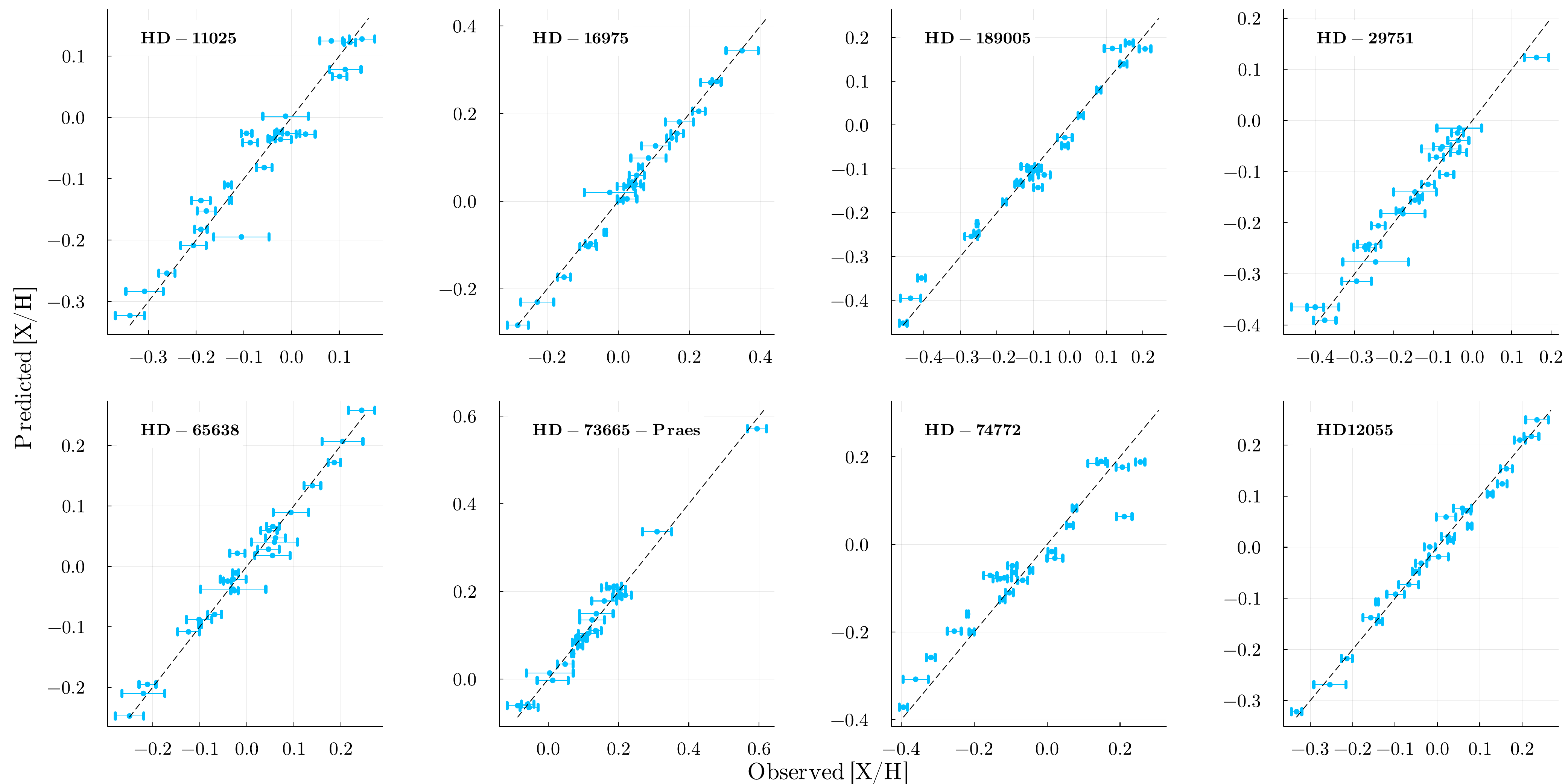}

    \caption{Measured abundances ([X/H]) for 22 elements compared to predicted abundances from the latent model in Ness et al. (submitted), with 6 latent channels that are shared among stars (and with fractional contributions that vary for each star). All elements in the catalogue are included. The eight stars were randomly selected to span our parameter space. This shows that a latent basis of $m=6$ can reconstruct the abundance patterns of the 22 elements.}
    \label{fig:latent model comp}
\end{figure*}

\noindent
\\
The contribution of each element used in the model to each of the 6 latent patterns is displayed in \ref{fig:latent contribution}. Each pattern has been normalised by dividing by the maximum element contribution (i.e., to set the element with the highest contribution to 1), so that it is clear which elements are grouped in each pattern. Darker segments indicate higher channel contributions for particular production sites and element families. Although mathematically only representative of the element covariance, the patterns in Figure \ref{fig:latent contribution} can also be tentatively linked to physical processes. Physical interpretation is tentative as co-varying structure in the latent space may also be driven by element abundance measurement systematics, relative uncertainties and outliers.

\begin{figure}
    \centering

    \includegraphics[width=\columnwidth]{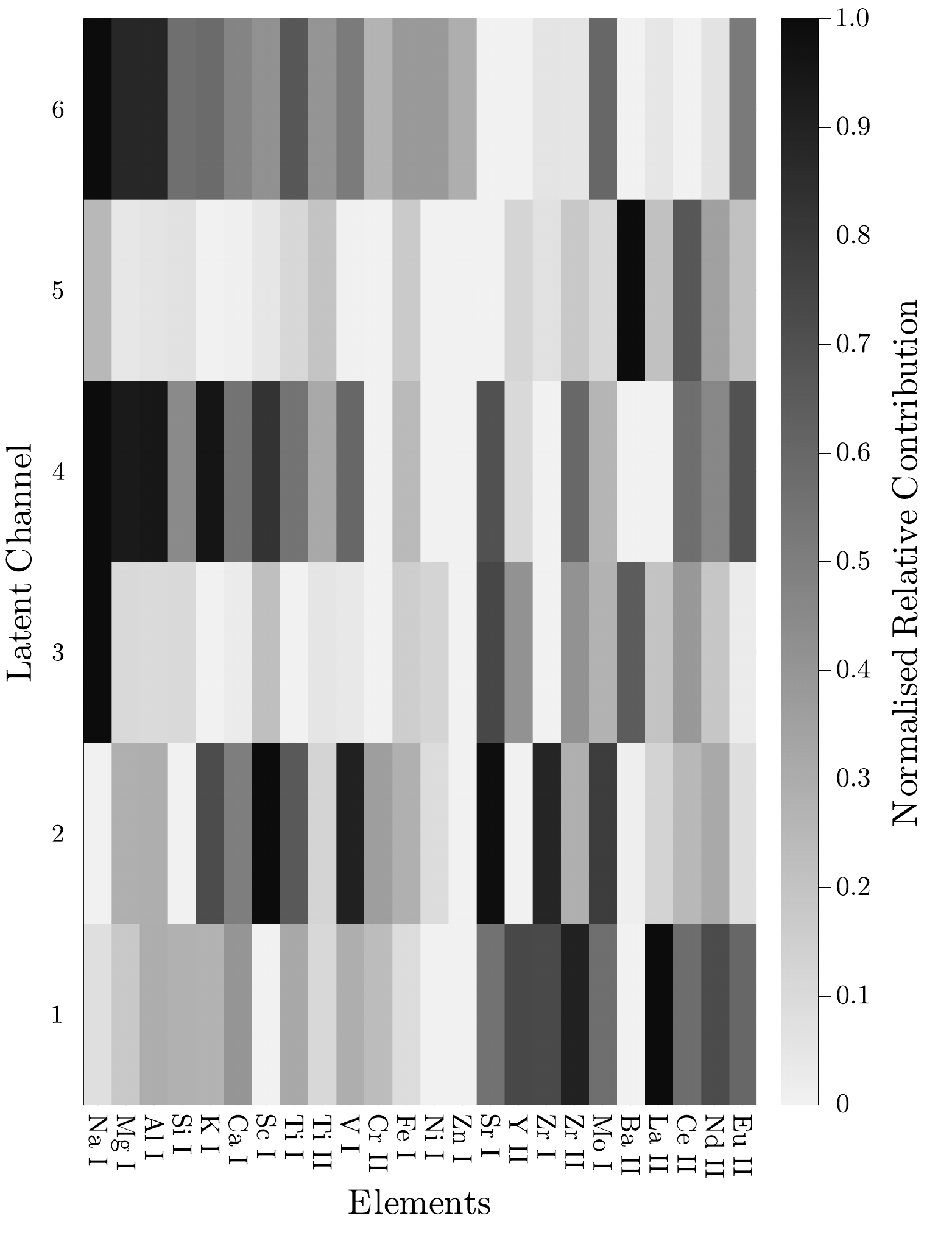}

    \caption{Normalised contribution to every element from each of the 6 channels in the latent model. The contributions are determined from the 214 stars with all elements reported. Darker segments indicate higher channel contributions for particular production sites and element families.}
    \label{fig:latent contribution}
\end{figure}

\noindent
\\
The groups of elements that contribute to channel patterns 1, 3, 5 and 6 trace physical similarities in production timescales that mirror the Pearson correlation matrix in Figure \ref{fig:correlation heatmap}. Channel 1 possibly reflects the smooth transition in the mass-dependent production timescales of light and heavy s-process elements, as both are produced in low-mass AGB stars. Channel 3 may reflect the metallicity dependence of odd-Z and s-process elements with contributions to Na I, Sr I and Ba II. Channel 5 reflects the production of low-mass AGB stars through contributions to heavy s-process elements. Channel 6 reflects the production of Type II supernovae through $\alpha$-element contributions as well as that of massive stars through odd-Z contributions. 
Channel patterns 2 and 4 can be linked to physical process, but appear to include additional non-physical signals, as elements from different production sites covary. Channel 2 may reflect production from intermediate to massive stars with contributions to Sc I, Ti I, V I, Sr I, Zr I, Mo I. However, this may also be tracing non-physical trends since the production timescales of these elements vary. We see little correlation between Zr I and iron-peak elements as well as between Zr I and $\alpha$-elements in Figure \ref{fig:correlation heatmap}. Mo I also shows litter correlation with iron-peak and $\alpha$-elements.} Channel 4 reflects that of Type Ia supernovae as we see contributions to iron-peak elements. Contributions to Al I and Mg I may be tracing advanced Hydrogen burning, while contributions to Na I are likely non-physical. We see little correlation between Na I and iron-peak elements in Figure \ref{fig:correlation heatmap}.
\\\\
We find no contributions to Eu II from channels 2, 3, and 5, which generally reflect the production of neutron-capture elements. Eu II production is indicated only by channels 1, 4, and 6, mirroring its correlation with $\alpha$-elements in the correlation matrix (Figure \ref{fig:correlation heatmap}). However, the Eu II contribution in channel 1 implies co-production with other neutron-capture elements across a specific stellar mass range. This link to $\alpha$-element production identifies massive stars and Type II supernovae as a production site of Eu II.

\section{Conclusions}
\label{sec:concl}

There has been a large increase in the number of stellar surveys over the last decade from echelle and fibre-fed spectrographs. Different surveys as well as the numerous assumptions implemented by different procedures can lead to variations in the derived stellar parameters and abundances for the same stars. It is therefore useful to have a catalogue of self-consistently derived parameters and abundances. In this first paper, we provide a self-consistent, high-precision catalogue of stellar parameters (\teff, \logg, [M/H], \vmic and \vsini) and 22 element abundances at a median precision $\sim$0.02~dex for 426 red giant stars in HARPS. Stellar parameters and line-by-line abundances were derived by using one of the most recently released spectral synthesis codes \textsc{korg}. We see systematic offsets between our stellar parameters and other analyses due to differences in methodology. However, such comparisons for stars in common with other analyses as well as the disk demonstrate overall good agreement indicative of the fidelity of our measurement. This high precision catalogue serves as a resource for understanding the origin of the elements, testing for signatures of planetary systems and for training data-driven models to derive element abundances.
\\\\
The systematic offset between our parameters and literature \citep{jofre_mh_2014, heiter_teff_2015, luck_2015, adibekyan_kgiants_2015, soubrian_gbs_2024} is reflected as offsets in $\alpha$- and iron-peak elements of red giant abundances when comparing to \citet{adibekyan_kgiants_2015}. Ti II is the only element to show deviations likely due to the impact of surface gravity on ionised species. It is difficult to make one-to-one comparisons with literature given the lack of consistently measured red giant star abundances. We compared to the abundance trends of main sequence stars and find good agreement for most elements. Na I and Zn I have an offset of 0.25~dex and 0.2~dex, respectively, from that of main sequence star abundances. We believe the offset between our work and the Zn I abundances from \citet{delgado_mena_chemical_2017} are explained by incorrect oscillator strengths. An incorrect oscillator strength is also the likely cause for the offset we observe between our Na I abundances and those from \citet{adibekyan_chemical_2012}. Discrepancies in abundances between ionisation states could arise from NLTE effects, which we have not included corrections for, as well as incorrect atomic data. More observations of thick disk stars with HARPS may disentangle deviations from evolutionary effects and would allow a comparison of the high- and low-$\alpha$ disk between red giant and main sequence stars. The inclusion of more metal-rich stars ([Fe/H] > 0.0) would also provide additional reference objects to further improve stellar models. Scatter in the line-by-line abundances could be reduced by using adaptive window sizes across our parameter space. This would account for the blending and weakening of absorption features with increasing and decreasing effective temperature, respectively. 
\\\\
We report inter-element gradients as key constraints on chemical evolution model and theory. These gradients quantify the rate of change of one element with respect to another across the disk population we examine. Gradients between all elements and elements chosen to reference each family revealed peaks of increasing magnitude located around each element family. We believe that the systematic mean increase in the inter-element gradient amplitudes with atomic number (numerator element) is due to the greater contrast in production timescales relative to the reference elements (denominator). The ensemble of peaks across atomic number themselves may reflect a varying metallicity dependence in the production of (denominator) individual elements within the same nucleosynthetic family. The correlation in element production efficiency between Eu II and elements in other nucleosynthetic families implies multiple sources for r-process elements. In general, we see elements from the same nucleosynthetic family have similar production efficiencies as quantified by these gradient calculations. The grouping of element families is also illustrated in the correlations in Figure \ref{fig:correlation heatmap}. Combined with the measurement of discriminating power, this means that the elements Ti I, Fe I, Zr II and Ba II, which have high discriminating power in this catalogue and derive from different nucleosynthetic families, could be used as key elements to probe the dimensionality of chemical space \citep{mead_regression_2025,griffith_kpm_2024}. Gradients and correlations in Figures \ref{fig:family slopes} and \ref{fig:correlation heatmap} also indicate that r-process elements are unlikely to provide unique information from that of other families. 
\\\\
Comparing these inter-element gradients to GCE model predictions is required to improve constraints on r-process and light s-process production sites. The 6-parameter latent model (see Figure \ref{fig:latent contribution}) re-parameterised individual elements as enrichment channels to provide a more direct connection to theoretical models of enrichment. Subsequently, with this catalogue we can map how these enrichment channels change over evolutionary state and metallicity. Further comparisons to GCE models could be made by adjusting the production site contributions in a GCE model to determine which best matches the observed abundance pattern \citep{galarza_gce_2010}. The Pearson correlations, latent patterns and the precision of the inter-element gradients will further aid in refining input parameters such as the size of the ${}^{13}$C pocket and potentially deepen our understanding of mixing processes.
\\\\
For the main-sequence catalogue, we will derive line-by-line abundances for the same 22 elements included in this red giant catalogue: Na I, Mg I, Al I, Si I, K I, Ca I, Sc I, Ti I/II, V I, Cr II, Fe I, Ni I, Zn I, Sr, Y II, Zr I/II, Mo I, Ba II, La II, Ce II, Nd II, and Eu II. Additionally, we will attempt to measure abundances for another 25 elements, including: Li, C, N, O, S, Mn, Co, Cu, Rb, Ru, Rh, Ag, In, Sb, Sm, Gd, Tb, Dy, Ho, Er, Tm, Ir, Au, Pb, and Th. These elements will provide valuable information for light elements crucial to modeling stellar atmospheres, the third s-process peak and a larger range of r-process elements. The pipeline we have developed for deriving stellar parameters and abundances can also be adapted for other high-fidelity stellar spectra such as Veloce \citep[$R=75000$,][]{Taylor_veloce_2024} and the High-Resolution Multi-Object Spectrograph \citep[HRMOS, $R=60000-80000$,][]{magrini_hrmos_2023}. Veloce focuses on M-dwarf stars which are difficult to analyse at lower resolution due to their cool atmospheres. It is predicted HRMOS will obtain 20-100 simultaneous observations from 3800~\AA{} to 8000~\AA{} and will allow for chemical abundance determinations at a precision of 0.01~dex. Both surveys will be crucial for tracing chemical evolution across the Milky Way. This series of catalogues deriving abundances with \textsc{korg} from high fidelity spectra serves to provide precision abundances to learn directly about the origin of the elements, constrain theory and build data-driven models.

\section*{Acknowledgments}
We are grateful to the anonymous reviewers for their careful reading of our manuscript and their many invaluable suggestions. We would also like to thank Thomas Nordlander for providing insights into different line lists and continuum normalisation. 
\\\\
This research has made use of data obtained from or tools provided by the portal exoplanet.eu of The Extrasolar Planets Encyclopaedia.
\\\\
Based on data obtained from the ESO Science Archive Facility with DOI: https://doi.org/10.18727/archive/33
\\\\
This publication makes use of data products from the Two Micron All Sky Survey, which is a joint project of the University of Massachusetts and the Infrared Processing and Analysis Center/California Institute of Technology, funded by the National Aeronautics and Space Administration and the National Science Foundation.
\\\\
This work makes use of data from the European Space Agency (ESA) space mission Gaia. Gaia data are being processed by the Gaia Data Processing and Analysis Consortium (DPAC). Funding for the DPAC is provided by national institutions, in particular the institutions participating in the Gaia MultiLateral Agreement (MLA). The Gaia mission website is https://www.cosmos.esa.int/gaia. The Gaia archive website is https://archives.esac.esa.int/gaia.

\section*{Data Availability}
All input data used are publicly available or can be requested from the lead author.

\bibliographystyle{mnras}
\bibliography{mnemonic,main} 

\appendix
\section{Validation of Fe stellar parameter lines}
\label{appendix:line fits}

\begin{figure}
    \centering

    \includegraphics[width=\columnwidth]{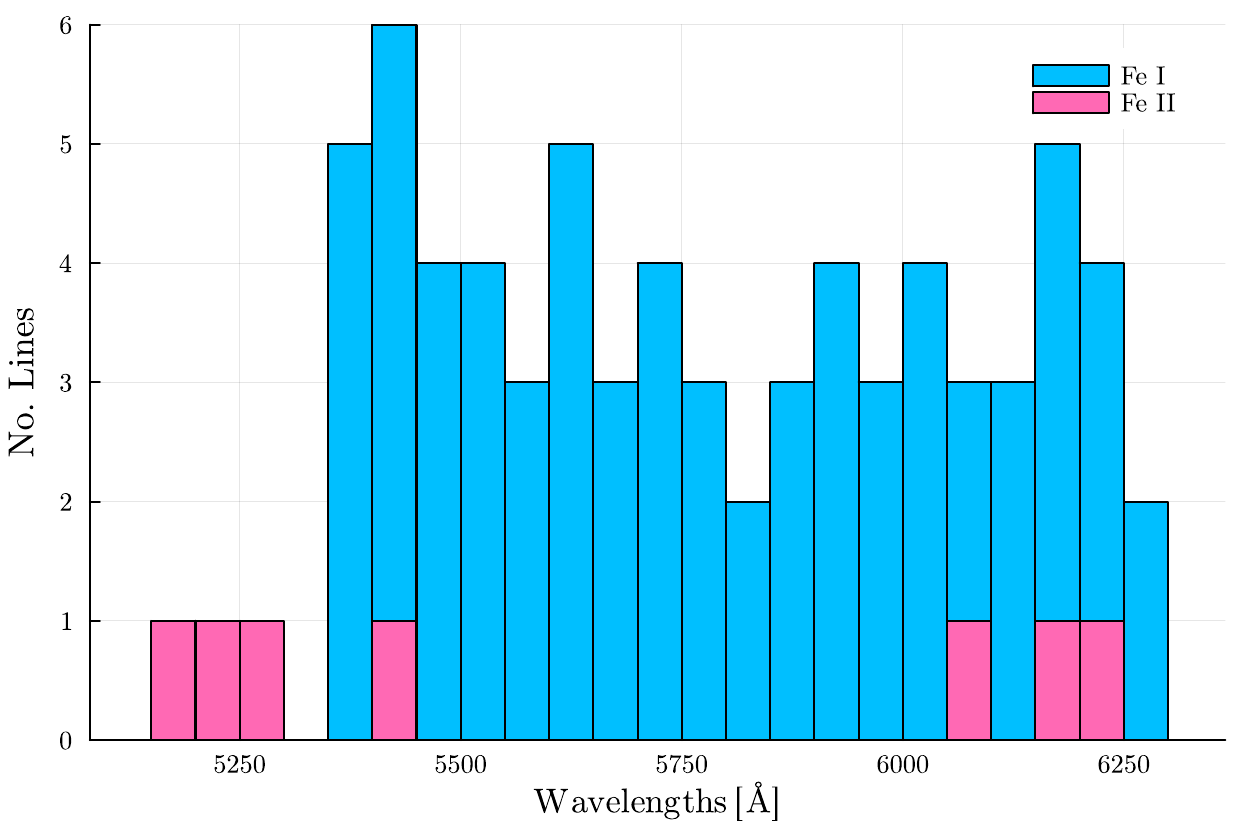}
    \caption{Wavelengths in the vacuum rest frame of the 60 Fe I lines and 7 Fe II lines we used to derive our stellar parameters. These Fe lines are a subset of those used to calculate stellar parameters in \citet{griffith_untangling_2023}.}
    \label{fig:fe param wl hist}
\end{figure}

\begin{figure}
    \centering

    \includegraphics[width=\columnwidth]{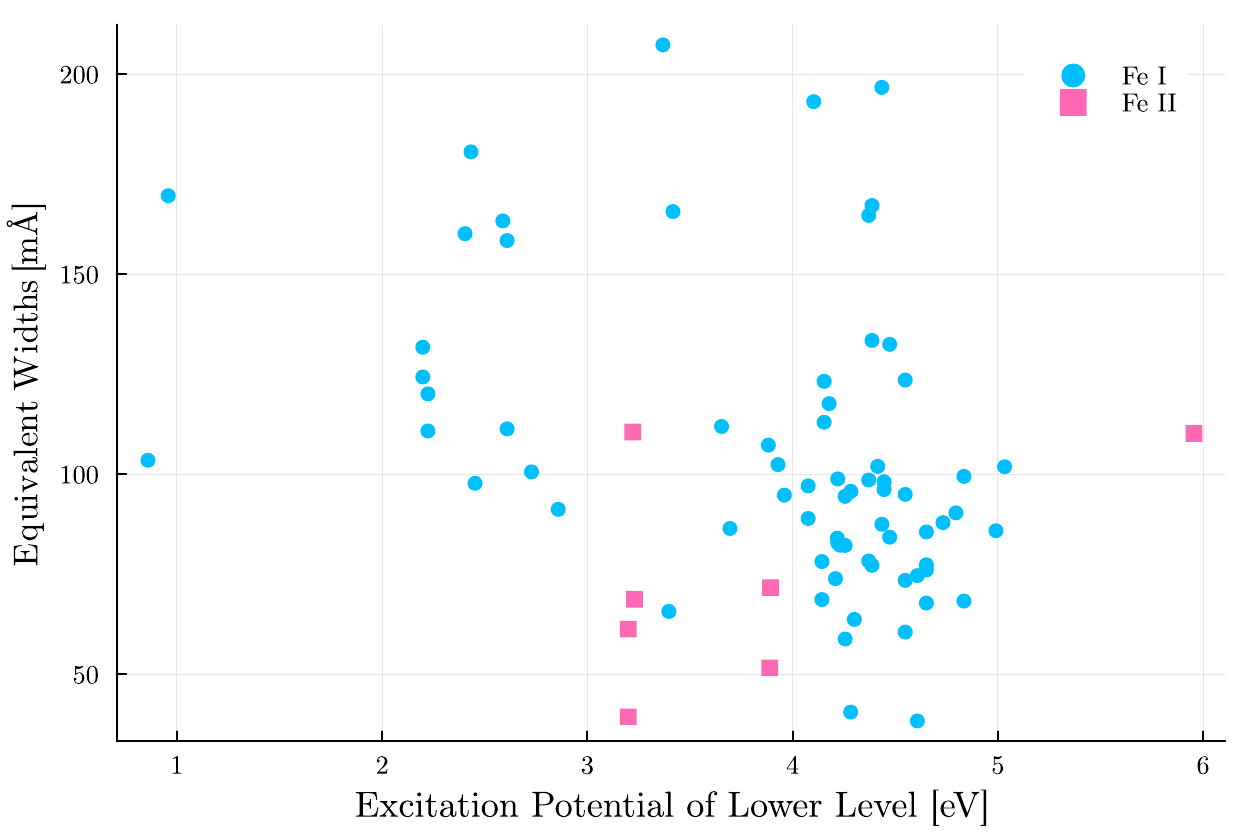}
    \caption{Equivalent widths and excitation potentials at the lower energy level (from the \textit{GAIA} ESO line list) for the 70 Fe I lines (blue circles) and 7 Fe II lines (pink squares) in $\lambda$-Pyxidis we used to derive our stellar parameters.}
    \label{fig:ew elow plot}
\end{figure}

To derive our stellar parameters (\teff,\logg,[M/H],\vmic \&\vsini) we used 77 Fe lines. These Fe lines are a subset of those used by \citet{griffith_untangling_2023} to calculate their stellar parameters. We demonstrate the range of Fe I and Fe II lines used in Figures \ref{fig:fe param wl hist}, which shows the range of wavelengths included. Figure \ref{fig:ew elow plot} shows the excitation potential of the lower energy level from the \textit{Gaia} ESO line list and the equivalent widths of each the Fe I and Fe II lines in the star $\lambda$~Pyxidis. Stars at lower metallicities in our sample show a similar distribution of equivalent widths that span 28~m\AA{} to 236~m\AA{}.
\\\\
Equivalent widths (EWs) for the selected Fe absorption features were determined via discrete numerical integration of the continuum-normalised spectra. The integration window of each line is $\pm 0.25$~\AA{}, the same window size used around each Fe line when deriving the stellar parameters. We then approximated the EW by summing the fractional line depth ($1 - f_\lambda$) at each pixel within this window and multiplying by the mean wavelength dispersion ($\Delta\lambda$) across the local pixel grid. We have 11 Fe lines that are notably strong ($\mathrm{EW} > 150\,\text{m\AA}$) that we retain in our analysis. Their reduced EWs do not exceed $-4.4$, placing them safely within the $\mathrm{-6.0 < REW < -4.2}$ threshold adopted by \citet{griffith_untangling_2023} to filter out severely saturated features.

\section{Formal Statistical Uncertainties and Model Sensitivities}
\label{appendix: accuracy test}

In this catalogue, we provided uncertainties on our stellar parameters and chemical abundances using repeat observations. This section details a supplementary evaluation. Here, we evaluated the isolated sensitivity of our model fits to \teff, \logg, and [M/H] by exploring the $\chi^2$ topology within the \textsc{korg} optimisation, independent of parameter covariances. We emphasise that model imperfections and the high SNR spectra led to minimum reduced $\chi^2$ values greater than unity when optimising the stellar parameters in Section \ref{sec:method-parameters}. As a result, the resulting $\Delta\chi^2=1$ intervals only provide a measure of the local parameter sensitivity of the fit and are not a formal uncertainty measurement.
\\\\
We selected three representative stars spanning the metallicity range of the catalog (HD~176704 at [M/H] = $+0.22$~dex, $\lambda$-Pyxidis at [M/H] = $-0.16$~dex, and HIP~20226 at [M/H] = $-0.67$~dex) to determine the sensitivity of the stellar parameters and line-by-line abundances using the following procedure. We first generated a grid of synthetic spectra in \teff, \logg, and [M/H], allowing each parameter to vary individually across a grid while fixing the remaining stellar parameters (\teff, \logg, [M/H], \vmic, and \vsini) to the best optimised values recorded in the catalogue. Centered on these best-fit values, the grids spanned $\pm 100$~K in steps of 20~K for \teff, $\pm 1$~dex in steps of 0.1~dex for \logg, and $\pm 0.5$~dex in steps of 0.1~dex for [M/H]. We utilised the same Fe lines, spectral windows and \textit{Gaia} ESO line list implemented to derive our recorded stellar parameters in Section~\ref{sec:method-parameters}. The synthetic spectra at each grid point was compared to the normalised HARPS spectra to calculate a reduced $\chi^2$. A parabola was fitted to the resulting reduced $\chi^2$ values across the parameter grid to interpolate the local shape of the $\chi^2$ surface. We then determined the two parameter values at which the fitted parabola reached $\chi^2_{\rm min}+1$, where $\chi^2_{\rm min}$ is the minimum reduced $\chi^2$ of the best-fitting model. Half of the difference between these two parameter values was adopted as the formal sensitivity for that parameter.
\\\\
We present our stellar parameter (\teff, \logg and [M/H]) sensitivities for our three representative stars (HIP 20226, $\lambda$~Pyxidis and HD 176704) across our parameter space in Table \ref{table: parameter sensitivity}. The stellar parameter sensitivities for all three stars are notably larger than the median empirical uncertainties discussed in Section \ref{results:parameters} ($\sigma_{\teff} = 21$~K, $\sigma_{\logg} = 0.04$~dex \& $\sigma_{\mathrm{[M/H]}} = 0.01$~dex). The disparity between our empirical precision and formal statistical uncertainties calculated here highlights the distinction between reproducibility and parameter sensitivity. The formal statistical uncertainties are derived from the shape of the $\chi^2$ surface for each parameter individually. As the minimum reduced $\chi^2$ is greater than unity ($\chi^2_{\mathrm{reduced}} = $13/49/45 for stars HIP 20226/$\lambda$-Pyxidis/HD 176704), these values should be interpreted as formal parameter sensitivities rather than strict $1\sigma$ confidence intervals. These larger $\chi^2$ values occur because our small flux uncertainties penalise minor model-data mismatches (despite the error inflation applied in Equation \ref{eqn:inflated error}), particularly regarding imperfect line wings or adjacent continuum features. If the $\chi^2$ statistic were dominated purely by statistical flux noise rather than these systematic model limitations, these formal sensitivities would likely be comparable to our empirical uncertainties. Therefore, these metrics quantify how sensitive the fit quality is to changes in each parameter. In contrast, the repeat-observation analysis measures the reproducibility of the complete analysis procedure from repeated observations of the same star. While a broad $\chi^2$ surface indicates that a range of parameter values provide similarly good fits to the spectrum, the small repeat-observation scatter demonstrates that the analysis procedure yields highly consistent stellar parameters and abundances in practice.

\begingroup{} 
    \setlength{\tabcolsep}{10pt} 
    \renewcommand{\arraystretch}{1.5} 
    \setlength{\extrarowheight}{2pt}    
    \begin{table}
        \centering
        \begin{tabular}{ c c c c}  
            \hline 
            \textbf{Star} & \textbf{$\boldsymbol{\sigma_{\teff}}$~[K]} & \textbf{$\boldmath{\sigma_{\logg}}$~[dex]} & \textbf{$\boldsymbol{\sigma_{[M/H]}}$~[dex]}\\
            \hline     
            {HIP 20226} & $51$ & $0.27$ & $0.04$\\
            {$\lambda$~Pyxidis} & $34$ & $0.17$ & $0.04$\\
            {HD 176704} & $60$ & $0.19$ & $0.07$\\
            
         \hline 
        \end{tabular}  
        \caption{Sensitivity of \teff, \logg and [M/H] for three representative stars HIP 20226 $\teff = 4722$~K, $\logg = 2.15$~dex, [M/H]~$= -0.67$~dex), $\lambda Pyxidis$ ($\teff = 5081$~K, $\logg = 2.97$~dex, [M/H]~$= -0.16$~dex) and HD 176704 ($\teff = 4457$~K, $\logg = 2.38$~dex, [M/H]~$= 0.22$~dex).}
        \label{table: parameter sensitivity}
    \end{table}
\endgroup{}

\noindent
\\\\
To propagate these parameter sensitivities to the chemical abundances, we individually perturbed the optimised \teff, \logg, and [M/H] values by their respective positive and negative sensitivities. For each perturbation, we re-derived the line-by-line abundances following the methodology outlined in Section~\ref{sec:method-line abundances}. This procedure yielded six distinct abundance variations (three positive, three negative) per spectral line for each of the three benchmark stars. The differences between these perturbed abundances and the best fit abundance ($\sigma_{X,l,\theta,+}$ for a given element $X$, line $l$ and parameter $\theta$) are combined in quadrature assuming that the parameter uncertainties are independent:

\begin{equation}
    \sigma_{X,l,+} = \sqrt{\sigma_{X,l,\teff,+}^2 + \sigma_{X,l,\logg,+}^2 + \sigma_{X,l,\mathrm{[M/H]},+}^2},
\end{equation}

\noindent
where $+$ refers to abundance differences obtained using the positive parameter perturbations. We performed the same calculation for negative parameter perturbations to obtain $\sigma_{X,l,-}$. This uncertainty per line neglects covariance between the stellar parameters and should be regarded as an approximation. 
\\\\
For elements with multiple measured lines in our catalogue, we combined all per line uncertainties from positive perturbations of that element as

\begin{equation}
    \sigma_{X,+} = \left(\Sigma_l^{N} \sigma_{X,l,+}^{-2}\right)^{1/2},
\end{equation}

\noindent
where $N$ is the number of measured lines for each element. We performed the same calculations for all negative perturbations to obtain a negative uncertainty ($\sigma_{X,-}$) per element. Positive and negative uncertainties for each of the 22 elements are reported in Table \ref{table: abundance sensitivity}.
\\\\
The large statistical sensitivities in the stellar parameters have translated to asymmetric chemical abundance sensitivities in Table \ref{table: abundance sensitivity} that are larger than the empirical measurement uncertainties reported using repeat visits. Only six element (Al I, K I, Ca I, V I, Ni I and Ba II) uncertainties for $\lambda$-Pyxidis remain within the empirical measurement uncertainties. The Fe I sensitivities also remain within the intrinsic scatter of the [Fe I/H]-[M/H] relation in Figure \ref{fig:median FeH vs MH}, demonstrating the robustness and high internal precision of the pipeline. Among individual species, heavy s-process (La II, Ce II and Nd II) and r-process elements (Mo I and Eu II) are more sensitive with shifts greater than 0.04~dex in all three stars. Nd II in particular spans 0.08~dex to 0.37~dex. Of our three representative stars, HD~176704 displays the largest sensitivities across all elements. This suggests that line formation is most sensitive to the adopted atmospheric structure in cool metal-rich stars--an effect that is amplified in heavy neutron-capture species.

\begingroup{} 
    \setlength{\tabcolsep}{10pt} 
    \renewcommand{\arraystretch}{1.5} 
    \setlength{\extrarowheight}{2pt}  
    \begin{table}
        \centering
        \begin{tabular}{ c c c c }  
            \hline 
            \textbf{Element} & \textbf{HIP 20226} & \textbf{$\lambda$~Pyxidis} & \textbf{HD 176704}\\
            \hline     
            {Na I} & ${}^{+0.04}_{-0.04}$ & ${}^{+0.03}_{-0.03}$ & ${}^{+0.06}_{-0.05}$ \\
            {Mg I} & ${}^{+0.03}_{-0.03}$ & ${}^{+0.02}_{-0.02}$ & ${}^{+0.04}_{-0.04}$ \\
            {Al I} & ${}^{+0.02}_{-0.03}$ & ${}^{+0.01}_{-0.02}$ & ${}^{+0.04}_{-0.06}$ \\
            {Si I} & ${}^{+0.03}_{-0.03}$ & ${}^{+0.02}_{-0.02}$ & ${}^{+0.04}_{-0.04}$ \\
            {K I} & ${}^{+0.04}_{-0.04}$ & ${}^{+0.03}_{-0.02}$ & ${}^{+0.06}_{-0.06}$ \\
            {Ca I} & ${}^{+0.02}_{-0.02}$ & ${}^{+0.01}_{-0.01}$ & ${}^{+0.02}_{-0.02}$ \\
            {Sc I} & ${}^{+0.04}_{-0.04}$ & ${}^{+0.02}_{-0.02}$ & ${}^{+0.04}_{-0.04}$ \\
            {Ti I} & ${}^{+0.02}_{-0.02}$ & ${}^{+0.02}_{-0.02}$ & ${}^{+0.04}_{-0.04}$ \\
            {Ti II} & ${}^{+0.08}_{-0.08}$ & ${}^{+0.06}_{-0.05}$ & ${}^{+0.07}_{-0.07}$ \\
            {V I} & ${}^{+0.02}_{-0.02}$ & ${}^{+0.01}_{-0.01}$ & ${}^{+0.02}_{-0.02}$ \\
            {Cr II} & ${}^{+0.04}_{-0.04}$ & ${}^{+0.02}_{-0.02}$ & ${}^{+0.04}_{-0.04}$ \\
            {Fe I} & ${}^{+0.01}_{-0.01}$ & ${}^{+0.00}_{-0.00}$ & ${}^{+0.01}_{-0.01}$ \\
            {Ni I} & ${}^{+0.04}_{-0.02}$ & ${}^{+0.03}_{-0.01}$ & ${}^{+0.05}_{-0.03}$ \\
            {Zn I} & ${}^{+0.04}_{-0.04}$ & ${}^{+0.02}_{-0.02}$ & ${}^{+0.04}_{-0.04}$ \\
            {Sr I} & ${}^{+0.00}_{-0.00}$ & ${}^{+0.04}_{-0.05}$ & ${}^{+0.07}_{-0.08}$ \\
            {Y II} & ${}^{+0.02}_{-0.04}$ & ${}^{+0.03}_{-0.03}$ & ${}^{+0.04}_{-0.04}$ \\
            {Zr I} & ${}^{+0.04}_{-0.05}$ & ${}^{+0.03}_{-0.03}$ & ${}^{+0.04}_{-0.06}$ \\
            {Zr II} & ${}^{+0.05}_{-0.05}$ & ${}^{+0.03}_{-0.03}$ & ${}^{+0.07}_{-0.07}$ \\
            {Mo I} & ${}^{+0.06}_{-0.06}$ & ${}^{+0.04}_{-0.04}$ & ${}^{+0.07}_{-0.07}$ \\
            {Ba II} & ${}^{+0.05}_{-0.06}$ & ${}^{+0.02}_{-0.02}$ & ${}^{+0.03}_{-0.04}$ \\
            {La II} & ${}^{+0.09}_{-0.09}$ & ${}^{+0.06}_{-0.06}$ & ${}^{+0.08}_{-0.09}$ \\
            {Ce II} & ${}^{+0.07}_{-0.07}$ & ${}^{+0.04}_{-0.03}$ & ${}^{+0.06}_{-0.06}$ \\
            {Nd II} & ${}^{+0.08}_{-0.18}$ & ${}^{+0.10}_{-0.17}$ & ${}^{+0.29}_{-0.37}$ \\
            {Eu II} & ${}^{+0.08}_{-0.00}$ & ${}^{+0.04}_{-0.04}$ & ${}^{+0.05}_{-0.05}$ \\
            \hline 
        \end{tabular}  
        \caption{Asymmetric abundance sensitivities for 22 elements in three representative stars across our parameter space.}
        \label{table: abundance sensitivity}
    \end{table}
\endgroup{}

\section{Validation of Linear Element Gradients}
\label{appendix:gradient validation}
\begin{figure}
    \centering
    \begin{subfigure}[b]{0.45\textwidth}
        \centering
        \includegraphics[width=\textwidth]{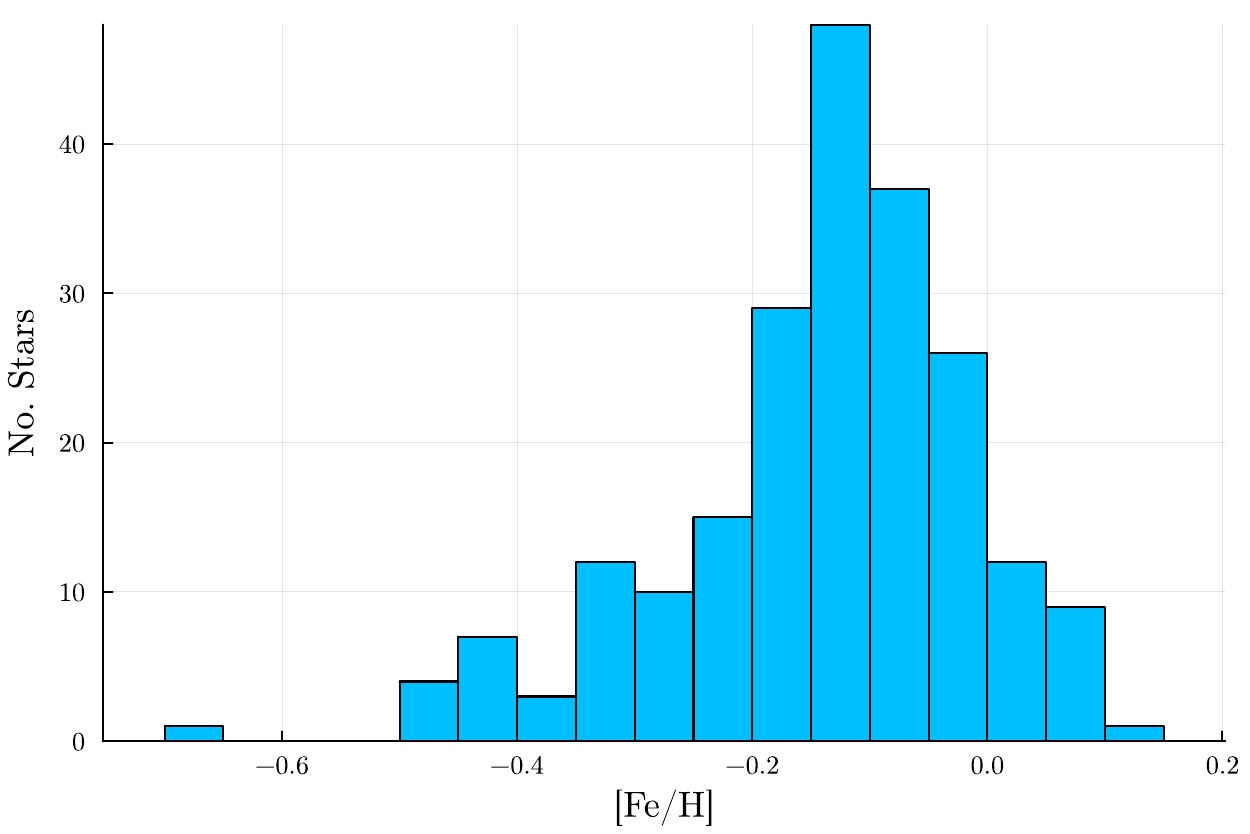}
    \end{subfigure}                                                                                            
    \hfill
    \begin{subfigure}[b]{0.45\textwidth}
        \centering
        \includegraphics[width=\textwidth]{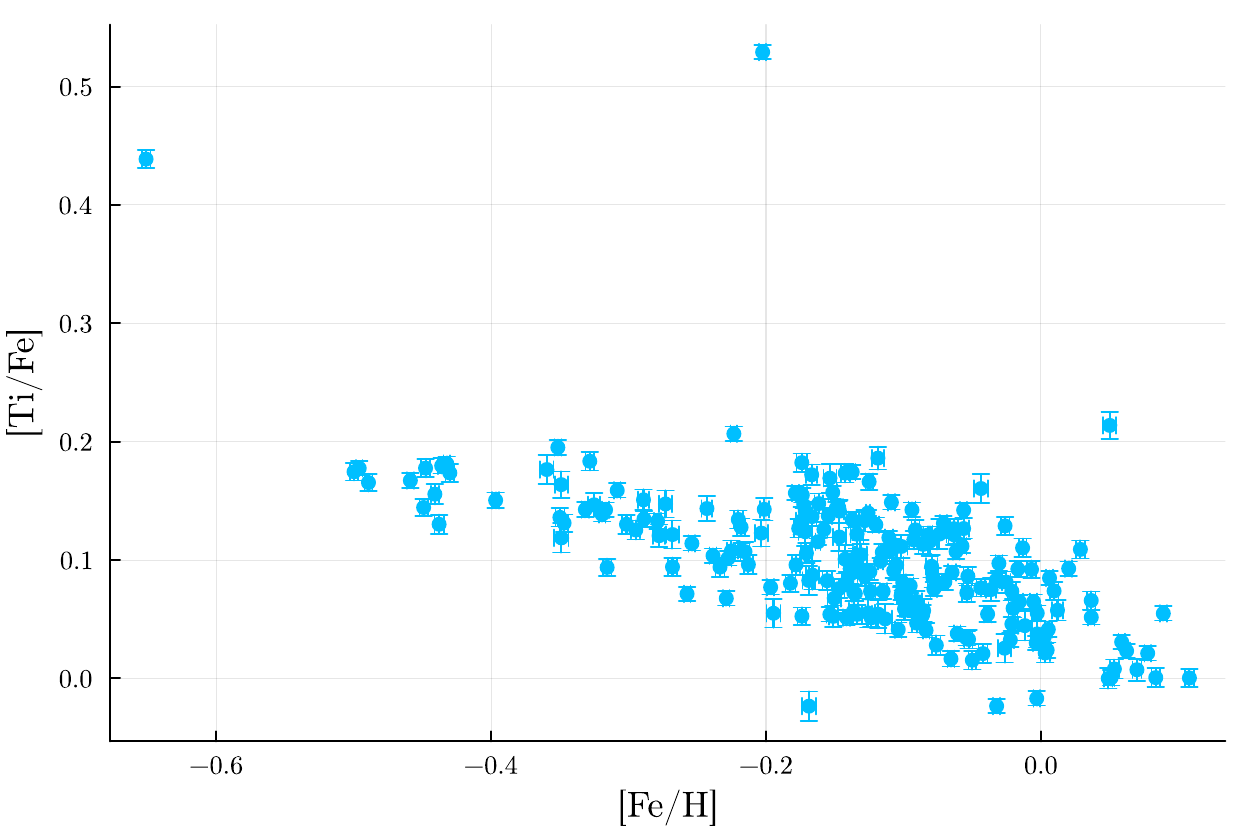}
    \end{subfigure}
    \caption{\textit{Top:} Metallicity distribution of the 214 stars used to derive our inter-element gradients. These stars are those with recorded abundances for the 22 elements included in this catalogue.
    \textit{Bottom:} [Ti I/Fe] abundances for the 214 stars used to derive our inter-element gradients. The range of Ti I abundances indicates that these stars are mostly thin disk.} 
    \label{fig: distribution of 214 stars}
\end{figure}

\begin{figure*}
    \centering

    \includegraphics[width=\textwidth]{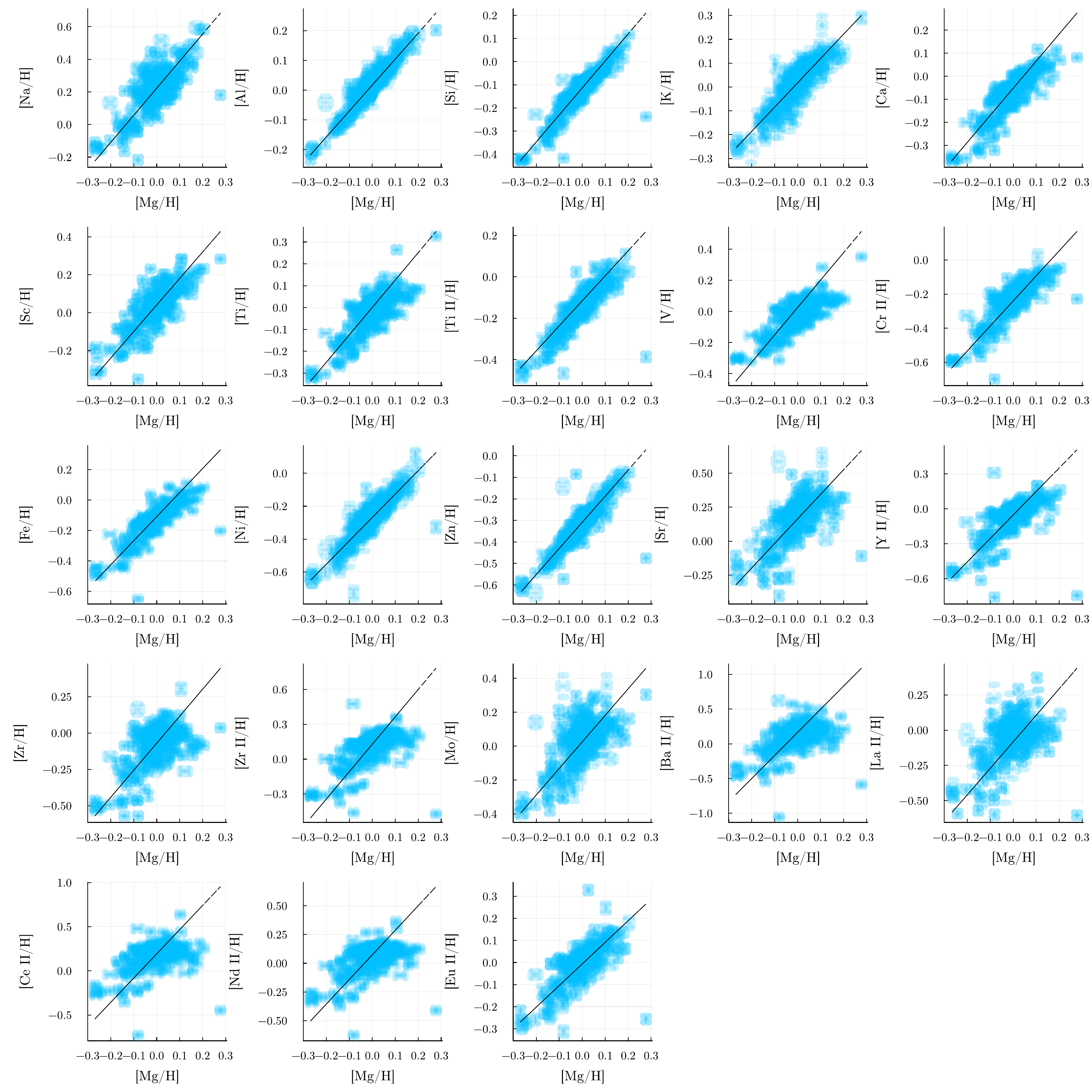}
    \caption{Our measured abundances for 21 elements with respect to our chosen representative element Mg I for hydrostatic $\alpha$-elements. The black line shows the linear fit to each relationship used to obtain our element-element gradients in Section \ref{sec:gradients}. We only included the 214 stars in our sample that have a measured abundance for each of the 22 elements.}
    \label{fig:Mg gradients}
\end{figure*}

We fitted gradients between our 22 measured element abundances and a chosen representative element for each nucleosynthetic family: Al I for odd-Z, Mg I for hydrostatic $\alpha$, Si I for explosive $\alpha$, Ni I for iron-peak, Y II for light s-process, Ce II for heavy s-process and Eu II for r-process elements. In these fits, we only include the 214 stars that have measured abundances for each of the 22 elements. We display the metallicity distribution and Ti I abundance trend for these 214 stars in Figure \ref{fig: distribution of 214 stars}. This demonstrates that these mostly thin-disk stars span a wide metallicity range required to calculate meaningful inter-element gradients. 
\\\\
To derive our inter-element gradients we assumed that all elements have a linear trend with respect to each of the representative elements. In Figure \ref{fig:Mg gradients} we show the abundance of each element with respect to our representative element Mg I to demonstrate the linearity between elements to validate this assumption. We visually inspected all element gradient fits and observed the same linearity.

\bsp	
\label{lastpage}
\end{document}